\documentclass[twocolumn,linenumber,trackchanges]{aastex631}

\newcounter{magicrownumbersCP}
\newcounter{magicrownumbersvetoed}

\newcommand\rownumber{\stepcounter{magicrownumbersCP}\arabic{magicrownumbersCP}}
\newcommand\rownumberspec{\stepcounter{magicrownumbersvetoed}\arabic{magicrownumbersvetoed}}

\usepackage{amsthm}

\usepackage{bbding}
\usepackage{longtable}
\usepackage{bm}
\usepackage{microtype}
\usepackage{graphicx,subfigure}
\usepackage{booktabs} 
\usepackage{enumitem}
\usepackage{amsmath}
\usepackage{multirow}
\usepackage{makecell}
\usepackage{slashbox}
\usepackage[linesnumbered, ruled, vlined]{algorithm2e}
\usepackage{algorithmic}
\usepackage{tabularx}
\newcolumntype{Y}{>{\centering\arraybackslash}X}

\newcounter{magicrownumbersCTOI}
\newcommand\rownumberctoi{\stepcounter{magicrownumbersCTOI}\arabic{magicrownumbersCTOI}}

\usepackage{array,etoolbox}

\setitemize{leftmargin=10pt,noitemsep,topsep=0pt,parsep=0pt,partopsep=0pt}
\usepackage{hyperref}

\usepackage[flushleft, para]{threeparttable}
\usepackage{booktabs}
\usepackage{array}

\newcommand{\ssymbol}[1]{$^{\color{red}\@fnsymbol{#1}}$}

\newcommand{\blue}[1]{{\color{blue} #1}}

\usepackage{scrextend}

\usepackage{comment}

\SetKwInput{KwInput}{Input}
\SetKwInput{KwOutput}{Output}

\usepackage{array}
\newcolumntype{P}[1]{>{\hspace{0pt}}p{#1}}

\newcommand{\kepler}{Kepler}

\newcommand{\TESSMission}{TESS Mission}
\newcommand{\vespa}{\texttt{vespa}} 
\newcommand{\ExoMiner}{\texttt{ExoMiner}}  
\newcommand{\ExoMinerplusplus}{\texttt{ExoMiner++}}  
\newcommand{\AstroNet}{\texttt{AstroNet}}  
\newcommand{\ExoNet}{\texttt{ExoNet}}
\newcommand{\GPC}{\texttt{GPC}}
\newcommand{\RFC}{\texttt{RFC}}

\newcommand{\Robovetter}{\texttt{Robovetter}}
\newcommand{\Autovetter}{\texttt{Autovetter}}

\newcommand{\TRICERATOPS}{\texttt{TRICERATOPS}}

\shorttitle{Transit Classification of TESS 2-min Data Using ExoMiner++}
\shortauthors{Valizadegan et. al.}

\begin{document}

\title{ExoMiner++: Enhanced Transit Classification and a New Vetting Catalog for 2-Minute TESS Data}

\correspondingauthor{Hamed Valizadegan}
\email{hamed.valizadegan@nasa.gov}

\author[0000-0001-6732-0840]{Hamed Valizadegan}
\affiliation{KBR Inc., Mountain View, CA 94043, USA}
\affiliation{NASA Ames Research Center (NASA ARC), Moffett Field, CA 94035, USA}

\author[0000-0002-2188-0807]{Miguel J. S. Martinho}
\affiliation{KBR Inc., Mountain View, CA 94043, USA}
\affiliation{NASA Ames Research Center (NASA ARC), Moffett Field, CA 94035, USA}

\author[0000-0002-4715-9460]{Jon M. Jenkins}
\affiliation{NASA Ames Research Center (NASA ARC), Moffett Field, CA 94035, USA}

\author[0000-0002-6778-7552]{Joseph D. Twicken}
\affiliation{The SETI Institute, Mountain View, CA  94043, USA}
\affiliation{NASA Ames Research Center (NASA ARC), Moffett Field, CA 94035, USA}

\author[0000-0003-1963-9616]{Douglas A. Caldwell}
\affiliation{The SETI Institute, Mountain View, CA  94043, USA}
\affiliation{NASA Ames Research Center (NASA ARC), Moffett Field, CA 94035, USA}

\author[0009-0003-0169-4841]{Patrick Maynard}
\affiliation{NASA Goddard Space Flight Center, Greenbelt, MD 20771, USA}
\thanks{Author contributed while interning at NASA ARC}

\author[0009-0001-4824-5766]{Hongbo Wei}
\affiliation{University of California, Berkeley, Berkeley, CA 94720, USA} 
\thanks{Author contributed while interning at NASA ARC}

\author[0000-0002-0157-7579]{William Zhong}
\affiliation{Yale University, New Haven, CT 06520, USA}
\thanks{Author contributed while interning at NASA ARC}

\author[0009-0008-1842-8837]{Charles Yates}
\affiliation{University of California, Berkeley, Berkeley, CA  94720, USA}
\thanks{Author contributed while interning at NASA ARC}

\author[0009-0007-7358-5723]{Sam Donald}
\affiliation{Pacific Northwest National Laboratory, Richland, WA 99354, USA}
\thanks{Author contributed while interning at NASA through $I\hat{\;}2$ program}

\author[0000-0001-6588-9574]{Karen A.\ Collins}
\affiliation{Center for Astrophysics \textbar \ Harvard \& Smithsonian, 60 Garden Street, Cambridge, MA 02138, USA}

\author[0000-0001-9911-7388]{David Latham}
\affiliation{Center for Astrophysics \textbar \ Harvard \& Smithsonian, 60 Garden Street, Cambridge, MA 02138, USA}

\author[0000-0003-1464-9276]{Khalid Barkaoui}
\affiliation{Astrobiology Research Unit, Université de Liège, 19C Allée du 6 Août, 4000 Liège, Belgium}
\affiliation{Department of Earth, Atmospheric and Planetary Science, Massachusetts Institute of Technology, 77 Massachusetts Avenue,\\ Cambridge, MA 02139, USA}
\affiliation{Instituto  de  Astrofísica  de  Canarias  (IAC),  Calle  Vía  Láctea  s/n, 38200, La Laguna, Tenerife, Spain}


\author[0000-0002-2830-5661]{Michael L.\ Calkins}
\affiliation{Center for Astrophysics \textbar \ Harvard \& Smithsonian, 60 Garden Street, Cambridge, MA 02138, USA}

\author[0009-0008-9182-7471]{Kylee Carden}
\affiliation{Department of Astronomy, The Ohio State University, 140 West 18th Ave., Columbus, OH 43210 USA}

\author[0000-0002-4070-7831]{Nikita Chazov}
\affiliation{Kourovka observatory \textbar \ Ural Federal University, 19 Mira street, Yekaterinburg, 620002, Russia}

\author[0000-0002-9789-5474]{Gilbert A.\ Esquerdo}
\affiliation{Center for Astrophysics \textbar \ Harvard \& Smithsonian, 60 Garden Street, Cambridge, MA 02138, USA}

\author[0000-0002-7188-8428]{Tristan Guillot}
\affil{Observatoire de la C\^ote d'Azur, UniCA, Laboratoire Lagrange, CNRS UMR 7293, CS 34229, 06304 Nice cedex 4, France}

\author[0000-0001-9388-691X]{Vadim Krushinsky} 
\affiliation{Laboratory of Astrochemical Research \textbar \ Ural Federal University, ul. Mira d. 19, Yekaterinburg, 620002, Russia}

\author[0000-0002-7031-7754]{Grzegorz Nowak}
\affiliation{Institute of Astronomy, Faculty of Physics, Astronomy and Informatics, Nicolaus Copernicus University, Grudzi\c{a}dzka 5, 87-100 Toru\'n, Poland}

\author[0000-0002-3627-1676]{Benjamin V.\ Rackham}
\affiliation{Department of Earth, Atmospheric and Planetary Sciences, Massachusetts Institute of Technology, 77 Massachusetts Avenue, Cambridge, MA 02139, USA}
\affiliation{Kavli Institute for Astrophysics and Space Research, Massachusetts Institute of Technology, Cambridge, MA 02139, USA}

\author[0000-0002-5510-8751]{Amaury Triaud} 
\affiliation{School of Physics \& Astronomy, University of Birmingham, Edgbaston, Birmingham, B15 2TT, UK}

\author[0000-0001-8227-1020]{Richard P.\ Schwarz}
\affiliation{Center for Astrophysics \textbar \ Harvard \& Smithsonian, 60 Garden Street, Cambridge, MA 02138, USA}

\author[0000-0003-4658-7567]{Denise Stephens} 
\affiliation{Department of Physics and Astronomy, Brigham Young University, N-486 ESC, Provo, UT 84602 USA}

\author[0000-0003-2163-1437]{Chris Stockdale}
\affiliation{Hazelwood Observatory, Australia}


\author[0000-0001-8621-6731]{Cristilyn N.\ Watkins}
\affiliation{Center for Astrophysics \textbar \ Harvard \& Smithsonian, 60 Garden Street, Cambridge, MA 02138, USA}

\author[0000-0003-2127-8952]{Francis P. Wilkin} 
\affiliation{Department of Physics and Astronomy, Union College, 807 Union St., Schenectady, NY 12308, USA}

\submitjournal{AJ}


\begin{abstract}
We present \ExoMinerplusplus, an enhanced deep learning model that builds on the success of \ExoMiner~\citep{Valizadegan_2022_ExoMiner} to improve transit signal classification in 2-minute TESS data. \ExoMinerplusplus\ incorporates additional diagnostic inputs, including periodogram, flux trend, difference image, unfolded flux, and spacecraft attitude control data, all of which are crucial for effectively distinguishing transit signals from more challenging sources of false positives. To further enhance performance, we leverage multi-source training by combining high-quality labeled data from the \kepler\ space telescope with TESS data. This approach mitigates the impact of TESS's noisier and more ambiguous labels. \ExoMinerplusplus\ achieves high accuracy across various classification and ranking metrics, significantly narrowing the search space for follow-up investigations to confirm new planets. To serve the exoplanet community, we introduce new TESS catalog containing \ExoMinerplusplus\ classifications and confidence scores for each transit signal. Among the 147,568 unlabeled TCEs, \ExoMinerplusplus\ identifies 7,330 as planet candidates, with the remainder classified as false positives. These 7,330 planet candidates correspond to 1,868 existing TESS Objects of Interest (TOIs), 69 Community TESS Objects of Interest (CTOIs), and 50 newly introduced CTOIs. 1,797 out of the 2,506 TOIs previously labeled as planet candidates in ExoFOP are classified as planet candidates by \ExoMinerplusplus. This reduction in plausible candidates combined with the excellent ranking quality of \ExoMinerplusplus\ allows the follow-up efforts to be focused on the most likely candidates, increasing the overall planet yield.
\end{abstract}

\section{Introduction}
\label{introduction}

The discovery and characterization of exoplanets have significantly expanded our understanding of planetary systems beyond the solar system. Among the primary methods for detecting these distant worlds is the transit method, which measures periodic dips in stellar brightness caused by an exoplanet crossing in front of its host star. This method provides crucial data on the exoplanet’s radius, orbital parameters, and can even offer constraints on atmospheric properties. NASA’s Kepler Space Telescope~\citep{borucki2010kepler} and the Transiting Exoplanet Survey Satellite~\citep[TESS;][]{ricker2015transiting} have been instrumental in advancing transit-based exoplanet research, providing large volumes of high-precision photometric data.

While Kepler focused on a single field to enable deep, long-term monitoring of stars, TESS conducts an all-sky survey targeting bright, nearby stars. TESS's shorter observation windows (27 days per sector) introduce unique challenges such as period aliasing, reduced signal-to-noise ratio (SNR), and greater susceptibility to instrumental noise. These differences necessitate advanced techniques for transit signal extraction and classification to fully exploit the scientific potential of TESS’s extensive, but shallower, dataset~\citep{Twicken-2020-TESS_Handbook}.

The volume and complexity of TESS data demand advanced analytical tools, with machine learning (ML), particularly deep learning techniques, emerging as crucial tools. Periodic transit-like features in light curves, called Threshold Crossing Events (TCEs), are automatically identified in TESS or \kepler\ data by the Science Processing Operations Center \citep[SPOC;][]{Jenkins2016SPIE} pipeline. However, not all TCEs are indicative of planets; many are false positives requiring further analysis. The task of classifying transit signals, which is the main focus of this work, involves distinguishing between signals attributable to planets and those arising from various false positive sources~\citep{Sullivan_2015_FPs}. Different machine learning classifier technologies have been employed for this task, including Random Forest~\citep{McCauliff_2015, armstrong-2020-exoplanet}, Gaussian Processes~\citep{armstrong-2020-exoplanet}, Deep Neural Networks~\citep{Valizadegan_2022_ExoMiner, shallue_2018, Ansdell_2018, Yu_2019_AstroNet, Tey_2023_astronet, Salinas_2023_transformer_TESS}, Logistic Regression~\citep{Valizadegan_2023_Multiplicity}, and others. 

Previous work~\citep{Valizadegan_2022_ExoMiner, Valizadegan_2023_Multiplicity, armstrong-2020-exoplanet, shallue_2018, Ansdell_2018, Osborn_2020_ExoNet, Tey_2023_astronet, Yu_2019_AstroNet} has demonstrated the efficacy of ML methods in analyzing TCEs for telescopes like \kepler\ or TESS. These models automate the classification of TCEs, streamlining the analysis pipeline and reducing the need for manual inspection. Beyond classification, machine learning approaches have been applied to validation~\citep{armstrong-2020-exoplanet, Valizadegan_2022_ExoMiner, Valizadegan_2023_Multiplicity, Tey_2023_astronet} and detection~\citep{Cui_2021_detection_TESS, hansen-2024-transit-detection-kepler} of transit signals. These methods play a crucial role in the validation of new exoplanets, particularly for TESS, where the mission has yielded a limited number of confirmed planet (CP) discoveries after six years.

Recent works, such as \ExoMiner~\citep{Valizadegan_2022_ExoMiner}, our previous CNN model, have made significant advancements by incorporating numerous diagnostic tests as inputs and validating hundreds of new exoplanets. Our present work represents a natural extension and adaptation of the successful \ExoMiner\ model to the TESS dataset. By leveraging the lessons learned from Kepler and incorporating them into the TESS framework, we aim to enhance the efficiency and accuracy of transit signal classification. For the first time, we include four new diagnostic tests as inputs to the model: difference image, flux trend, flux periodogram, and spacecraft reaction wheel desaturation event (``momentum dump'') data. These tests have proved crucial for human vetting and may prove useful for machine classification of transit signals but have seldom been adopted for machine use due to their complexity. This work not only refines the transit signal classification process for TESS, addressing its unique observational challenges, but also sets the stage for applying these methodologies to future transit surveys.

This paper is organized as follows: Section~\ref{sec:archeciture} introduces a general framework for machine learning classification of transit signals, discussing how existing and future classifiers can be viewed as specific instances of this framework. In Section~\ref{sec:exominer++}, we present \ExoMinerplusplus\ as a concrete implementation of this framework. Section~\ref{sec:data-processing} details the dataset used in this study and describes the data processing pipeline employed to prepare the inputs for \ExoMinerplusplus. The hyper-parameter optimization (HPO) process used to tune the parameters of \ExoMinerplusplus\ is covered in Section~\ref{sec:hpo}. Section~\ref{sec:performance} evaluates the classification and ranking performance of \ExoMinerplusplus, using its performance on \kepler\ as a baseline. Given that the model performs better on \kepler\ data, Section~\ref{sec:TESS_difficulty} explores the challenges that make the classification of TESS transit signals more difficult compared to those of \kepler. Section~\ref{sec:catalog} presents the \ExoMinerplusplus\ catalog, including labels and confidence scores for all TCEs and their corresponding TOIs. Finally, we present our conclusions in Section~\ref{sec:conclusion}.

\begin{figure*}[htb!]
\begin{center}
\centerline{\includegraphics[width=\linewidth]{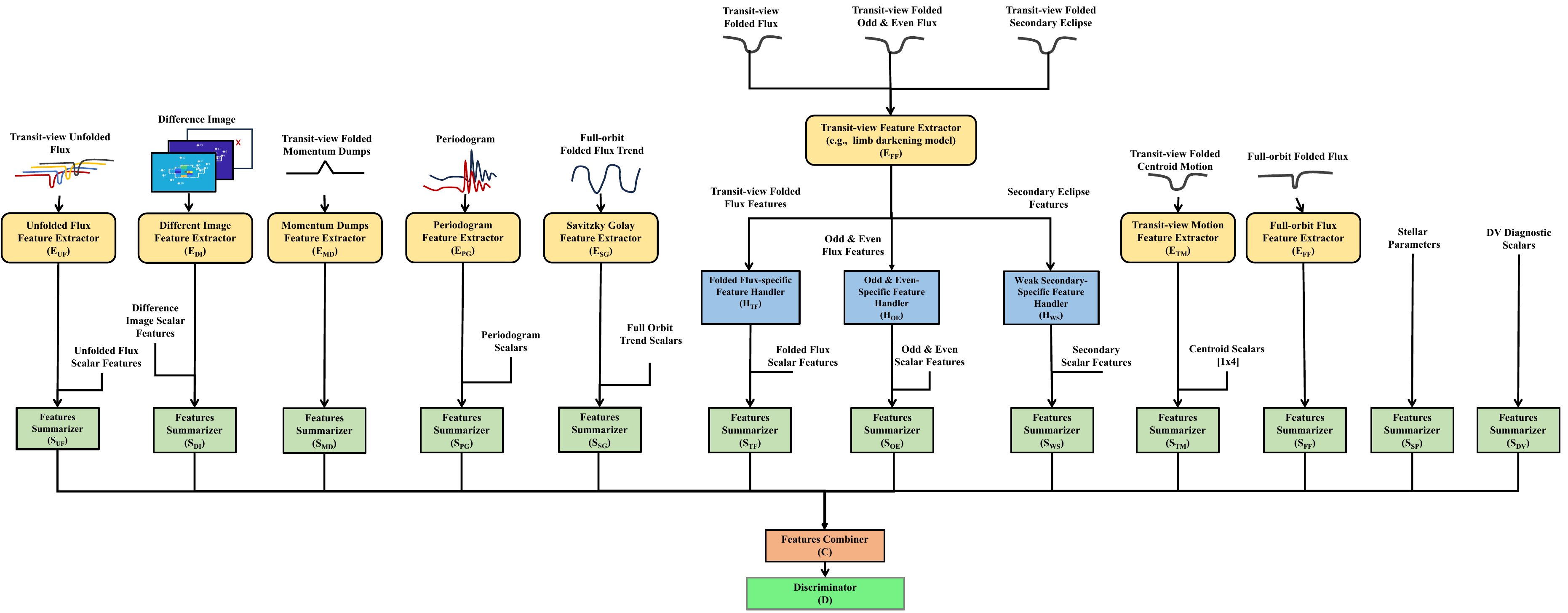}}
\caption{Classification Architecture.}
\label{fig:architecture}
\end{center}
\vskip -0.3in
\end{figure*}

\section{Machine Learning Classification of Transit Signals}
\label{sec:archeciture}

We developed \ExoMiner~\citep{Valizadegan_2022_ExoMiner} based on the observation that domain experts rely on various components of the \kepler\ and TESS Data Validation~\citep[DV;][]{Twicken_2018_DV, Li2019KeplerDataValidation2} reports to classify transit signals. We posited that an effective ML model should be guided by this expert-driven process. DV reports contain a series of diagnostic tests designed to identify different types of false positives (FPs), which are essential inputs for any ML model tasked with transit signal classification. Because \ExoMiner\ is a sophisticated deep learning model, it may be challenging for domain experts to fully interpret its decision-making process. In this section, we offer a high-level overview of how machine learning systems, informed by domain expertise, should be structured in general. We then explain how \ExoMiner\ follows such structural design.

To design a system capable of vetting a transit signal, it is necessary to process various types of data represented in the DV report, which include: 1) transit-view of the unfolded flux time series, 2) transit-view of the phase-folded flux time series, 3) full-orbit-view of the phase-folded flux time series, 4) transit-view of the phase-folded odd and even flux time series, 5) transit-view of the phase-folded weak secondary flux time series, 6) transit-view of the phase-folded flux-weighted centroid motion time series, 7) difference image, 8) transit-view of the phase-folded momentum dump time series, and 9) several other scalar values such as stellar parameters and DV tests.

The first step in using complex diagnostic inputs, such as time series and images, is to extract features that can discriminate between the classes of interest. This can be done manually by subject matter experts (SMEs) or automatically through a data-driven approach. For example, to extract features from the transit-view of diagnostic tests such as the phase-folded flux, one might use a limb-darkened transit fit model~\citep{Mandel_limbdarkened_2002}. To determine whether the odd and even test indicates a false positive scenario, one can estimate the transit depth again from a transit fit model and test whether the difference in transit depth between odd and even transits is statistically significant. Once the features/statistics/parameters are extracted and represented as scalar values for different diagnostic tests, they can be combined to classify the transit signals.

Figure~\ref{fig:architecture} shows our proposed architecture, demonstrating a general process that can be used to classify transit signals. As enumerated in the previous paragraph, this system has different diagnostic tests\footnote{A DV report does not include the periodogram but includes flux before de-trending.} as its input. It consists of five main building blocks to process different diagnostic tests and perform classification, discussed below:

\begin{enumerate}
    \item {Feature extractor:} This building block is necessary to extract features from complex data types such as images and time series. Different diagnostic tests may require their own specific feature extractors; however, diagnostic types of the same kind might benefit from shared feature extractors. In our model, transit-view of unfolded flux, difference image, transit-view of phase-folded centroid motion time series, full-orbit-view of phase-folded flux, and transit-view of phase-folded momentum dump each have their own feature extractor model. Conversely, the transit-view of phase-folded flux, weak secondary, and odd \& even time series share the same feature extractor because we are interested in the characteristics of the transit for these three diagnostic tests. We will discuss this choice in more detail later in Section~\ref{sec:exominerplusplus}.
    \item{Test-specific feature handling:} This component processes features extracted for specific diagnostic tests if special treatment is required. This is only done for transit-view of phase-folded flux, weak secondary, and odd \& even time series that share the same feature extractor. For example, after estimating  parameters such as transit depth using a limb-darkened transit fit model for odd and even separately, we need to subtract them to determine whether they are statistically different or not.
    \item{Feature summarization:} This component summarizes the features extracted by the feature extractor block. This is necessary to obtain more complex features on top of those obtained from the feature extractor block and provides a way to combine relevant scalar values with features extracted from complex data types such as time series or images. For example, in the presence of a weak secondary transit, we need to know whether it is due to an eclipsing binary (EB) star or a large exoplanet exhibiting thermal emission, Doppler boosting, and/or ellipsoidal variations. The feature summarization component combines the features extracted from the weak secondary test with scalar values such as the secondary geometric albedo and planet effective temperature comparison statistics~\citep{Twicken_2018_DV, Jenkins2020KeplerHandbook}, and other required variables to build the final features for this test.
    \item{Feature Combiner:} This component receives the features summarized by the feature summarizers of various diagnostic tests and identifies any high-level complex relationships (linear or nonlinear) between features. It generates a series of final features in which the two classes of interest (exoplanet or false positive) are linearly separable. This component could be a simple model such as principal component analysis~\citep[PCA; ][]{Jolliffe-2016-PCA} or a more complex model such as a neural network with millions of parameters~\citep{NIPS_2012_alexnet}.
    \item{Classification:} This could be a simple linear classifier that receives the highly discriminating features from the previous component and differentiates between the two or more classes of interest. 
\end{enumerate}

Traditionally, feature extraction and classification have been treated as separate tasks. First, a pool of candidate features, such as diagnostic tests and their statistics, are proposed by SMEs or extracted by automated feature extractors designed by data engineers. Then, a classifier is applied to these extracted features to distinguish between different classes. However, these two steps are closely interconnected. The discriminative power of the extracted features influences the choice of classifier, and in turn, more effective features can be designed with the classifier's characteristics in mind.

Bayesian classifiers, such as \vespa~\citep{Morton-2016-vespa} and \TRICERATOPS~\citep{Giacalone_2020_TRICERATOPS}, which decompose the posterior into prior and likelihood, are theoretically optimal for sets of independent input variables~\citep{Bishop:2006:PRM:1162264}. However, they face practical challenges. They rely heavily on assumptions about the data, including the accuracy of priors, the form of the likelihood (e.g., Gaussian), and the independence of features. These assumptions make Bayesian classifiers less practical for high-dimensional data, where the number of features exceeds a few, unless strong assumptions are made.

In contrast, modern machine learning classifiers aim to address these challenges by optimizing the posterior directly, without decomposing it into prior and likelihood. This approach has led to the development of powerful discriminative classifiers, such as Support Vector Machines (SVMs)\citep{Vapnik-statisticallearningtheory-1998}, Random Forest\citep{Breiman-2001-randomforest}, and Gradient Boosting~\citep{Schapire-1990-boosting}. These classifiers are more flexible and better suited for high-dimensional data, but they still assume that feature extraction and classification are separate steps.

To address this limitation, an integrated approach can be employed, where both feature extraction and classification are optimized simultaneously. This is achieved by defining a parametric functional form for each component of the model architecture, resulting in a complex, nested function. By defining an objective function, such as a surrogate for the error rate, the parameters of this function can be optimized using a training set, automating the process of feature extraction and classification.

Before the advent of the backpropagation algorithm~\citep{schmidhuber2022-annotatedhistorymodernai}, optimizing such nested architectures was a significant challenge. The issue lies in the fact that the optimal output (the ground truth) is only available for the final output of the architecture, not for internal components like feature extractors. Backpropagation solves this by utilizing the chain rule to propagate the gradient from the final output backward through the network, enabling simultaneous optimization of the entire model, including both feature extraction and classification. This breakthrough underpins the success of modern deep learning models.

The form of the parametric function for each component is another important consideration. Neural networks, which are supported by the Universal Approximation Theorem~\citep{HORNIK-1989-NN-universal}, offer a powerful approach. This theorem ensures that a neural network with at least one hidden layer can approximate any function to an arbitrary degree of accuracy, provided it has sufficient units. Different types of neural network layers can be applied to each component in the architecture, and we will explore this further in Section~\ref{sec:deeplearningimplmentatopn}.

\subsection{Existing Classifiers}
\label{sec:existing-classifiers}
In this section, we demonstrate how the general architecture (or its minor variation) introduced in Section~\ref{sec:archeciture} includes the existing machine classifiers. Existing classifiers are different in two major ways: 1) the inclusion of different diagnostic tests represented by different branches and 2) the implementation of different components of this architecture. For example, while some models such as deep learning based classifiers do feature extraction and classification simultaneously, other classifiers assume that such features are already extracted before doing the task of classification. Table~\ref{table:classifiers-summary} demonstrates this difference. If the underlying function is not a deep neural network, a separate feature extractor needs to be employed a-priori before the classification is done; the approach employed by most existing exoplanet transit signal classifiers~\citep{Coughlin2017robovetter, Jenkins-Autovetter-2014IAUS, McCauliff_2015, armstrong-2020-exoplanet, Morton-2016-vespa, Giacalone_2020_TRICERATOPS}. These classifiers differ in the type of functions they employ for the classification which includes expert system if-then rules~\citep{Coughlin2017robovetter}, random forest~\citep{McCauliff_2015, armstrong-2020-exoplanet}, and Bayesian classifiers~\citep{Morton-2016-vespa, Giacalone_2020_TRICERATOPS}. 

As shown in Table~\ref{table:classifiers-summary}, the existing classifiers also differ in the range of diagnostic tests they use.

\begin{table*}[htb]
\centering
\caption{Comparison of different models. \Asterisk~means that the model uses the input diagnostic test directly. \PlusThinCenterOpen~ means that the model uses extracted features (i.e., statistics and other parameters) in the form of scalar values from the diagnostic test.}
\label{table:classifiers-summary}
\scriptsize\setlength{\tabcolsep}{3pt}
\begin{threeparttable}
\begin{tabularx}{\linewidth}{c|ccccccccccc}
\toprule
& \vespa & \Robovetter & \Autovetter & \AstroNet-2018 & \ExoNet & \GPC & \RFC & \TRICERATOPS & \ExoMiner & \AstroNet-2023  & ExoMiner++\\
\midrule
Stellar parameters  & \PlusThinCenterOpen & \Asterisk & \Asterisk & &\Asterisk &\Asterisk &\Asterisk &\PlusThinCenterOpen & \Asterisk & \Asterisk & \Asterisk\\
DV Diagnostic scalars & & \Asterisk & \Asterisk & & & \Asterisk  &\Asterisk & &\Asterisk &  & \Asterisk \\
Other Scalar Values & &  &  & & &  &  &  &  & \Asterisk \\
\midrule
Unfolded Flux & &  &  &  &  &  & & & & \Asterisk & \Asterisk\\
Phase-folded Flux & \PlusThinCenterOpen & & & \Asterisk & \Asterisk & \Asterisk & \PlusThinCenterOpen & \PlusThinCenterOpen & \Asterisk & \Asterisk & \Asterisk\\
Odd \& Even Flux & & \PlusThinCenterOpen & \PlusThinCenterOpen &  &  & \PlusThinCenterOpen &  \PlusThinCenterOpen & &\Asterisk & \Asterisk & \Asterisk \\
Weak secondary Flux & & \PlusThinCenterOpen & \PlusThinCenterOpen & &  & \PlusThinCenterOpen & \PlusThinCenterOpen& & \Asterisk & \Asterisk & \Asterisk \\
Centroid Motion & & \PlusThinCenterOpen & \PlusThinCenterOpen & & \Asterisk & & & & \Asterisk &  & \Asterisk\\
Difference image & & \PlusThinCenterOpen & \PlusThinCenterOpen & & & \PlusThinCenterOpen  & \PlusThinCenterOpen & & \PlusThinCenterOpen & & \Asterisk \\
\midrule
Momentum Dump & &  &  & &  & & &  &  & & \Asterisk\\
Periodogram & &  &  & &  & & &  &  & & \Asterisk\\
Flux Trend & &  &  & &  & & &  &  & & \Asterisk\\
\bottomrule
\end{tabularx}
\end{threeparttable}
\end{table*}

\section{ExoMiner++}~\label{sec:exominer++}
\label{sec:exominerplusplus}
Here we present \ExoMinerplusplus, an enhanced version of \ExoMiner\ with multiple improvements over the original model. \ExoMinerplusplus\ incorporates additional data inputs from DV reports, including unfolded flux, difference image, and momentum dump data, along with new inputs such as flux periodogram and flux trend data to improve false positive identification. In addition to these expanded inputs, \ExoMinerplusplus\ features architectural modifications informed by domain expertise, further refining its classification accuracy and interpretability.


\subsection{Deep Learning Implementation}
\label{sec:deeplearningimplmentatopn}
\begin{figure*}[htb!]
\begin{center}
\centerline{\includegraphics[width=\linewidth]{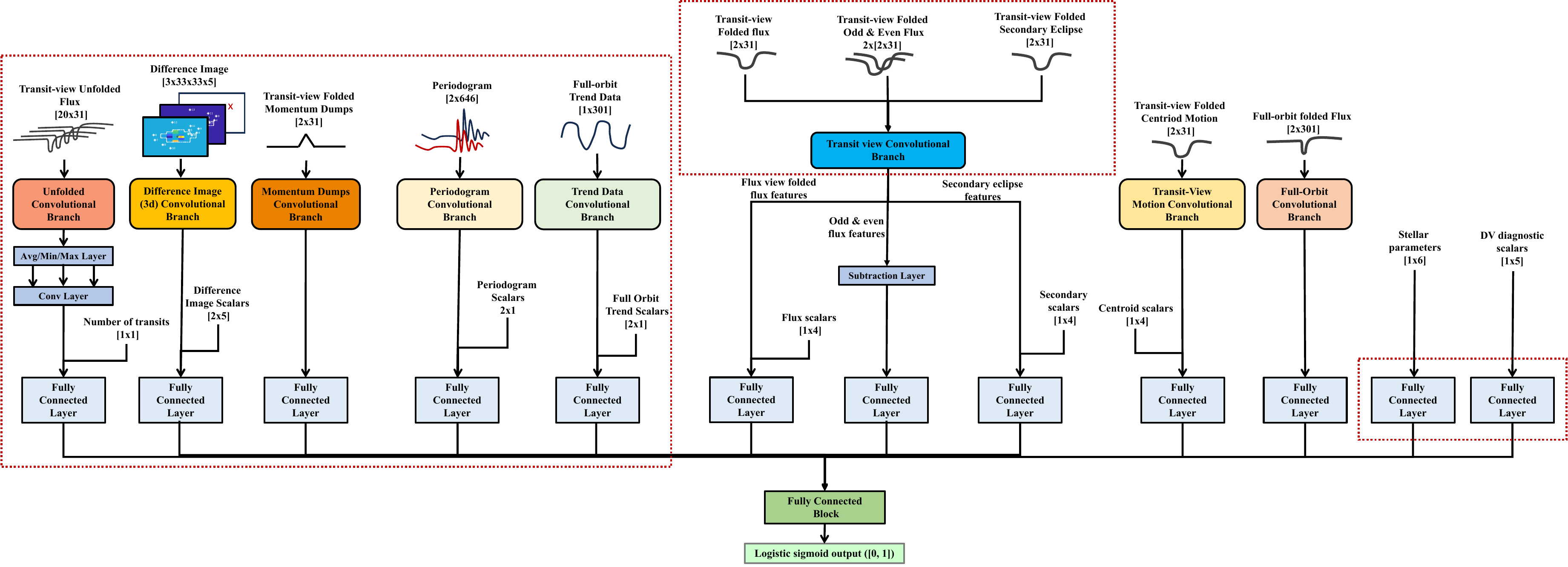}}
\caption{Deep Learning Implementation of the Classification Design in Figure~\ref{fig:architecture}. The dotted red boxes indicate the difference between \ExoMinerplusplus\ and original \ExoMiner~\citep{Valizadegan_2022_ExoMiner}.}
\label{fig:DNN-architecture}
\end{center}
\vskip -0.3in
\end{figure*}

After introducing the general architecture of Figure~\ref{fig:architecture}, we need to fix the details of the components of this architecture. Similar to our previous \ExoMiner\ model~\citep{Valizadegan_2022_ExoMiner}, we use neural network layers for different components of this architecture. The new architecture, however, has multiple important improvements over the earlier version. Figure~\ref{fig:DNN-architecture} shows the new architecture where the major changes from the original one are highlighted using red dotted rectangles. These changes include:
\begin{enumerate}
    \item We introduced five additional branches to the framework that include: Transit-view Unfolded Flux, Difference Image, Transit-view Folded Momentum Dumps, Periodogram, and Full-orbit-view Folded Flux Trend branches (Figure~\ref{fig:DNN-architecture}). For each branch, we performed multiple iterations over the design of a minimal stand-alone DNN model (without the use of other branches) in order to design the most suitable branch. Detailed explanations of these branches are provided in Section~\ref{sec:preprocessing}. 
    \item We integrated segments of the convolutional branches from three distinct original branches: Transit-view Folded Flux, Secondary Eclipse Flux, and Odd \& Even Flux (middle part of the architecture in Figure~\ref{fig:DNN-architecture}). All three of these diagnostic tests involve using as input the transit-view of their respective phase-folded flux time series. Furthermore, in all three cases we are interested in capturing potential transit signatures (or their absence). As the convolutional part of their processing branch functions as a low-level feature extractor, merging them ensures that the same type of low-level features (e.g., existence and shape of the transit) are shared across similar diagnostic tests. Despite this convergence, each diagnostic test retains its individual high-level feature extractor unit. These units remain capable of extracting unique features tailored to the specific requirements of each test. 
    \item We added a measure of uncertainty in the estimated average value of each bin. This uncertainty is computed during the binning process for each phase-folded time series. This added another channel to the time series inputs. For example, in the original model, the input for the Transit-view Folded Secondary Eclipse Flux branch was $1 \times 31$ whereas in the new model it is $2 \times 31$. This uncertainty time series has the same dimensionality as the average binned time series and it is fed to the model as an additional second channel. The dimensionality of these phase-folded and binned time series inputs is shown in Figure~\ref{fig:DNN-architecture}. We discuss this in more detail in Section~\ref{sec:preprocessing}. 
    \item We removed the transit depth from the Transit-view Folded Flux branch due to the significant difference in transit depth distributions between the two missions, which could lead to challenges in transfer and multi-source learning caused by distribution shift. However, since self-normalization~\citep{Valizadegan_2022_ExoMiner} results in the loss of transit depth information, we introduced the minimum value of the Transit-view Folded Flux as a scalar input to this branch to retain crucial depth information and the original amplitude of the signal before normalization.
    \item We removed the weak secondary depth from the Transit-view Folded Secondary Eclipse Flux branch and replaced it with the minimum value of this time series for the same reason as what was done for the primary transit counterpart. As a result, this branch now includes four scalar values: the minimum value, secondary Multiple Event Statistic (MES), geometric albedo and planet effective temperature comparison statistics~\citep{Twicken_2018_DV, Jenkins2020KeplerHandbook}.
    \item We removed multiple scalar values from the Transit-view Folded Centroid Motion branch because they are either not accurate or not used for the TESS mission. These are the flux-weighted centroid motion statistics and the difference image centroid offset from the out-of-transit (oot) centroid and its uncertainty. The flux-weighted centroid motion statistic is not computed in the Data Validation (DV) module of the SPOC pipeline for the TESS mission, while the centroid offset from the oot centroid computed from the difference image is not reliable for TESS as it was for \kepler~(due to substantial crowding in the photometric apertures of many TESS targets). We instead added the following two new scalar values in this branch: 1) Target star magnitude, $mag$, required to identify saturated stars for which the centroid diagnostic test is invalid, 
    and 2) Renormalized Unit Weight Error~\cite[RUWE, ][]{Lindegren-2021-GAIA}, which is an indicator of the potential stellar multiplicity of the target. 
    This leads to a total of 4 scalar values for this branch.
    \item We removed the rolling band count statistic \citep{Cleve-KeplerInstrumentHandbook-2016} for level zero from the DV Diagnostic branch since the TESS cameras do not experience rolling band noise. We instead added MES, Max Single Event Statistic (SES), Robust Statistic, and Goodness of Fit Chi Square Statistic~\citep{Twicken_2018_DV, Jenkins2020KeplerHandbook,Seader_2013,Seader_2015}, all of which are key diagnostics within the transiting planet search (TPS) used to quantify the validity of the putative signal. This led to a total of 9 scalar values for this branch.
    \item We updated the architecture by adding a fully connected (FC) layer at the end of both stellar parameters and DV diagnostic branches to make sure the contribution of each branch is learned through training instead of being dictated by the architecture. Similar to the other convolutional branches, these FC layers have 4 units each.
    \item We use a Savitzky-Golay filter to detrend time series data. In the original ExoMiner model, we used a spline fit to remove low-frequency signals such as stellar variability, following the approach in~\cite{shallue_2018}. However, we found that there are many TESS targets with higher frequency signals including sinusoidal stellar variability~\citep{Fetherolf_2023}. Such signals can prevent the model from examining the transits hidden in the data. Thus, we replaced the spline filter with the more manageable Savitzky-Golay filter~\citep{Savitzky-1964-filter},  based on local least-squares fitting of the data by polynomials, to better remove higher frequency stellar variability. 

\end{enumerate}
Some of the minor changes (Items 3-7) in this architecture are reflected in a recent version of \ExoMiner\ used to classify multi-planet systems from Kepler~\citep{Valizadegan_2023_Multiplicity}.

The most important change to the \ExoMiner\ model~\citep{Valizadegan_2022_ExoMiner} is Item 1 above for which we provide some details below:
\begin{itemize}
    \item Added Transit-view Unfolded Flux branch: This branch is tasked with processing and extracting information from unfolded flux data. Unfolded flux data are used by SMEs to study the consistency of the transits over different periods. We employed both  a Transformer~\citep{vaswani-2023-transformer} and a conventional branch for this purpose. After implementing and testing both approaches, we determined that the convolutional branch outperformed the Transformers. Our convolutional branch for unfolded flux data includes a set of convolutional blocks for feature extraction at each phase, an assessment block to evaluate the differences in the extracted features, and a final convolutional layer to extract features on top of the previous assessment block. In the assessment block, we use specific layers to compute the maximum, minimum, and mean values of the features. The details of this branch are depicted in Figure~\ref{fig:DNN-architecture}. We discuss this in more detail in Section~\ref{sec:preprocessing}.
    
    \item Added Difference Image branch: In-transit versus out-of-transit difference imaging is essential for detecting offsets in the location of the transit source relative to the target star that may indicate the influence of background objects or nearby stars. A significant difference image centroid offset often suggests that the observed light variation is not due to a genuine planetary transit on the target, but rather to contamination from other sources. Figure~\ref{fig:diff_img_data_dvm_example} illustrates the usefulness of difference imaging for identifying transits occurring in a nearby star. In this case, the source of TESS SPOC TCE TIC 309787037-1-S35\footnote{To refer to a TESS SPOC TCE, we use the naming form of x-y-Sz where x is the target TIC ID, y is the TCE planet number set by the SPOC pipeline, and z is the sector run (if TCE comes from a multi-sector run, then Sz1-z2, where z1 and z2 are the start and end sectors of the sector run, respectively) where it was detected.} is not the target star TIC 309787037, but a known nearby fainter star that is located in an adjacent pixel that is included in the photometric aperture. This ability to identify and quantify centroid offsets enhances TESS's and \kepler's capacity to filter out FPs, thereby improving the reliability of exoplanet candidate confirmations and providing deeper insights into their characteristics. Thus, we added the Difference Image branch to process difference image data. 

\begin{figure}[htb!]
\begin{center}
\centerline{\includegraphics[width=\columnwidth]{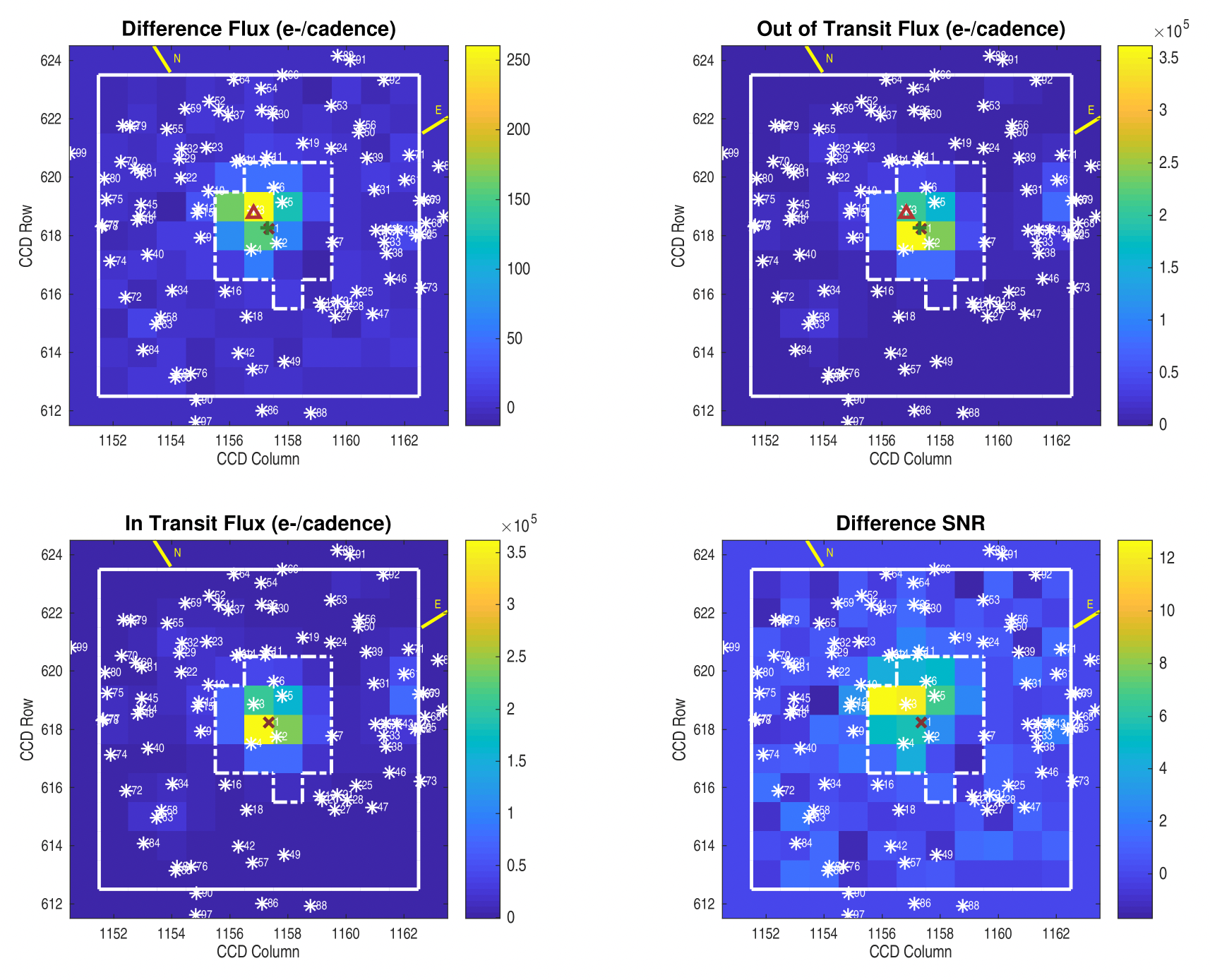}}
\caption{Difference image data for TESS SPOC TCE TIC 309787037-1-S35 (TOI 1046.01, a nearby EB), as shown in the DV mini-report. From left to right, top to bottom, the images show the difference flux, out-of-transit flux, in-transit flux, and difference SNR flux images. The red triangle marks the estimated transit source location, the red cross indicates the TIC's coordinates in the CCD frame, the green plus represents the out-of-transit centroid estimate, and the white asterisks denote the positions of neighboring stars in the postage stamp. The photometric aperture is outlined with a dashed white line, while the target mask is indicated by a solid white line.}
\label{fig:diff_img_data_dvm_example}
\end{center}
\end{figure}

    \item Added Periodogram branch: We included the periodogram of the Pre-search Data Conditioning SAP (PDCSAP) flux~\citep{smith2012kepler} as a tool to detect and analyze periodic signals in time series data. A periodogram measures the power or variance of the data at different frequencies (or periods), providing insights into power vs. frequency, identification of periodic signals, dominant frequencies, signal strength, and noise harmonics. We included the periodogram to introduce additional data that might not be captured by other diagnostics (e.g., possible information regarding the presence and characteristics of other transits in the system), aiming to enhance performance, particularly in distinguishing brown dwarfs (BD), single-lined spectroscopic binaries (SB1), and double-lined spectroscopic binaries (SB2). 
    
    In single-planet systems, exoplanets generally produce cleaner, single periodogram peaks with lower amplitude variations compared to BDs or binaries, which tend to show stronger and more complex signals. Due to their higher-masses, BDs, SB1, and  SB2 systems can display multiple periodogram peaks or harmonics, as they induce larger ellipsoidal variations in their parent star, and potentially exhibit Doppler beaming and higher thermal self-emission. However, in multi-planet systems, the presence of additional transits may also lead to a more complex periodogram, with multiple peaks corresponding to the different orbital periods of the planets. In such cases, while the overall signal might be more intricate, the presence of multiple transits can actually indicate that the target transit is more likely to be planetary. This makes the periodogram a useful tool in identifying and analyzing the signals from different types of objects, particularly in more complex systems.
    

    \item Added Full-orbit-view Folded Flux Trend branch: Given that detrending might remove some informative part of the signal~\citep{Stumpe_2012_detrending, Twicken-2010-detrending, Morris-2020-detrending}, we build a new branch that receives the Savitzky-Golay fit component as input. The idea is to provide information to the model about components within the flux time series that occur at the time scale of the detected orbital period of the corresponding transit signal, and were fitted during the detrending process. Figure~\ref{fig:pdcsap_flux_trend_correctly_classified_example} showcases the flux time series for TIC 167526485 in sector 6.  We can observe a sinusoidal signal (at half period of the transit event) caused by the ellipsoidal variations  that happen in this system due to the gravitational interaction of the binary components.  For most exoplanets, these variations will not be in phase with the transit, and so the phase-folded trend will not show such variations. Exceptions to this are massive planets such as hot Jupiters, and also other sources of FPs like BDs. Other potential signals that can show up in the flux trend time series include unmasked transits. Although we mask transits from detected TCEs in the light curve before detrending the time series, uncertainties in ephemerides might lead to the total or partial exclusion of real transits from the detrended time series. Furthermore, unidentified transits are not masked and might be fitted by the detrending model, thus showing up in these data.

\begin{figure*}[htb!]
{\includegraphics[width=\linewidth]{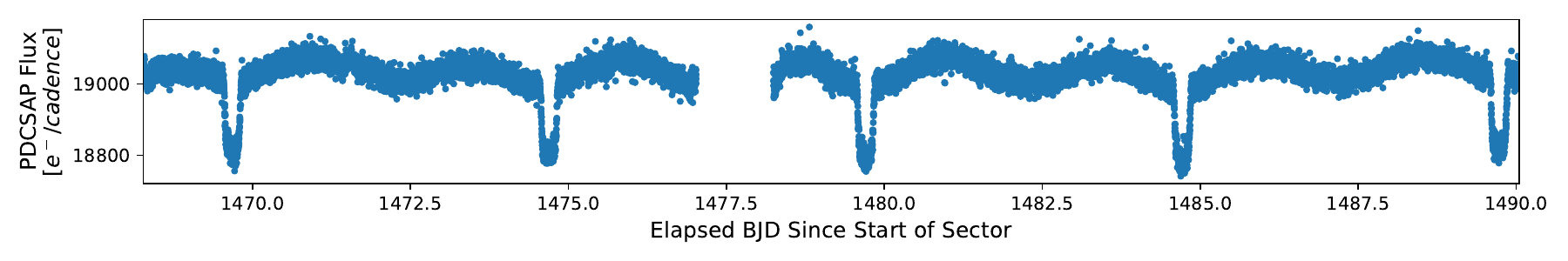}}
\caption{PDCSAP flux for TIC 167526485 in sector 6.}
\label{fig:pdcsap_flux_trend_correctly_classified_example}
\end{figure*}
    
    \item Added Momentum Dump branch: This branch is dedicated to processing momentum dump data. Spacecraft reaction wheel momentum management cycles are known to introduce artifacts in the data due to changing pointing behavior. These artifacts can occasionally be identified by the SPOC pipeline as TCEs.  Figure~\ref{fig:momentum_dump_pdcsap_flux_example} highlights the momentum dump events for the flux time series for TIC 82707763 in sector 37. These events are closely aligned with the transits. We implemented a convolutional branch, akin to those used for processing transit view data in the original \ExoMiner\ model for this particular data input. We discuss this in more detail in Section~\ref{sec:preprocessing}.

\begin{figure*}[htb!]
{\includegraphics[width=\linewidth]{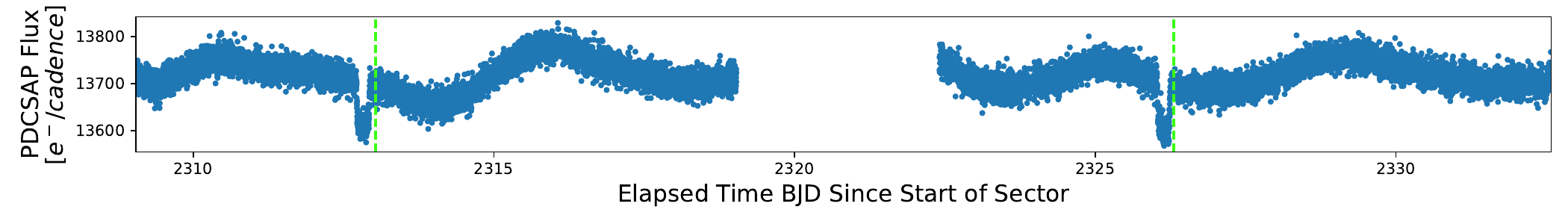}}
\caption{PDCSAP flux for TIC 82707763 in sector 37. The occurrence of momentum dump events is shown as vertical dashed green lines.}
\label{fig:momentum_dump_pdcsap_flux_example}
\end{figure*}

\end{itemize}

We will study the effect of each of these new branches in the performance of the model in Section~\ref{sec:new_branches}. 

\section{Data Processing}
\label{sec:data-processing}

\subsection{Data Preparation}
\label{sec:data-preparation}
In the processing of TESS 2-minute data, the workflow comprised two phases. Initially, data collection spanned Year 1 through 4, encompassing sectors 1--55. This yielded approximately 125,000 SPOC TCEs from 71 transit searches, including 55 single-sector and 16 multi-sector pipeline runs. These data formed the basis for the initial set of experiments for model and experiment development. In the second phase, data from Year 5 sectors 56--67 were incorporated, totaling nearly 208,000 SPOC TCEs across 85 transit searches, with 18 being multi-sector runs.

We leveraged \kepler\ labeled data to address annotation limitations in TESS data, but still found it necessary to use TESS-specific data for model training and evaluation. Consequently, we sought reliable and credible sources of TESS labels to minimize label noise and potential interference with model training and evaluation. The labels of the TESS 2-minute SPOC TCE dataset were regularly updated using the following procedure:

\begin{enumerate}
    \item We first obtain reliable labels from the publicly available Exoplanet Follow-up Observing Program (ExoFOP) TESS Object of Interest (TOI) \href{https://exofop.ipac.caltech.edu/tess/}{catalog}\footnote{\url{https://exofop.ipac.caltech.edu/tess/}} in January 2024. This catalog is regularly updated with new TOIs alerted by the TESS Science Office \citep{guerrero2021TOI} and then dispositioned by the TESS Follow-up Observing Program Working Group (TFOPWG) based on a variety of observations, including but not limited to TESS data (e.g., radial velocity measurements). The dispositions in this catalog vary in certainty. For instance, planet candidate (PC) and ambiguous planet candidate (APC) TOIs are not confirmed as planets or non-planets, making them unusable for our training and evaluation purposes. In contrast, known planets (KP) — i.e., planets identified outside the TESS mission — and confirmed planets (CP) — i.e., planets identified and confirmed during the TESS mission — offer a high degree of certainty, minimizing label noise in these categories. The CP category also includes BDs. However, to accurately assess our classifier's performance, we label BDs as non-planets in our dataset, ensuring a more precise evaluation in the presence of BDs.

    The FP category in the ExoFOP catalog is based on evidence from both photometric and spectroscopic data. Typically, one of these sources provides conclusive evidence leading to an FP disposition. In summary, we include TCEs labeled as KP, CP, or FP from ExoFOP and exclude those labeled as PC or APC. TCEs without an ExoFOP label are carried forward to the next steps for further labeling.
    \item As the next step for the TCEs without any labels (including PC or APC), we use the TESS EB catalog from \href{https://archive.stsci.edu/hlsp/tess-ebs}{Villanova}\footnote{\url{https://archive.stsci.edu/hlsp/tess-ebs},\\ \url{https://iopscience.iop.org/article/10.3847/1538-4365/ac324a}}~\citep{Prsa_2022-EB-catalog}. This is a catalog of about 4,500 EBs observed during the first two years of the TESS survey. These EBs were manually classified as likely being caused by an on-target EB signal based on light curve shape alone. Thus, the labels are not perfect. 
    \item For those TCEs without a label in the previous two catalogs, we label them as non-planets if they did not pass the TESS-ExoClass (\href{https://github.com/christopherburke/TESS-ExoClass}{TEC}\footnote{\url{https://github.com/christopherburke/TESS-ExoClass}}) flux triage step. These TCEs are deemed non-transiting phenomena (NTP). This catalog was built on the results of TEC up to sector 41. 
\end{enumerate} 

To utilize labels from the enumerated catalogs above, we performed matching between TESS SPOC TCEs and the transit signals listed in these catalogs based on their ephemerides (i.e., matching TCEs and transit objects from these catalogs associated with the same TIC ID using their orbital period, epoch of the first detected transit, and transit duration). To achieve this, we followed the same procedure mentioned in~\citep{Twicken2018keplerDV}. We created a periodic pulse train for each TCE/object based on their ephemerides. These time series span the duration of the sector run under analysis. For each target star in a given sector run, we computed the cosine similarity between each TCE and object for that target star. A TCE was considered matched to an object if its matching score exceeded $0.75$ and it had the highest score across all TCEs for that object. This second condition was employed to prevent the matching of objects to secondary transits or residual TCEs. This strict matching procedure minimizes the number of spurious matches, thereby reducing the label noise in our labeled dataset.

Table~\ref{table:2min_tce_counts_labeled_set} shows the counts and relative percentage of TCEs in the data set created by our pipeline at the end of the second stage for the TESS 2-min data for six major categories: KP, CP, BD, and FP from ExoFOP, EB from Villanova EB catalog, and NTP from TEC. Those TCEs that could not be labeled were dispositioned as unknown (UNK) and were not used to train nor to evaluate the models. Out of the initial set of 208k TCEs, about 205k were preprocessed successfully. Those that were not preprocessed had been primarily labeled either as NTPs or UNKs.

Multiple TCEs may correspond to the same event due to the multi-sector nature of the TESS mission. For TCEs labeled from the ExoFOP and the Villanova EB catalog, we identify TCEs from the same event using their matched TOIs/objects in these catalogs. However, for the NTP category, we do not group TCEs into the same event for two reasons: 1) the non-transiting nature of NTPs means we do not expect to find ephemeris matches in multiple sectors in general, and 2) ephemeris matching for NTPs is computationally expensive due to the lack of reference points, such as a TOI list or the Villanova EB catalog. Table~\ref{table:2min_event_counts_labeled_set} shows the number of events for each category after grouping the TCEs listed in Table~\ref{table:2min_tce_counts_labeled_set}. 

The process of matching TCEs to objects from different catalogs based on their ephemerides is imperfect for several reasons. First, weak secondary events are sometimes detected as separate TCEs, leading to the TCEs corresponding to the weak secondary being left unlabeled (i.e., classified as unknown) despite being known signals. For example, in the case of TIC 16740101, multiple TCEs are detected: some correspond to the primary TESS SPOC TCE TIC 16740101-1-$Sz_1$, while others correspond to the weak secondary TESS SPOC TCE TIC 16740101-1-$Sz_2$ for different sector runs (denoted by different values of $z_1$ and $z_2$). While the TCEs for the primary are matched to TOI 1150.01, the weak secondaries remain unmatched to any TOIs and are labeled as UNKs. Although this scenario does not lead to mislabeling, such TCEs should not be classified as UNKs.

Second, the orbital period values can vary across different sectors. In extreme cases, the TCE period and the TOI/object period may differ by factors of 2 or 3. A small error in the period can propagate into substantial mismatches of transit locations over longer sector runs. This issue is further exacerbated by errors in other ephemeris parameters such as duration and epoch, making the matching process even more complex. The following examples illustrate these challenges: 1) TESS SPOC TIC 82308728-1-S14-50: The period is twice that of the planet TOI 1821.01 (KP) because two out of four transits are missing in sectors 22 and 49. This incorrect period results in failed ephemeris matching, leaving the TCE labeled as UNK, 2) TESS SPOC TIC 23434737-1-S1-65: The estimated orbital period of 25.5 days differs slightly from the planet TOI 1203.01 (CP) period of 25.52 days. Although the error is small, it propagates over the data span (sector 1 through sector 65), leading to substantial misalignment, failed matching to the TOI, and an UNK label, 3) TESS SPOC TIC 307210830-2-S35: The transit duration is significantly different from that of the corresponding TOI 175.02 (CP) (about 1.8 times longer than the TOI duration), leading to a failed ephemeris match and UNK label, and 4) TESS SPOC TIC 283722336-1-S17: The transit midpoint location differs significantly from the corresponding planet TOI 1469.01 (KP) due to epoch mismatches. The TCE epoch originates from sector 17, while the TOI epoch comes from sector 57.

Additionally, data gaps caused by phenomena such as scattered light and momentum dumps further exacerbate these issues. These examples highlight the various scenarios where ephemeris matching can fail, complicating the classification process.

Given that the label we use for TESS comes from different sources that sometimes are not very reliable,
we  also incorporate \kepler\ data, which has higher-quality labels, to assist in training our machine learning model. A summary of the \kepler\ data used in our study is provided in Table~\ref{table:kepler_labeled_set}.

\begin{table*}[!t]
\footnotesize
\centering
\caption{TESS 2-min TCE Counts. We consider BD, EB, and FP together as one group of Astrophysical FPs or AFPs.}
\begin{threeparttable}
\begin{tabular}{c|cccccc }
\toprule
Classes & \multicolumn{2}{c|}{Exoplanets} & \multicolumn{4}{c}{Non-planets} \\
\cline{1-7}
Sub-classes & KP & \multicolumn{1}{c|}{CP} & BD & EB & FP & NTP    \\
\hline
Count (Percentage) &  1835 (3.21\%) & \multicolumn{1}{c|}{1846 (3.23\%)} & 32 (0.06\%) & 12738 (22.28\%) & 1702 (2.98\%) & 39009 (68.24\%)\\
\hline
Total (Percentage)  & \multicolumn{2}{c|}{3681 (6.44\%)} & \multicolumn{4}{c}{53481 (93.56\%)}   \\
\bottomrule
\end{tabular}
\end{threeparttable}
\label{table:2min_tce_counts_labeled_set}
\end{table*}

\begin{table*}[!t]
\centering
\caption{TESS 2-min Event Counts. We consider BD, EB, and FP together as one group of Astrophysical FPs or AFPs.}
\begin{threeparttable}
\begin{tabular}{c|cccccc }
\toprule
Classes & \multicolumn{2}{c|}{Exoplanets} & \multicolumn{4}{c}{Non-planets} \\
\cline{1-7}
Sub-classes & KP & \multicolumn{1}{c|}{CP} & BD & EB & FP & NTP    \\
\hline
Count (Percentage) &  486 (1.13\%) & \multicolumn{1}{c|}{371 (0.88\%)} & 10 (0.02\%) & 2557 (5.96\%) & 442 (1.03\%) & 39009 (90.98\%)\\
\hline
Total (Percentage)  & \multicolumn{2}{c|}{857 (2.01\%)} & \multicolumn{4}{c}{42018 (97.99\%)}   \\
\bottomrule
\end{tabular}
\end{threeparttable}
\label{table:2min_event_counts_labeled_set}
\end{table*}

\begin{table}[!t]
\footnotesize
\centering
\caption{\kepler\ Counts.}
\begin{threeparttable}
\begin{tabular}{c|cc }
\toprule
Classes & Exoplanets & Non-planets \\
\midrule
Count (Percentage) &  2654 (8.58\%) & 28287 (91.42\%)\\
\bottomrule
\end{tabular}
\end{threeparttable}
\label{table:kepler_labeled_set}
\end{table}

\subsection{Preprocessing Pipeline}
 \label{sec:preprocessing}
The preprocessing pipeline was subjected to regular changes as part of the iterative process of data processing and model development. Below, we describe those changes compared to the original model~\citep{Valizadegan_2022_ExoMiner}:
\begin{itemize}
    \item Momentum Dump pipeline: We extracted information relative to the momentum dumps from the data quality arrays contained in the light curve data for each target~\citep{Twicken-2020-TESS_Handbook}. We then phase-folded this binary time series using the estimated orbital period for each TCE and binned the signal through mean averaging to the same size as the flux and centroid motion time series. Thus, each bin value reflects the fraction of cadences in the bin affected by the momentum dump according to the respective data quality flag. The location of the peaks in these time series tells us whether the momentum dump events align with the transits of the TCE; the amplitude of the signal tells us the fraction of cadences per bin that were affected by this phenomenon. Figure~\ref{fig:momentum_dump_data_example} shows the phase-folded momentum dump flag data using the detected orbital period for TESS SPOC TCE TIC 82707763-1-S37 (top), and its binned version ready to feed to the model for TOI 1991.01.

\begin{figure}[htb!]
{\includegraphics[width=\linewidth]{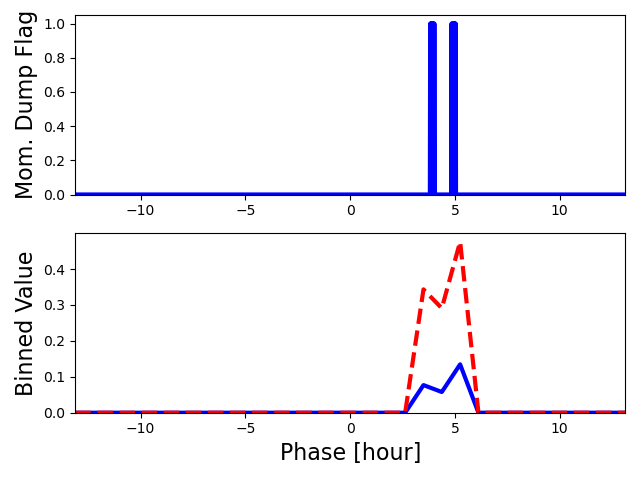}}
\caption{Momentum dump data for TESS SPOC TCE TIC 82707763-1-S37 (TOI 1991.01, classified as a spectroscopic EB). The top panel displays the phase-folded time series for the momentum dump flag array, while the bottom panel presents the corresponding binned version obtained through the preprocessing steps described in Section~\ref{sec:preprocessing}. The solid blue and dashed red lines indicate the average binned value and one standard deviation, respectively.}
\label{fig:momentum_dump_data_example}
\end{figure}

\begin{figure*}[htb!]
\begin{center}
\centerline{\includegraphics[width=0.9\linewidth]{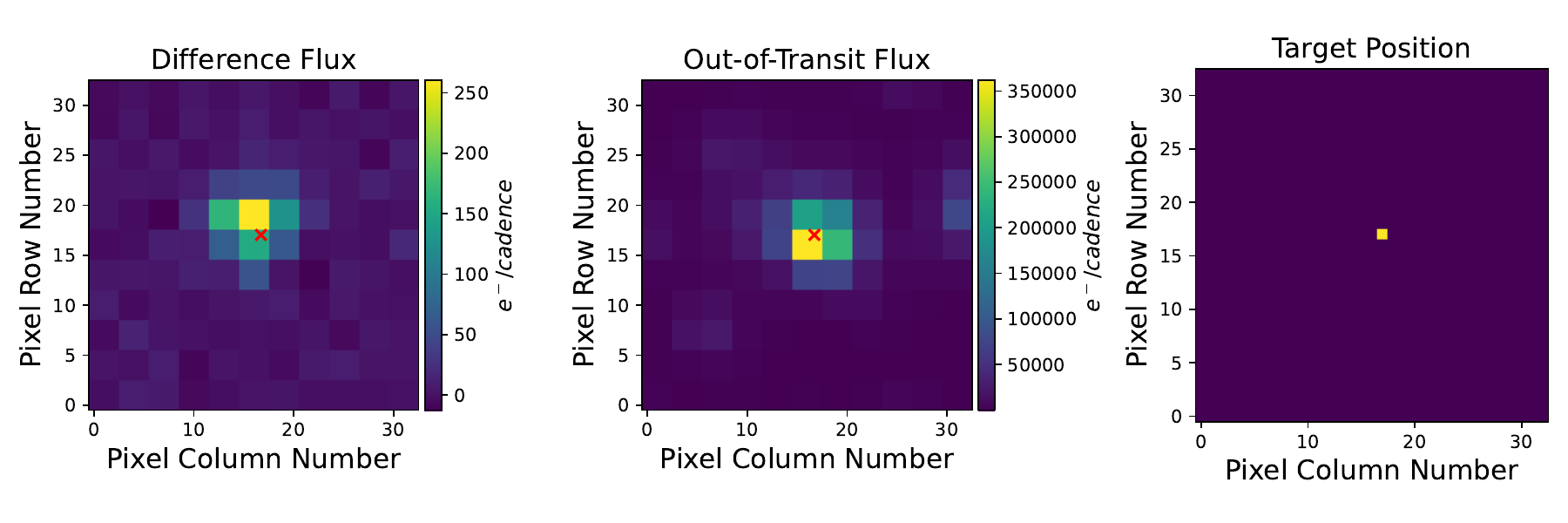}}
\vskip -0.2in
\caption{Preprocessed difference image data for TESS SPOC TCE TIC 82707763-1-S37 (TOI 1046.01, a nearby EB), following the steps outlined in Section~\ref{sec:preprocessing} without normalization. From left to right, the panels display the difference image, out-of-transit image, and target position image. In the difference and out-of-transit images, the red cross marks the location of the target, TIC 309787037, with its coordinates mapped to the CCD frame. In the target position image, the pixel containing the target is highlighted in yellow.}
\label{fig:diff_img_data_example}
\end{center}
\end{figure*}

    \item Difference Image pipeline: we developed a preprocessing pipeline that first extracts the difference image data from the target DV xml files\footnote{The format of the archival SPOC products including DV xml file are described in the TESS Science Data Products Description Document at \url{https://archive.stsci.edu/missions/tess/doc/EXP-TESS-ARC-ICD-TM-0014-Rev-F.pdf.}} for each TCE in the different sector runs, and then preprocesses these data into a format amenable to be fed into our models. This was an involved process  requiring several iterations and feedback from SMEs on our team to make the data suitable to be ingested by our models, and useful in terms of the information and patterns to be used by the models to distinguish false positives due to transit source offsets from the target star. From the difference image data for a single sector, we create the following features:
    \begin{enumerate}
        \item Out-of-transit image: mean image created using all image data for the out-of-transit (oot) cadences. This image is created in the DV module of the SPOC pipeline.
        \item Difference image: mean image created by subtracting the in-transit image from the oot image. This image is created in the DV module of the SPOC pipeline.
        \item Target position: we map the TIC coordinates of the target to the CCD pixel frame. The target pixel is set to one while all other pixels are set to zero (see detailed description of the preprocessing steps below).
        \item Quality metric: this metric is computed in the DV module of the SPOC pipeline and is the correlation between the difference image and the pixel response function (PRF) centered on the centroid of the difference image thresholded at 75\% (i.e., correlations $\ge$ 75\% are deemed ``good'', while correlations $<75$\% are deemed ``bad'').
    \end{enumerate}

    The number of images available per TCE depends on the number of sectors the respective target was observed and the number of clean transits that were observed for the TCE. To make the difference image data input size uniform to the model across examples (i.e., TCEs), we sample with replacement a set of 5 images (i.e., 5 sectors of data in the case of TESS; 5 quarters of data in the case of \kepler)\footnote{The same instance is repeated five times for single-sector cases.}. This gives the model the chance to see (when there are multiple sectors of data available) a diverse set of images for a given TCE. Hence, the dimensionality of any image data (i.e., difference, oot, and target images) is $33\times 33 \times 5 \times 3$, where the first two dimensions are the height and width of the images, the third dimension is the number of sampled sectors/quarters, and the last dimension is the number of images, one for out-of-transit image, another for difference image, and the last for the target position. For the quality metric the dimensionality is then $1\times5$, since this feature is a scalar value for each sector of sampled difference image data.
    
    On average, \kepler\ and TESS targets have postage stamps with dimensions $5\times6$ px and $11\times11$ px, respectively. Brighter targets and cases in which the target is close to the edge of the CCD lead to changes in the size and shape of the aperture, thus leading to out-of-transit and in-transit images with different sizes. Since the model requires uniform dimensions across all examples, we need to go through a preprocessing step to make all images the same size. Given that the majority of TESS targets have $11\times11$ px, we transform all images to that size, thus minimizing the amount of transformation performed to the images for TESS targets, and ensuring that the majority of \kepler\ targets' images are filled to that size without resorting to any image cropping. Furthermore, less than 20\% of the TESS TCEs in our dataset come from targets whose images are larger than $11\times11$ px, and this subset of cases comes mostly from bright stars that are more likely to saturate the image data. In such scenario, it is less likely that the images contain useful information for estimating the transit source offset.

    After sampling the set of images for a given TCE, we execute multiple steps to address missing values and size differences. We set negative out-of-transit pixels to missing in both out-of-transit and difference images, and fill out missing values using nearest neighbors with a $3\times3$ px window with the same weights for all non-missing pixels. If the images are smaller than $11\times11$ px, we pad them by extending the edges. Given that the target location is provided at the subpixel level, we resize the images through nearest neighbor interpolation using a factor of 3 so all images are at least $33\times33$ px after this step. By doing this, each pixel becomes a $3\times3$ grid that we can use to more accurately describe the subpixel location of the target. Finally, for images that are larger than the desired final size, we crop them relative to the center pixel so they become $33\times33$ px. The image data is standardized using the median and standard deviation statistics computed across all pixels and images from the training set examples.
    Figure~\ref{fig:diff_img_data_example} shows the three input channels of different image data for TESS SPOC TCE TIC 309787037-1-S35 that are fed to the difference image branch after normalization. The difference image suggests that the transit signal does not originate in the target star (TIC 309787037 with $T_{mag}=10.012$), but from a neighboring star, observed in a different pixel. A known nearby, fainter star, TIC 309787035 ($T_{mag}=16.480$), is located at this location.

    \item Unfolded Flux pipeline: 
    The number of observed phases for a detected TCE is a function of the observation time for the respective target, the orbital period of the TCE, and the number and size of gaps in the data during that observation. For this reason, the number of transits across TCEs can vary significantly. To make these data uniform across TCEs, so that it can be amenable to be ingested by our models, we sample with replacement a set of 20 transit signatures, with each binned to a fixed size (more concretely, 31 bins for the transit view). These phases are normalized using the same median and minimum statistics computed from the corresponding phase folded and binned transit view time series. To give the model information about the number of valid transits that are observed in the data, we add as a scalar input feature to this branch the number of transits observed by the SPOC DV pipeline for any given TCE. For completeness, we also add the number of transits expected by the SPOC DV pipeline, which could provide some information about missing data. Figure~\ref{fig:flux_local_unfolded_data_example} shows the non-normalized unfolded flux data that are fed to the unfolded branch of \ExoMinerplusplus\ after normalization.

\begin{figure}[htb!]
{\includegraphics[width=\linewidth]{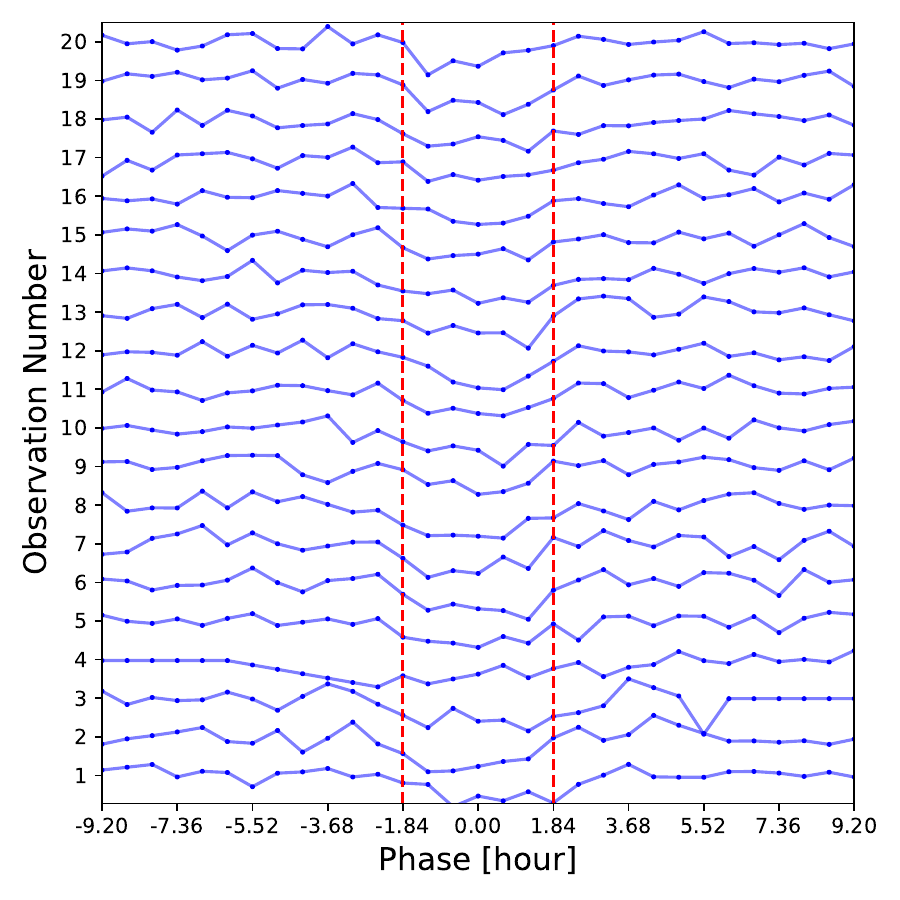}}
\caption{Unfolded binned flux data for TESS SPOC TCE TIC 98545929-1-S1-36 (TOI 1916.01, KP). The red dashed lines highlight the transit duration estimate of the detected TCE ($\sim3.68$ hours) centered on the estimated transit midpoint.}
\label{fig:flux_local_unfolded_data_example}
\end{figure}

    \item Periodogram pipeline: We applied the Lomb-Scargle periodogram~\citep{lomb1976least,scargle1982studies} to estimate the spectral density of the flux time series prior to detrending. This approach allows the periodogram to capture additional signals present in the time series, such as stellar variability, which are typically removed during the detrending process. The periodograms are computed for a period range that spans 0.04 days to 54 days. This allows us to capture information for even the shortest period TCEs ($\sim0.2$ days; so the range extends to at least 5 harmonics for any TCE) and covers approximately two sectors of observation. About 11.5\% TCEs in our dataset show a period longer than 54 days\footnote{For many of these the true period is ambiguous and may be (significantly) shorter.}. Out of those, we count only 12 unique planets and 7 unique EBs. We chose this observation time as a trade off between periodogram feature complexity and the number of long-period TCEs whose transit periodicity falls outside of the range set for the periodogram. Nevertheless, even for those scenarios the periodogram can reveal information about other signals in the light curve data. The frequency values are defined linearly across the range using a downsampling factor of 4 to reduce the dimensionality of the feature array. The periodogram is normalized to show the amplitude values, and then smoothed to remove noise using a box kernel filter of width 2. Finally, the periodogram is normalized by its maximum amplitude so all examples fed to the model are defined in [0, 1] interval. Information about the maximum power is conveyed to the model as a scalar feature. This feature is standardized using training set statistics following the same methodology applied to all other scalar features. To provide to the model information about the location and characteristics of the TCE signal in the frequency domain, we create another periodogram using the same methodology described for the one created from the flux data. Using the estimated period, transit duration, secondary offset, and transit depths of the primary and secondary events, a simple transit model is created by setting primary and secondary in-transit cadences to the corresponding transit depth, with out-of-transit cadences set to the median value of the original flux time series. By design, the periodogram of this time series will show peaks for the period and harmonics of the detected TCE, thus revealing to the model where in the frequency domain there is relevant information related to the TCE transit signal. The model can use that information to compare the features in the periodogram of the data to a periodogram specific to the transit model associated with the TCE. Figure~\ref{fig:periodogram_data_example} shows an example of the periodogram data that are normalized and then fed to the model. 

\begin{figure}
  \centering
    \subfigure[Frequency scale.]{\label{fig:periodogram_data_example_freq}\includegraphics[width=\linewidth]{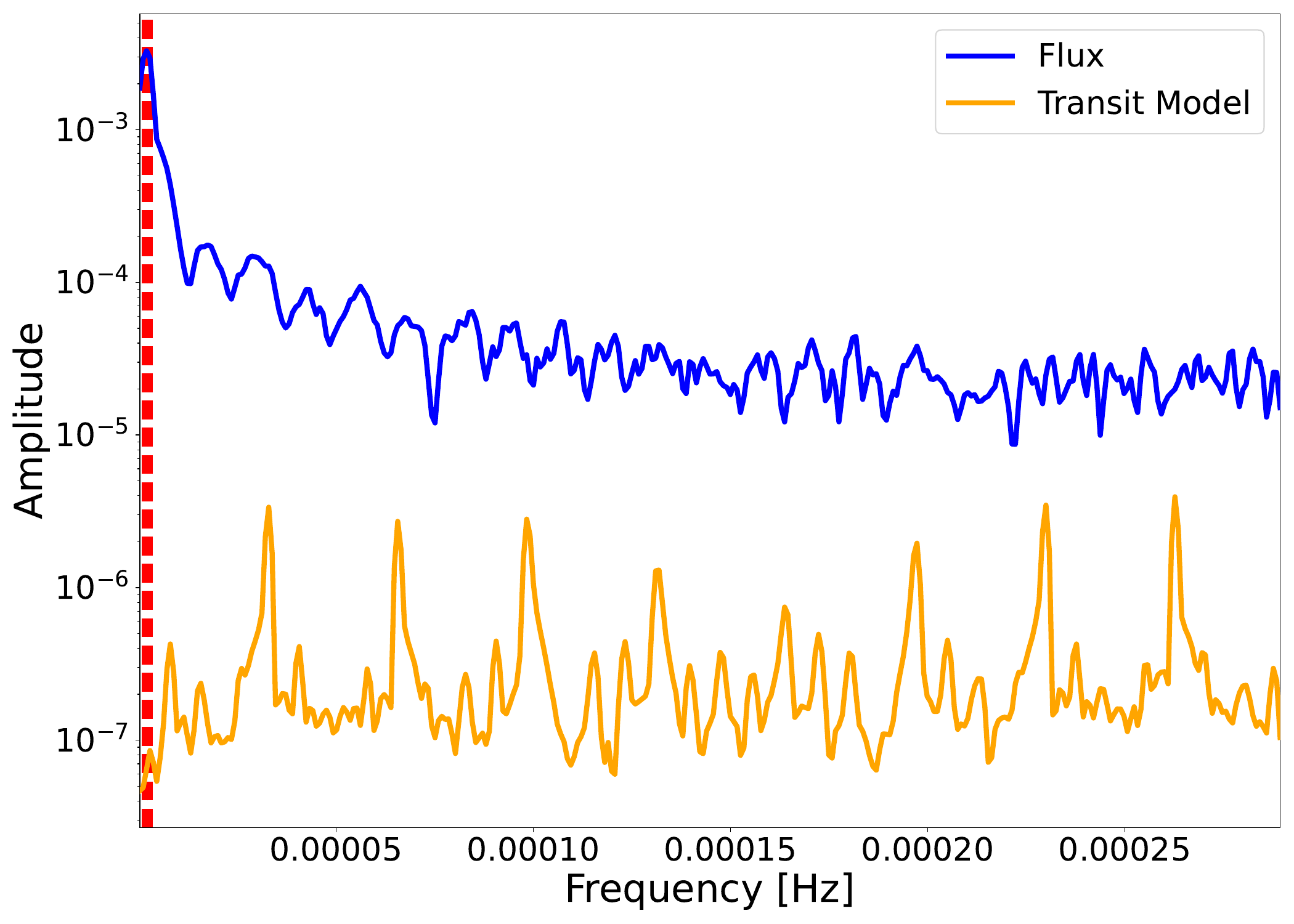}}
    \subfigure[Period scale.]{\label{fig:periodogram_data_example_period}\includegraphics[width=\linewidth]{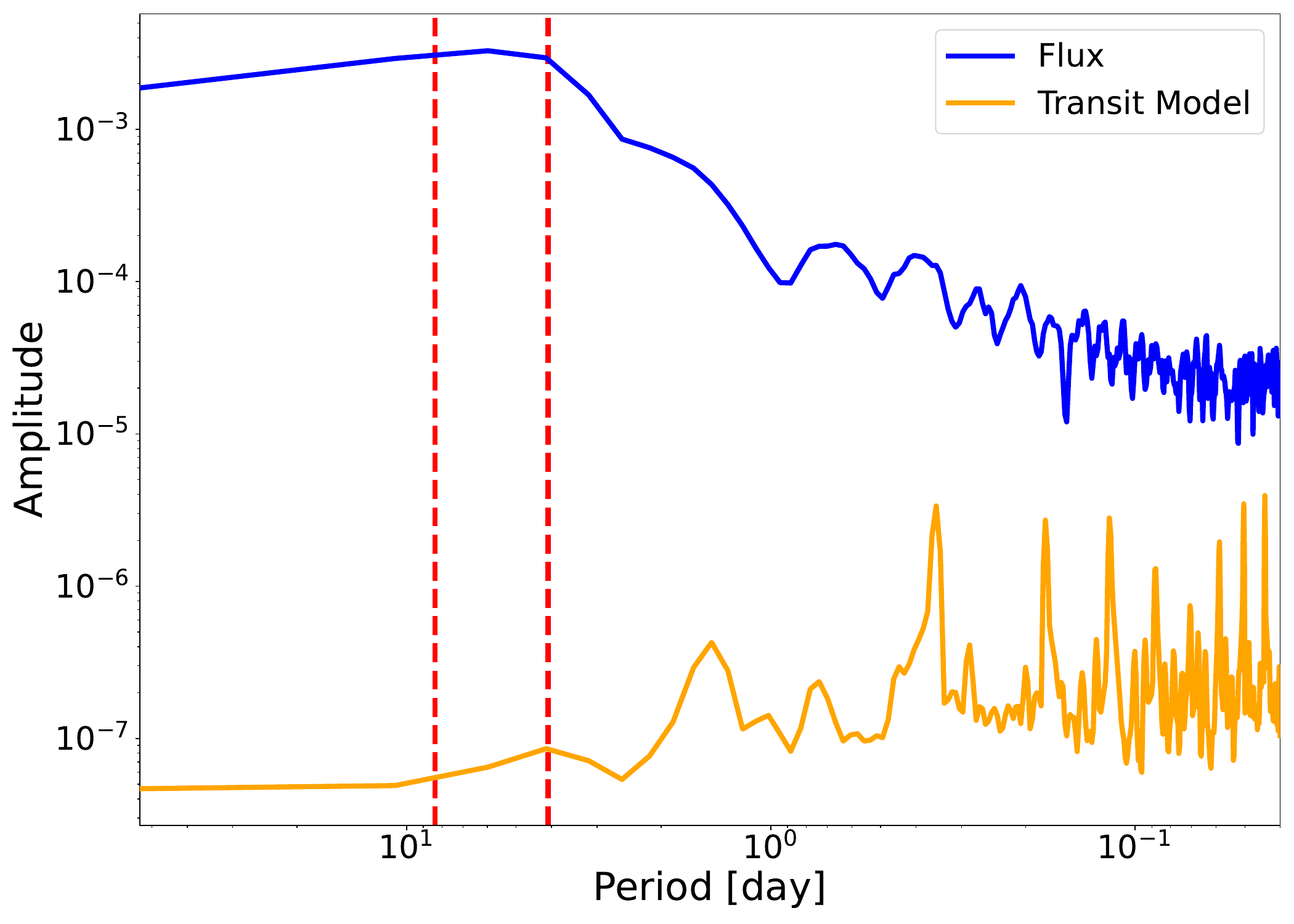}}
\caption{Periodogram data for TESS SPOC TCE TIC 232568235-1-S24 (TOI 2260.01, CP) after preprocessing, as described in Section~\ref{sec:preprocessing}, without normalization. The figures display the flux and transit model periodograms. TOI 2260.01 is classified as a CP orbiting a variable star. The red dashed lines indicate the estimated periods of a double-sinusoidal function fitted to the photometric variability of TIC 232568235, derived from \cite{fetherolf2023variability}.}
\label{fig:periodogram_data_example}
\end{figure}

    \item Detrending pipeline: we changed the detrending method from employing a spline in the original \ExoMiner\ to using a Savitzky-Golay filter to remove low-frequency trends in the data. We used a window length of 1.2 days, and chose the best fitting model (up to maximum 8th degree polynomial order) using Bayesian information criterion~\cite[BIC; ][]{schwarz1978estimating, Stoica-2004-BIC} with a penalty weight of 1 for model selection. We masked the in-transit cadences of all detected transits in the light curve before fitting any model, and the raw time series was split into smaller segments if the gap between two consecutive samples was larger than 5 times the nominal cadence duration. A sigma value of 5 was used to remove outliers from the detrended flux time series. A similar approach was used to detrend the centroid motion time series. Figure~\ref{fig:flux_trend_correctly_classified_example} shows an example of flux trend data before it is normalized and fed to the model. In this case, the ellipsoidal variations occurring at half period of the EB in TIC 167526485 are made even more clear by phase-folding the flux trend time series using the orbital period of the corresponding TCE.

    \item Uncertainty for the phase-folded and binned time series: for each bin in the phase-folded time series, the standard deviation of the values is computed along with the median value. The standard error of the mean is then computed by normalizing the standard deviation by the square root of the number of points within the bin to provide information to the model regarding the uncertainty in the average estimation.
    
\begin{figure}[htb!]
{\includegraphics[width=\linewidth]{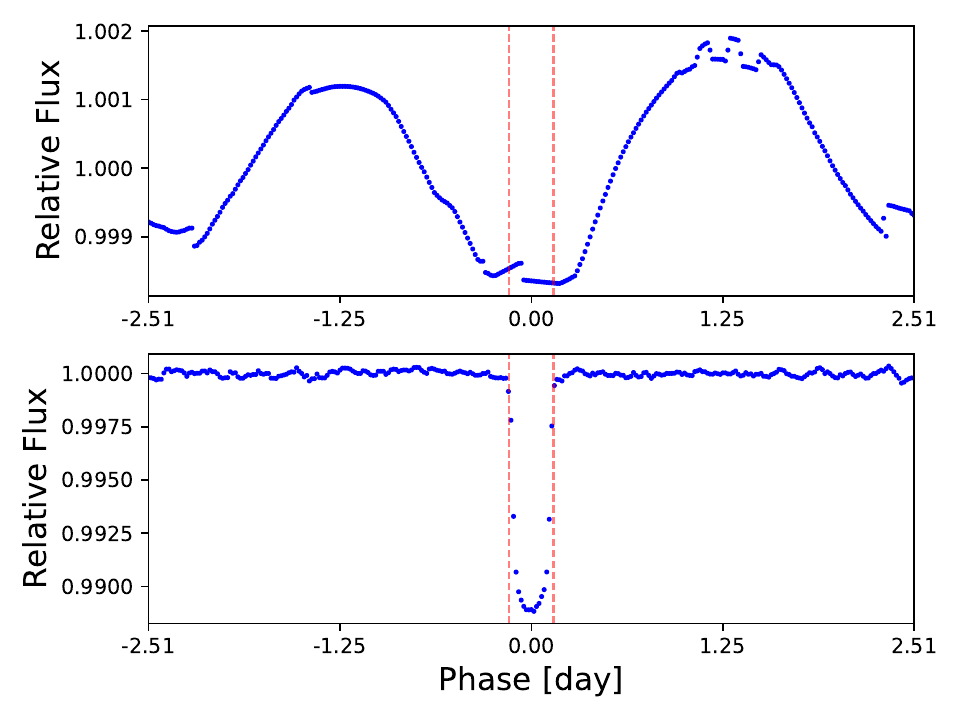}}
\caption{The top and bottom figures show the phase-folded, binned trend and detrended flux time series for TESS SPOC TCE TIC 167526485-1-S6 (dispositioned as an EB in ~\cite{Prsa_2022-EB-catalog}), respectively, without normalization. The red dashed lines indicate the estimated transit duration for this TCE, centered around the estimated transit midpoint.}
\label{fig:flux_trend_correctly_classified_example}
\end{figure}

\end{itemize}

\section{Hyperparameter Optimization and Model Training}
\label{sec:hpo}
Early in our work, we conducted a hyperparameter optimization (HPO) study with initial data from the \TESSMission. The goal was to find a configuration of \ExoMinerplusplus\ that would be optimized for TESS data, as opposed to the previous \ExoMiner\ developed for \kepler. Using a Bayesian/Hyperband method~\citep{falkner2018bohb}, we evaluated a total number of 177 configurations over the course of 3 days and using a V100 GPU node from the NASA Advanced Supercomputing Facility. For each configuration evaluated, an average score ensemble of three models was trained on the provided budget, and the precision-recall area under curve (PR AUC) on the validation set for this ensemble was set as the optimization metric for the HPO framework (i.e., a configuration with a higher PR AUC on the validation set is ``better"). 

The set of hyperparameters from the best configuration found in this HPO run was used in the \ExoMinerplusplus\ for the experiments described in Section~\ref{sec:performance}. Appendix~\ref{sec:optimized_exominer} describes the details of the architecture and optimization parameters.

To evaluate the performance of our optimized \ExoMinerplusplus\ model, we used 10-fold cross-validation, following the approach in~\cite{Valizadegan_2022_ExoMiner}. The target stars were divided into 10 equal subsets, with the TCEs from one subset serving as the test set while the remaining 9 subsets were used as the training and validation sets (1 subset was reserved for validation). This process was repeated for all folds, ensuring that each subset was used once as the test set. 

The motivation behind cross-validation is to ensure that the model performs well not only on the data seen during training but also on unseen data, thereby simulating real-world conditions. By dividing the data into multiple subsets and iteratively testing on each subset, the model’s performance is evaluated on every part of the dataset. This approach reduces the risk of overfitting, where the model performs well on training data but fails on new data. Cross-validation provides a robust measure of the model's ability to generalize to new target stars, making it a critical step in assessing its overall reliability~\citep{Hastie-2009-statistical_learning}.

For each cross-validation iteration, a set of 10 models was trained, and an ensemble model was created by averaging the scores of the 10 models. This procedure minimizes the effects of stochasticity in the training process. By training multiple models with different weight initializations, the ensemble captures variations in local minima encountered during the optimization of the loss function, resulting in a more robust representation of the model’s behavior and performance. 

All models were trained for 300 epochs, optimized using binary cross-entropy as the loss function and the Adam optimizer~\citep{kingma2014adam} ($\beta_1=0.900$, $\beta_2=0.999$, $\epsilon=1\mathrm{e}{-8}$). Early stopping with a patience of 20 epochs was employed to prevent overfitting. If no improvement in the validation PR AUC was observed after 20 epochs, the training was stopped, and the model instance at that point was selected as the final model.

Model training for the cross-validation experiments was conducted using a compute node at NAS HECC with four NVIDIA V100 GPUs. Each model took on average 80 seconds per epoch to train on a single GPU. To speed up training, four models were trained simultaneously across the four GPUs.

Table~\ref{table:classification_results} provides a detailed list of all TCEs in our dataset, including their ephemeris information and the scores obtained from all models evaluated in this study using 10-fold cross-validation.

The \ExoMinerplusplus\ code is publicly available on GitHub\footnote{\url{https://github.com/nasa/ExoMiner}}.

\begin{table}[htb]
\footnotesize
\centering
\caption{Scores of different classification model used in this work on the TCEs that are assumed labeled (Table~\ref{table:2min_tce_counts_labeled_set}) The model label can be obtained by score $> 0.5$. This table describes the available columns. The full table is available online.}
\label{table:classification_results}
\begin{threeparttable}
\begin{tabularx}{\linewidth}{lp{5cm}}
\toprule
Column & Description \\
\midrule
uid & unique id that includes TCE TIC ID and sector run\\
target\_id & TCE TIC ID\\
tce\_plnt\_num & TCE planet number\\
TOI & TOI number\\
fold & a value between 0 and 9 indicating the cross-validation fold for that TCE\\ 
tce\_period & SPOC TCE period\\
tce\_duration & SPOC TCE duration\\
tce\_prad & SPOC TCE planet radius (Earth Radii)\\
MES & SPOC TCE MES\\
original\_label & KP, CP, BD, EB, FP, or NTP\\
binary\_label & 1 for KP and CP, 0 for BD, EB, FP, or NTP\\
tess-individual & Score by the model trained on TESS\\
tess-aggregate & Score by the model trained on TESS but aggregated using strategy in Section~\ref{sec:multi-sector-aggregation}\\
tess+kepler-individual & Score by the model trained on TESS+\kepler\ (Multi-source learning idea in Section~\ref{sec:transfer-learning})\\
tess+kepler-aggregate & Score by the model trained on TESS+\kepler\ but aggregated using strategy in Section~\ref{sec:multi-sector-aggregation}\\
DV full report & The URL to the DV full report in the MAST\\
DV summary report & The URL to the DV summary report in the MAST\\
DV mini report & The URL to the DV mini report in the MAST \\
\bottomrule
\end{tabularx}
\end{threeparttable}
\end{table}


\section{Performance Study}
\label{sec:performance}




\subsection{Evaluation Metrics}
\ExoMinerplusplus\ not only assigns a binary label to each TCE — 1 for planets and 0 for non-planets — but also provides a confidence score\footnote{The confidence scores output by the model are not calibrated probabilities and should not be interpreted as such. These scores primarily serve as relative indicators for ranking and comparison between instances. Calibrating the confidence values remains an important but an area for future work.} between 0 and 1 associated with the label. This score reflects the model's certainty about its classification, with values closer to 1 indicating a higher confidence in the planetary nature of the TCE, and values closer to 0 signaling stronger confidence in the non-planet classification. These scores are particularly useful for ranking and prioritizing TCEs for further human inspection and for identifying cases that may require additional scrutiny due to intermediate confidence levels. By combining both labels and confidence scores, \ExoMinerplusplus\ provides a comprehensive framework for automating transit classification while enabling nuanced analysis of the results. 

To measure the overall performance of the model on the binary classification problem of exoplanet versus non-exoplanet, we use the following metrics: 

\begin{itemize}
    \item \textbf{Accuracy}: 
    Accuracy measures the proportion of correctly classified instances out of the total instances. It is defined as:
    \[
    \text{Accuracy} = \frac{\text{True Positives} + \text{True Negatives}}{\text{Total Population}}
    \]
    It is useful when the classes are balanced but can be misleading if the dataset is imbalanced.

    \item \textbf{Precision}: 
    Precision quantifies how many of the instances classified as positive are actually positive. It focuses on the relevance of the positive predictions and is calculated as:
    \[
    \text{Precision} = \frac{\text{True Positives}}{\text{True Positives} + \text{False Positives}}
    \]
    Precision is important when the cost of false positives is high.

    \item \textbf{Recall} (Sensitivity or True Positive Rate): 
    Recall measures how many actual positive instances were correctly identified. It focuses on detecting all positive instances and is given by:
    \[
    \text{Recall} = \frac{\text{True Positives}}{\text{True Positives} + \text{False Negatives}}
    \]
    Recall is critical when missing positive instances has a high cost.

    \item \textbf{PR-AUC} (Precision-Recall Area Under Curve): 
    PR-AUC is the area under the Precision-Recall curve, which plots precision against recall at different thresholds. It is a better metric than accuracy for imbalanced datasets where the positive class is rare because it focuses on the performance on the minority class.

    \item \textbf{ROC-AUC} (Receiver Operating Characteristic Area Under Curve): 
    ROC-AUC is the area under the ROC curve, which plots the True Positive Rate (recall) against the False Positive Rate at different thresholds. A higher AUC indicates a better-performing model. It is widely used for binary classification problems and is effective for evaluating models across different decision thresholds.
\end{itemize}

Besides classification, we are also interested in the quality of the scores generated by \ExoMinerplusplus\ for ranking TCEs. High-quality scores imply a more effective catalog that can help prioritize exoplanet candidates. A classifier that ranks exoplanets higher than false positives is preferred, as it enables more efficient follow-up efforts. To assess the model’s ranking performance, we use Precision@k or P@k:

\begin{itemize}
    \item \textbf{Precision@k or P@k:} P@k measures the fraction of relevant instances (exoplanets) among the top $k$ predictions:

    \[
    \text{P@k} = \frac{\text{Number of relevant items in the top k results}}{k}
    \]
    where $k$ is a predefined number of top results to evaluate.
\end{itemize}

These classification and ranking metrics together provide a comprehensive view of the model's performance from multiple perspectives, depending on the specific needs of the problem (e.g., handling imbalanced classes or managing the cost of false positives and false negatives). Additionally, we report the model's performance across various sub-classes — e.g., KP, CP, EB, BD, FP, and NTP — using recall, which reflects the percentage of each sub-class correctly classified.

\subsection{Multi-Source Learning} 
\label{sec:transfer-learning}
Due to the lack of gold-standard labels for TESS, we leveraged \kepler\ data to enhance performance. Initially, we experimented with various transfer learning approaches~\citep{NG-2016}, such as training on \kepler\ data and fine-tuning certain layers of the model using TESS data. However, as TESS data grew in size and label quality, a simpler approach of combining \kepler\ and TESS data to create a larger training set proved more effective. In this approach, we incorporated \kepler\ data into the training set of all cross validation iterations, ensuring that the validation and testing were performed only on TESS data.

\subsection{Multi-Sector Aggregation of the results}
\label{sec:multi-sector-aggregation}
Given the multi-sector nature of the TESS mission and the fact that many target stars are observed across multiple sectors, the same event is often detected as different TCEs in both single-sector and multi-sector runs. With more data for an event, we achieve a higher SNR and a more reliable signal for those TCEs. Naturally, the longest sector runs (i.e., the TCEs with the largest number of transits observed) provide better data for classifying that event. One could argue that only the longest sector runs should also be used to train the classifier. However, since a significant portion of target stars are not observed in multiple sectors, and the classifier must perform well on those too, we include all TCEs detected by the pipeline in the training. Including shorter-run TCEs (even for events with longer sector runs) exposes the classifier to more low-SNR signals, ensuring it performs well across all targets and generalizes better to low-SNR cases. This results in classifying the same event multiple times—once for each TCE associated with that event. To consolidate this, we generate a single model score/label in post-processing for all TCEs of the same event based on the result of the model for the longest sector run, which ultimately improves overall performance, as we see in Section~\ref{sec:exominer++-results}.

\begin{table*}[!t]
\footnotesize
\centering
\caption{Results of \ExoMinerplusplus\ on labeled TESS data set. The best performer is highlighted by bold. }
\resizebox{.9\linewidth}{!}{
\begin{threeparttable}
\begin{tabularx}{\linewidth}{@{}Y@{}}
\begin{tabular}{ccc|cccc|cc|cccc }
\toprule
 \multirow{2}{*}{Training} & \multirow{2}{*}{Test} & \multirow{2}{*}{Strategy} & \multicolumn{4}{c|}{\multirow{2}{*}{Binary results}} & \multicolumn{6}{c}{Recall for subclasses}  \\
\cline{8-13}
 \multicolumn{3}{c|}{\empty} & \multicolumn{4}{c|}{\empty} & \multicolumn{2}{c|}{Exoplanets}  & \multicolumn{4}{c}{Non-planets}  \\
\hline
Training & Test & Strategy & Precision \& Recall & PR AUC & ROC AUC & Accuracy & KP & CP & BD & EB & FP & NTP\\
\midrule
    \multirow{2}{*}{TESS}  & \multirow{2}{*}{TESS}  & Individual  &  0.918 \&     0.917 &     0.966 &     0.995 &     0.989 &     0.939 &     0.894 &     0.531 &     0.995 &     0.886 &     \textbf{0.999}    \\
    \empty & \empty &  Aggregate  &  0.926 \&     0.945 &     0.967 &     0.995 &     \textbf{0.992} &     0.953 &     0.938 &     0.469 &     0.996 &     0.890 &     \textbf{0.999}   \\
\midrule
    \multirow{2}{*}{TESS+Kepler}  &  \multirow{2}{*}{TESS} & Individual   &  0.924 \&     0.927 &     0.970 &     0.997 &     0.990 &     0.945 &     0.909 &     \textbf{0.719} &     0.996 &     0.887 &     \textbf{0.999}\\
    \empty & \empty & Aggregate  &   \textbf{0.933} \&     \textbf{0.951} &     \textbf{0.976} &     \textbf{0.998} &     \textbf{0.992} &     \textbf{0.957} &     \textbf{0.944} &     0.562 &     \textbf{0.997 }&     \textbf{0.896} &     \textbf{0.999}\\
\midrule
\midrule
    Kepler  &  Kepler & N/A   &  0.976 \& 0.965 &     0.994 &     0.999 &     0.995 &     \multicolumn{6}{c}{N/A}\\
\bottomrule
\end{tabular}
\end{tabularx}
\end{threeparttable}
}
\label{table:performance_metrics}
\end{table*}

\subsection{Classification performance}
~\label{sec:exominer++-results}
Table~\ref{table:performance_metrics} presents the performance results of \ExoMinerplusplus\ on TESS data across various models and strategies, using standard binary classification metrics. Each row corresponds to a different training set (either TESS only or TESS+\kepler\ for multi-source learning) and indicates whether score aggregation was applied after obtaining predictions from \ExoMinerplusplus. For comparison, we also report the baseline results of \ExoMinerplusplus\ when trained and tested on \kepler\ data. 

First, note that the multi-source learning and multi-sector aggregation methods, introduced in Sections~\ref{sec:transfer-learning} and~\ref{sec:multi-sector-aggregation}, improve overall performance across all binary class metrics; overall, the model trained on the combined \kepler\ + TESS data using the multi-sector aggregation strategy achieved the best results. 

Second, both multi-source learning and multi-sector aggregation strategies enhance performance across nearly all sub-classes. However, for BD, these strategies have a negative effect. Two possible reasons explain this: 1) only 10 BDs are observed accounting for 32 TCEs, and the small number of instances increases uncertainty in the results, and 2) photometric data for BDs is generally indistinguishable from exoplanets, so a more accurate model relying solely on photometric data may perform worse on the BD sub-class.

\begin{figure}
  \centering
  \subfigure[Orbital Period (days, log scale)]{
    \label{fig:score_period}
    \includegraphics[width=0.95\columnwidth]{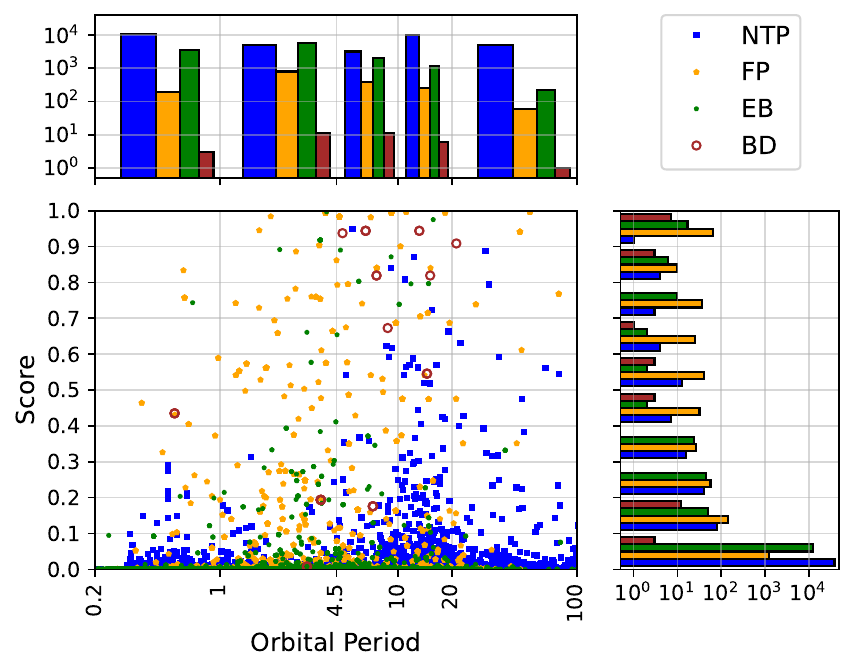}
  }\\
  \subfigure[Planet Radius ($R_\oplus$, log scale)]{
    \label{fig:score_radius}
    \includegraphics[width=0.95\columnwidth]{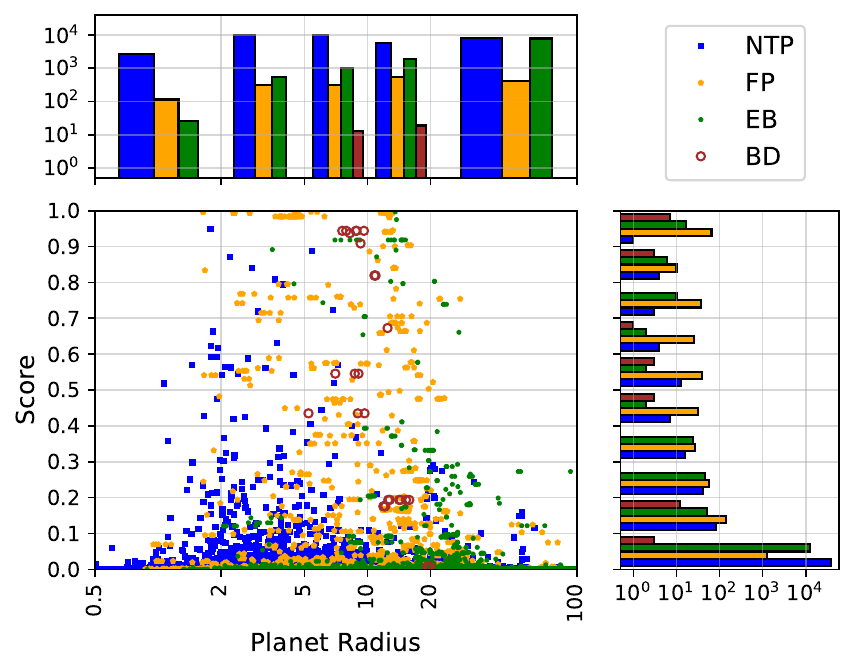}
  }
  \caption{Score distribution versus orbital period (top) and planet radius (bottom) of non-exoplanet categories for the TESS+Kepler-Aggregate model.}
  \label{fig:period_radius_subclasses_histogram}
  \vspace{0.5em}
  {\footnotesize \textit{Note:} The width of bars does not carry any meaning.\par}
\end{figure}

To further investigate the regions in parameter space where \ExoMinerplusplus\ tends to misclassify different non-exoplanet sub-classes, we present in Figure~\ref{fig:period_radius_subclasses_histogram} a scatter plot of scores versus orbital period and planet radius. An ideal model gives a small score to all the instances of these sub-categories. Our model does a reasonable job on EB and NTP sub-classes. For BDs, \ExoMinerplusplus\ performs reasonably well when $R_p > 10 R_\oplus$, likely due to the larger size of these objects, which aids classification. Additionally, the score distribution for the FP sub-class spans the entire range from 0 to 1, highlighting the inherent difficulty of correctly classifying FPs for \ExoMinerplusplus. However, no clear pattern emerges regarding the model's difficulty within specific regions of planet radius and orbital period parameter space.

\begin{figure}
  \centering
    \subfigure[PR-Curve]{\label{fig:pr_curve}\includegraphics[width=0.95\columnwidth]{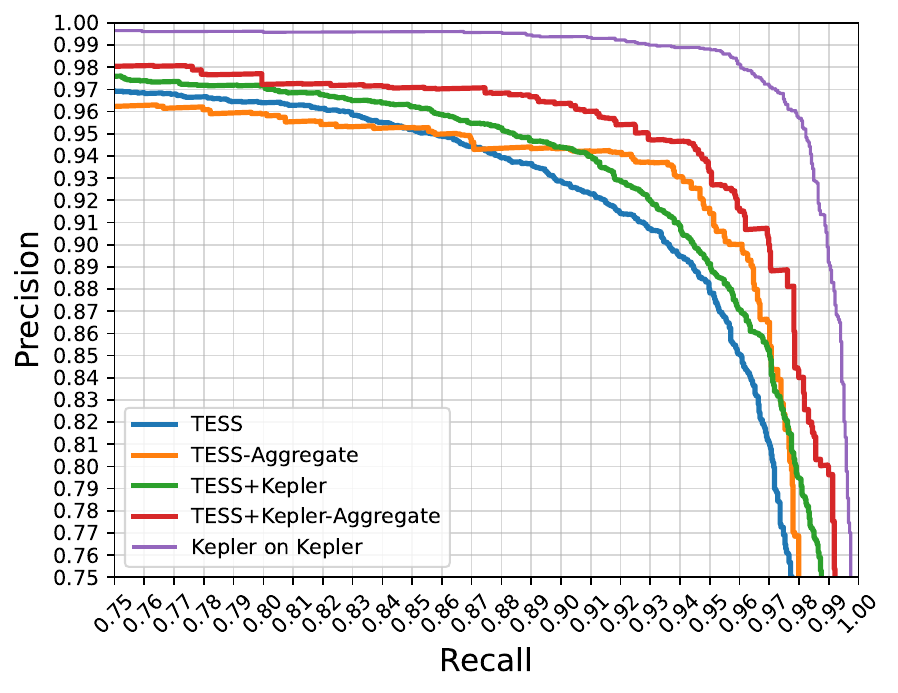}}
    \subfigure[Models Confidence]{\label{fig:confidence}\includegraphics[width=0.95\columnwidth]{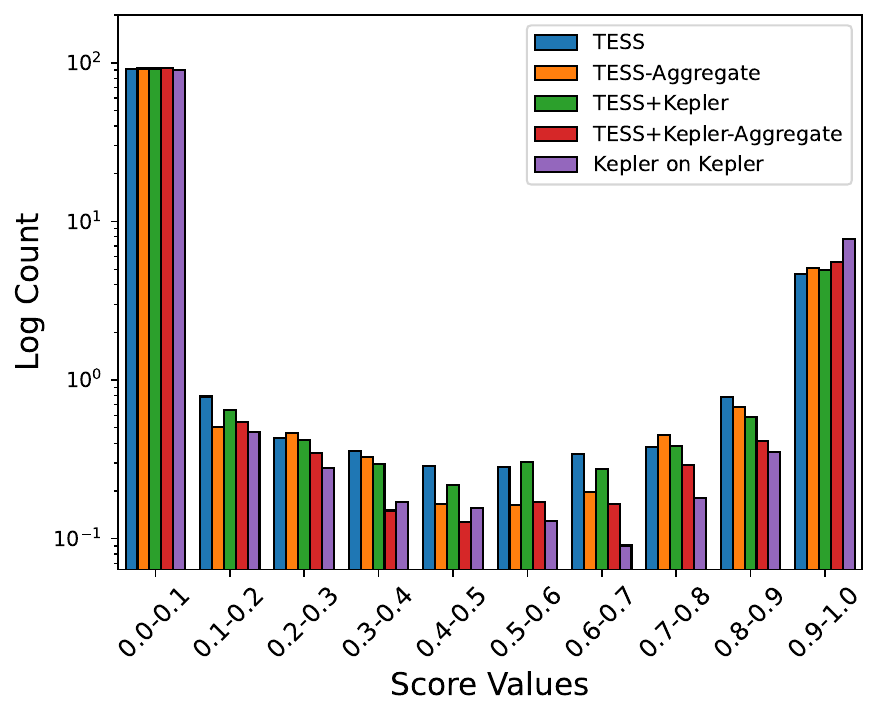}}
\caption{PR-Curve and Confidence of different models.}
\label{fig:PR_plots}
\end{figure}

Table~\ref{table:performance_metrics} summarizes the precision and recall values at a fixed cutoff threshold of 0.5 for the classifier's output. To visualize the trade-off between precision and recall across different cutoff thresholds, the PR curve is shown in Figure~\ref{fig:pr_curve}. As evident from the curve, aggregate models consistently outperform individual models across a range of threshold values. Furthermore, multi-source learning significantly improves performance across thresholds. Figure~\ref{fig:confidence} illustrates the score distributions for various models. A model that places the majority of instances near the extremes (close to zero or one) demonstrates higher confidence in its predictions. \ExoMinerplusplus\ exhibits the highest confidence on \kepler\ data, with both multi-source learning and aggregation strategies further enhancing \ExoMinerplusplus's confidence in its classifications.

\begin{figure}
  \centering
    \subfigure[AFP]{\label{fig:AFP}\includegraphics[width=0.95\columnwidth]{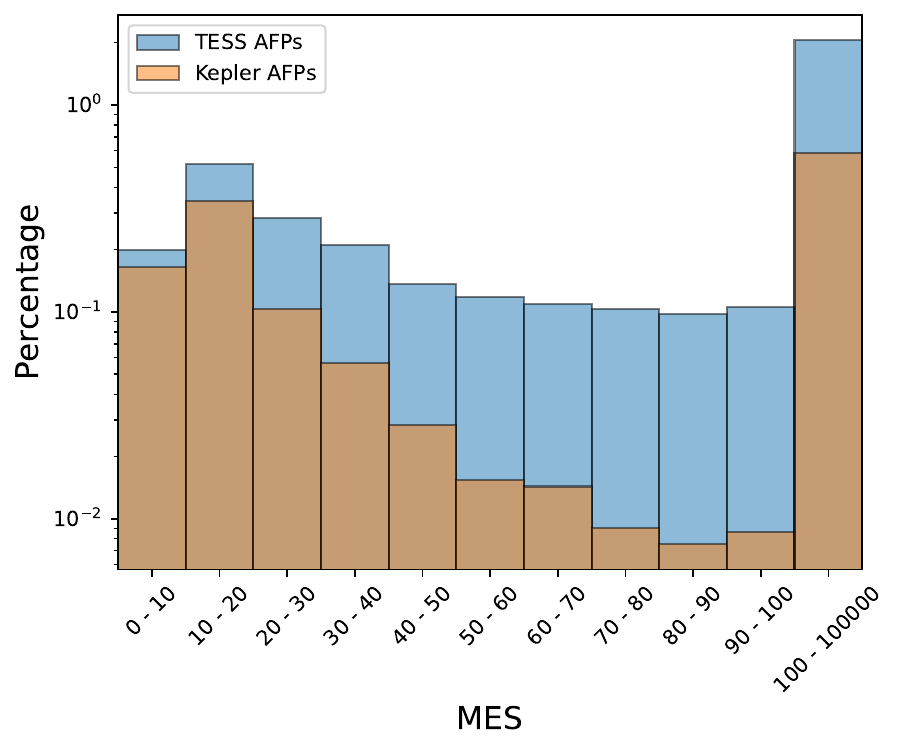}}
    \subfigure[Exoplanets]{\label{fig:Exoplanets}\includegraphics[width=0.95\columnwidth]{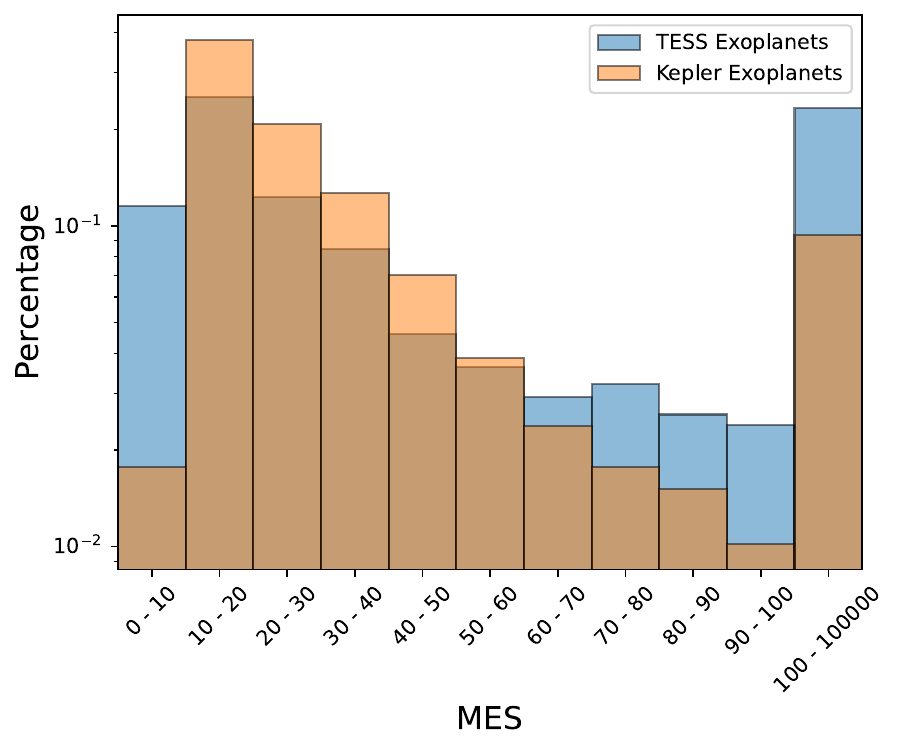}}
\caption{\kepler\ versus TESS histograms of MES with a focus on values $>10$ for exoplanets (KP+CP) and AFPs (EB+FP+BD).}
\label{fig:mes-kepler-TESS}
\end{figure}

\begin{figure}
  \centering
    \subfigure[AFP]{\label{fig:AFP}\includegraphics[width=0.95\columnwidth]{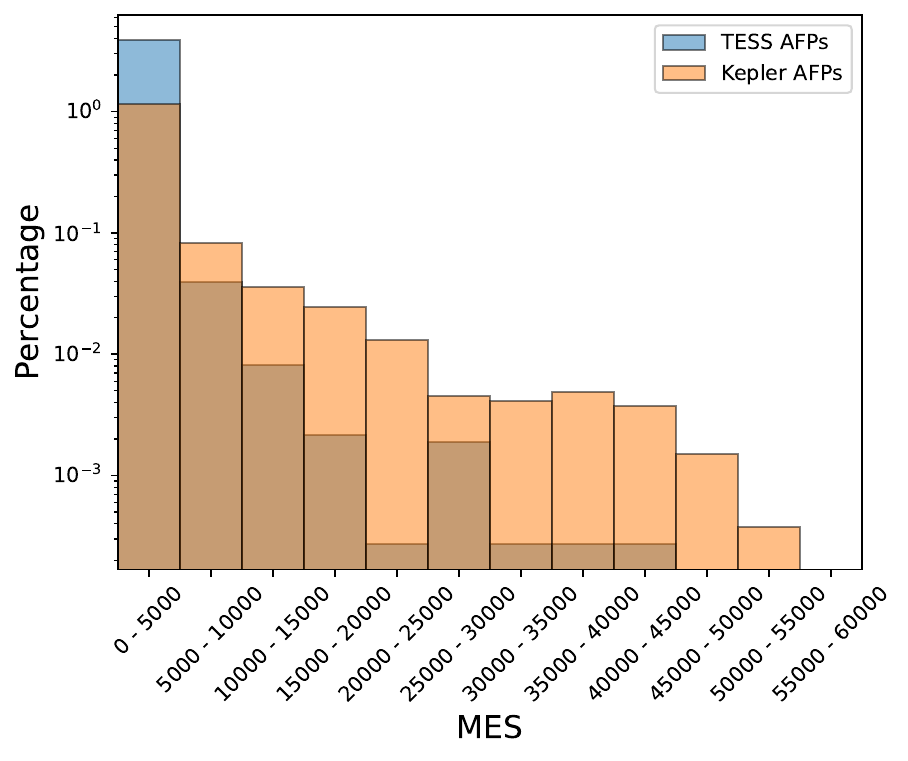}}
    \subfigure[Exoplanets]{\label{fig:Exoplanets}\includegraphics[width=0.95\columnwidth]{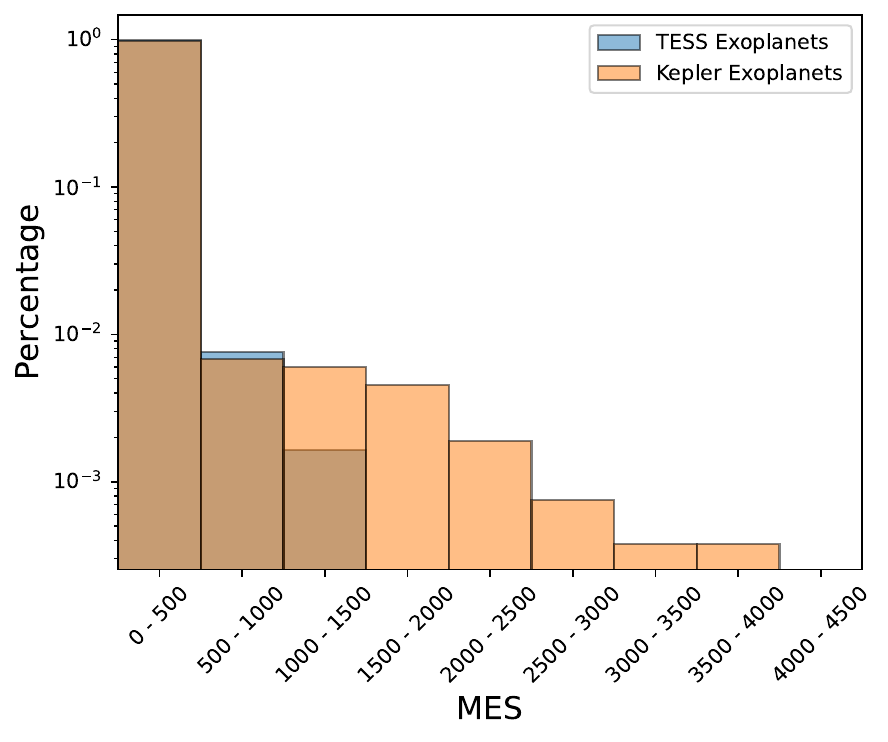}}
\caption{\kepler\ versus TESS histograms of large values of MES for exoplanets (KP+CP) and AFPs (EB+FP+BD).}
\label{fig:mes-kepler-TESS-high}
\end{figure}

\begin{figure*}
  \centering
    \subfigure[Precision]{\label{fig:mes-period-precision}\includegraphics[width=0.49\linewidth]{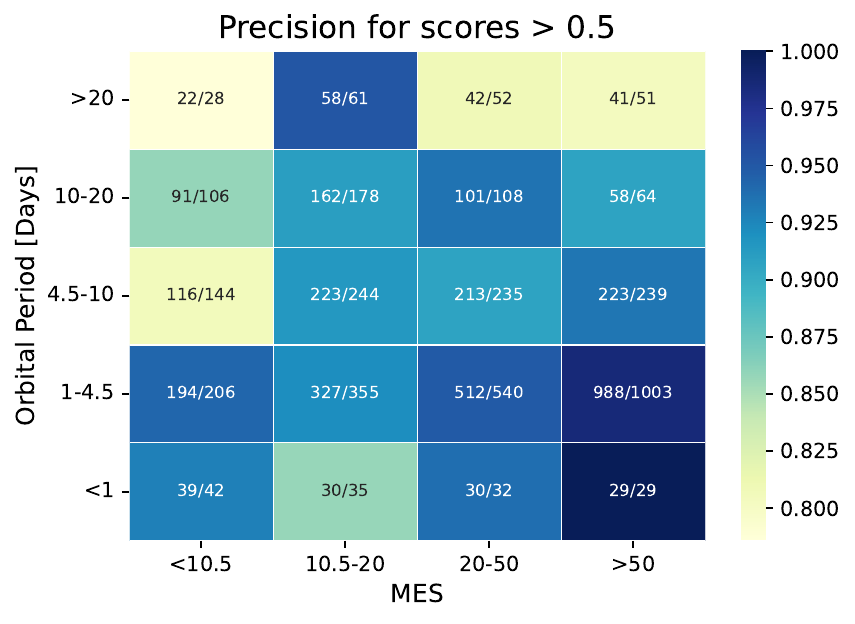}}
    \subfigure[Recall]{\label{fig:mes-period-recall}\includegraphics[width=0.49\linewidth]{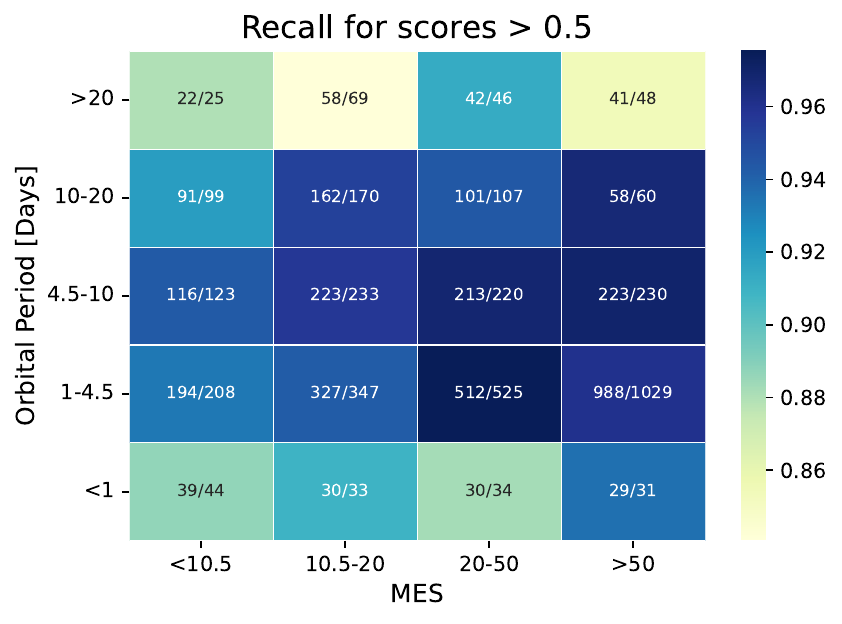}}
    \subfigure[Precision]{\label{fig:mes-period-precision}\includegraphics[width=0.49\linewidth]{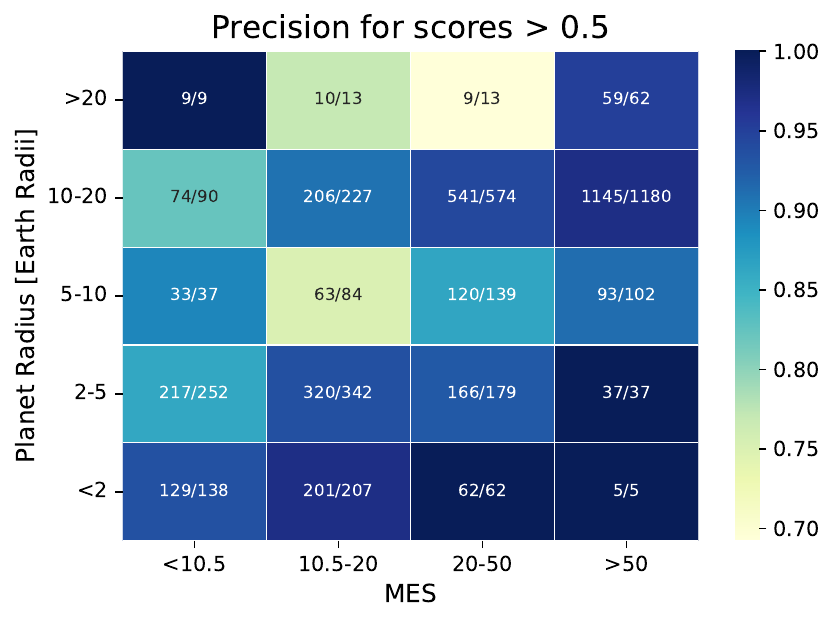}}
    \subfigure[Recall]{\label{fig:mes-period-recall}\includegraphics[width=0.49\linewidth]{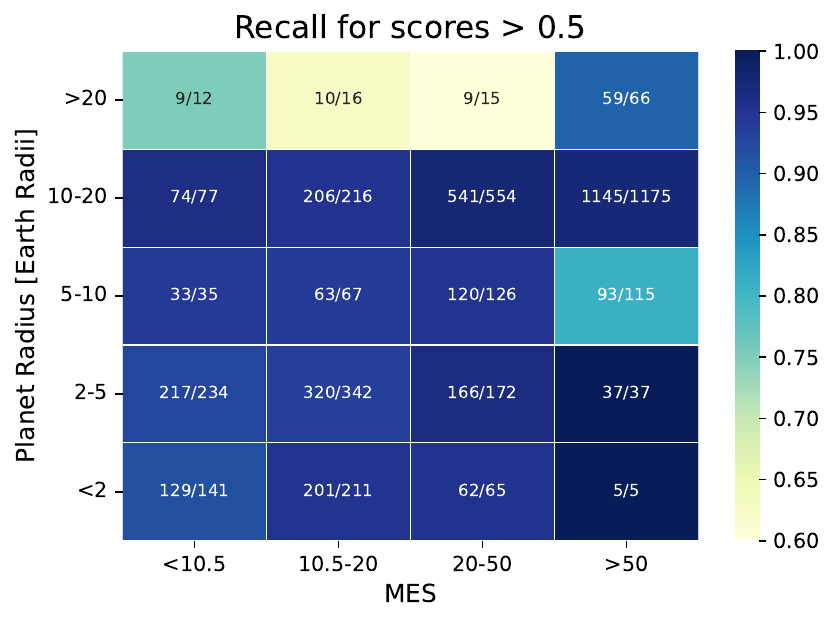}}
\caption{Precision and recall heatmaps as functions of MES, orbital period (days), and planet radius (Earth radii). White cells indicate regions with insufficient data, where the denominator is zero.}
\label{fig:mes-period-kepler-TESS-performance}
\end{figure*}

The previously reported results indicate that \ExoMinerplusplus\ performs significantly better on \kepler\ data compared to TESS. A possible explanation for this disparity might be TESS’s shorter observation windows, which could result in lower signal-to-noise ratios (SNR) for TCEs. To investigate this, we plotted the MES value distributions for \kepler\ and TESS. MES, or Multiple Event Statistic, is a measure used in transit photometry to quantify the statistical significance of a transit signal detected in the light curve of a star. It represents the SNR of a candidate transit event, aggregated over multiple transit events. The MES is calculated by summing the squared SNRs of individual transits, weighted by their uncertainties, and taking the square root. Figure~\ref{fig:mes-kepler-TESS} shows the MES values for both exoplanet and AFP populations in TESS SPOC TCEs (i.e., EBs, FPs, and BDs). Notably, there are more exoplanets with $\rm{MES}<10$ in TESS than in \kepler, primarily due to two factors: 1) most \kepler\ exoplanets were validated using a $\rm{MES}>10$ or $\rm{MES}>10.5$ threshold, and 2) some exoplanets in TESS’s KP category, previously detected by other telescopes, exhibit lower MES values in TESS data due to its relatively small aperture (10~cm) and shorter observation periods. 

Beyond this, \kepler\ shows a greater number of TCEs in the lower MES bins, while TESS dominates the higher MES bins in Figure~\ref{fig:mes-kepler-TESS}. Still, \kepler\ was able to produce more TCEs with very high MES values. To illustrate this, Figure~\ref{fig:mes-kepler-TESS-high} presents a histogram of exceptionally high MES values, showing that \kepler\ yields significantly more TCEs with very high MES scores. However, MES values $>10.5$ are generally considered high enough for reliable planetary candidate identification. Thus, one might say the main reason for the inferior performance of the model on TESS compared to \kepler\ is the existence of numerous exoplanets with low MES values in TESS.

To examine this hypothesis, in Figure~\ref{fig:mes-period-kepler-TESS-performance} we plotted the precision and recall values of different part of the MES versus orbital period and planet radius parameter space for the TESS+\kepler-Aggregate model. There is no clearn pattern indicating that the model has lower performance in small MESS values compared to high MESS values. 
To further confirm this, we removed the exoplanet TCEs with MES $<10.5$ from the data set and remeasured the model performance. This resulted in the lower values of precision and recall. 

Therefore, the question remains: Why does the model perform better for \kepler\ than for TESS? Section~\ref{sec:TESS_difficulty} explores this key question, with implications for model trust and validation.

\subsection{Ranking Performance}
\label{sec:ranking_performance}

\begin{figure}
  \centering
    \includegraphics[width=0.95\columnwidth]{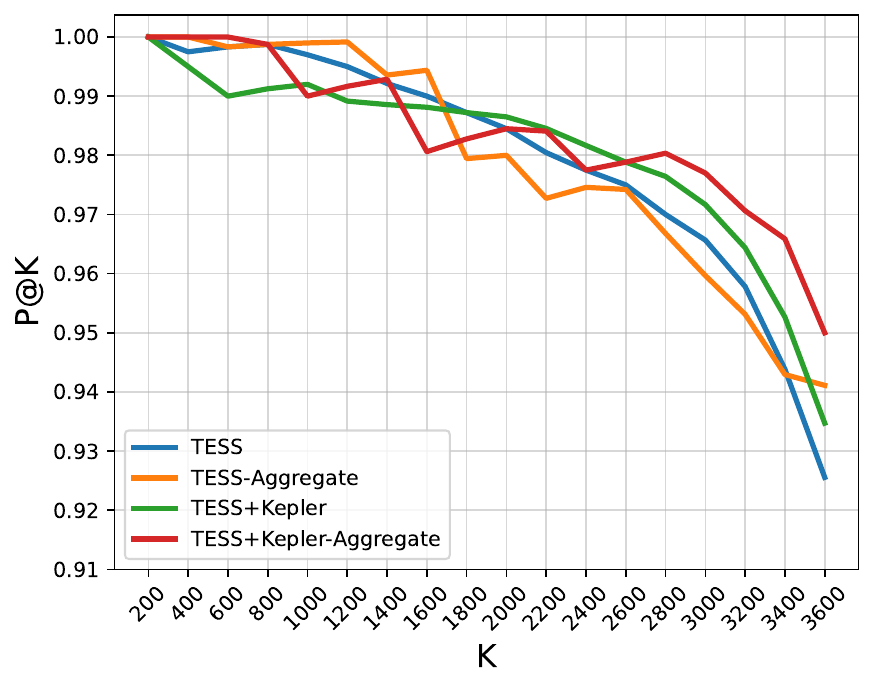}
\caption{Precision@$k$ for the top $k$ TCEs for different values of $k$ and different models. }
\label{fig:p_k_plot}
\end{figure}

Figure~\ref{fig:p_k_plot} shows Precision@k for various $k$ values and models. Given that our dataset contains 3681 exoplanet TCEs, we plot Precision@k up to $k = 3600$. The models perform well in ranking exoplanets near the top. Notably, the top 600 ranked by TESS+Kepler-Aggregate model and top 200 TCEs ranked by all models are exoplanets. At $k = 1000$, all models achieve a Precision of at least 0.99, meaning 990 out of the top 1000 TCEs are exoplanets. By $k = 3000$, Precision remains 0.977 for the TESS+Kepler-Aggregate model, indicating that only around 70 out of the top 3000 are not exoplanets. These indicate that \ExoMinerplusplus\ is highly reliable for follow-up study or validation to find new exoplanets.

\begin{figure}
  \centering
    \subfigure[NTPs]{\label{fig:p_k_NTP}\includegraphics[width=0.7\columnwidth]{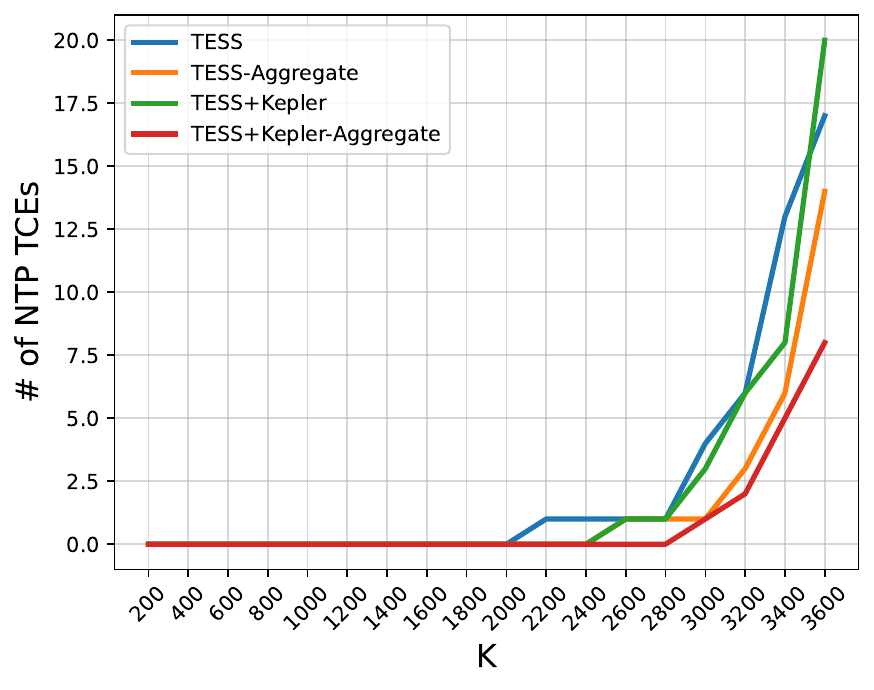}}
    \subfigure[BDs]{\label{fig:p_k_BD}\includegraphics[width=0.7\columnwidth]{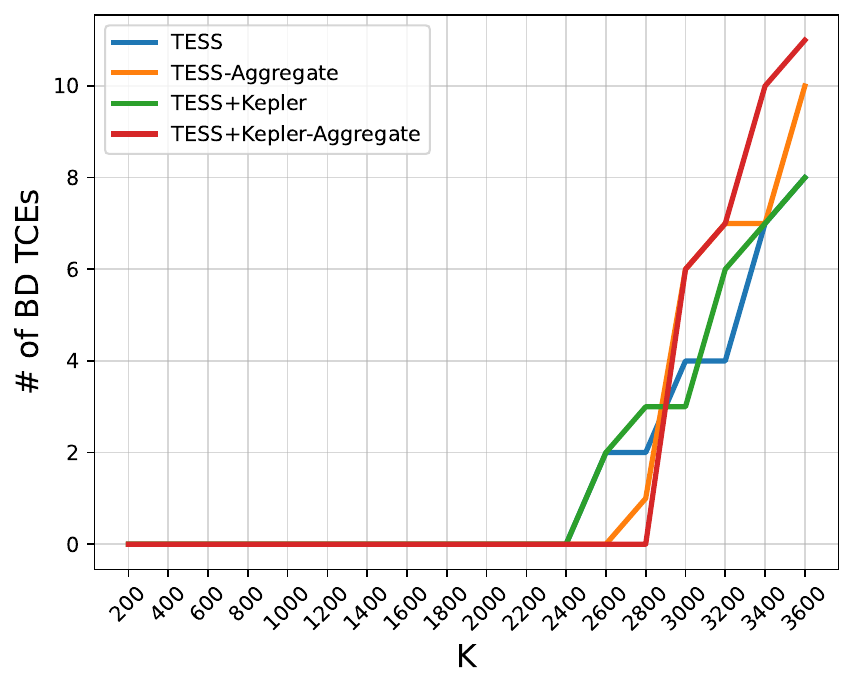}}
    \subfigure[EBs]{\label{fig:p_k_EB}\includegraphics[width=0.7\columnwidth]{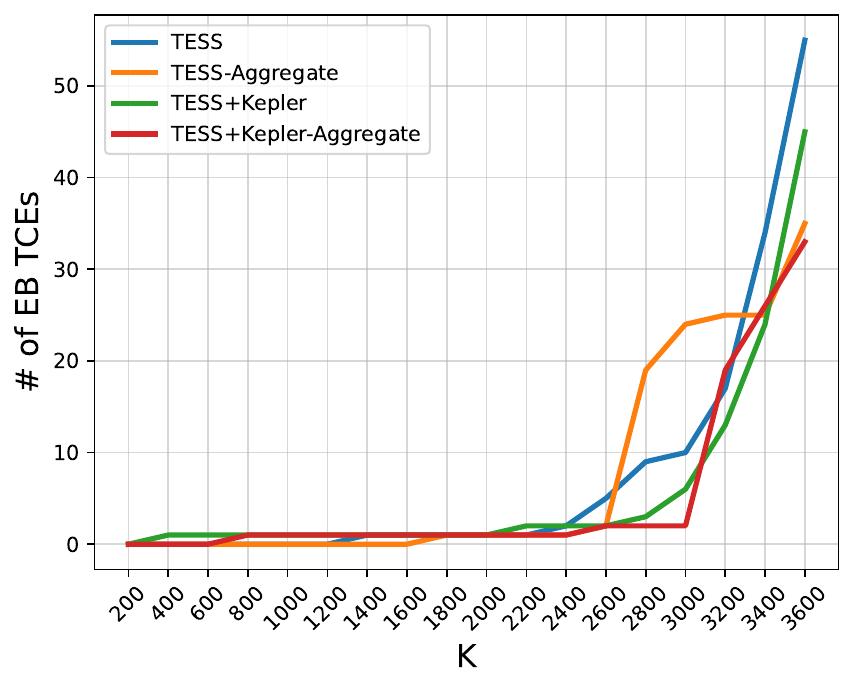}}
    \subfigure[FPs]{\label{fig:p_k_FP}\includegraphics[width=0.7\columnwidth]{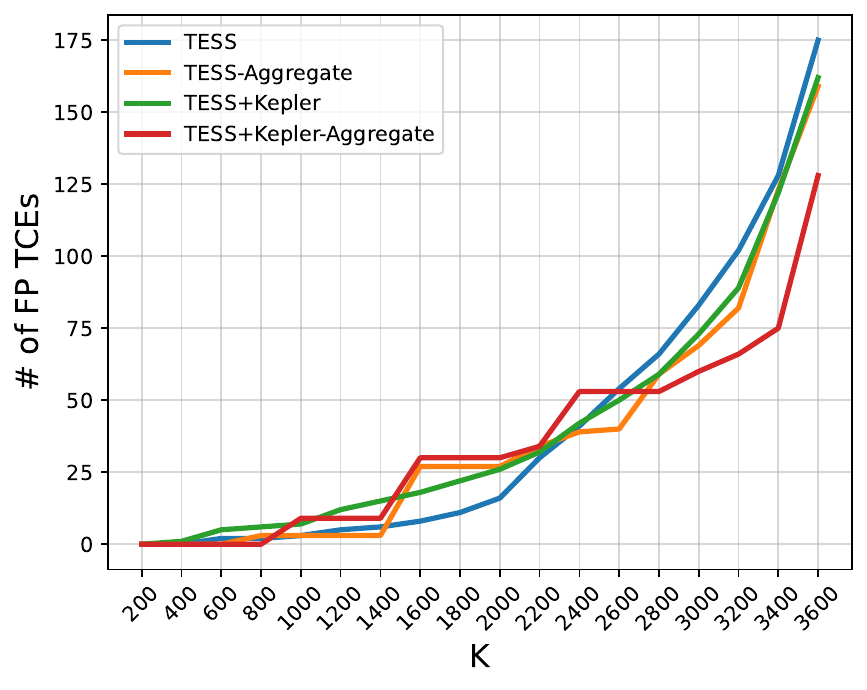}}
\caption{Number of non-planets in the top k TCEs as ranked by \ExoMinerplusplus\ using TESS+Kepler-Aggregate strategy.}
\label{fig:p_k_plots_subclasses}
\end{figure}

It is also important to emphasize that, in our dataset, the label quality of the positive class (exoplanets) is much higher than that of FPs. This suggests that the Precision@k for different $k$ values is likely even better than what is reported here. Overall, this demonstrates that our models are highly effective in selecting the most likely exoplanet candidates for follow-up studies or validation.

To further examine the distribution of misclassified instances among the top $k$ TCEs, we plot in Figure~\ref{fig:p_k_plots_subclasses} the count of misclassified TCEs within each category (NTPs, BDs, EBs, and FPs) from Table~\ref{table:2min_event_counts_labeled_set}. The majority of top-ranked misclassified cases belong to the FP class, followed by a smaller number from EBs, with even fewer from the NTP and BD categories. This trend aligns with Table~\ref{table:performance_metrics}, which shows the FP class as the most challenging and the NTP subclass as the least. Despite the limited number of BDs in the dataset, it is notable that the first BD TCE appears at position 2800 for the TESS+Kepler-Aggregate model, indicating that \ExoMinerplusplus\ effectively assigns lower scores to BDs compared to exoplanets, even though some BDs are still misclassified (score $>0.5$). We will discuss this further in Section~\ref{sec:TESS_difficulty}. 

\subsection{ExoFOP's FP Subcategories Performance}

The ExoFOP disposition is derived from two initial dispositions: photometric and spectroscopic. The mapping from these two dispositions to the final ExoFOP disposition depends on the evidence present in the available datasets. For example, if strong evidence of a false positive is found in the photometric data (such as a confirmed nearby EB), the ExoFOP disposition will be classified as FP. Similarly, 
if there is strong evidence of a stellar companion orbiting the target star at the TESS ephemeris, then the master disposition will take on the spectroscopic SEB1 disposition, which is then converted to FP on ExoFOP. Typically, one of the photometric or spectroscopic evidences is clearly more conclusive. In Sections~\ref{sec:photo-dis} and~\ref{sec:spec-dis} below, we provide a performance study of \ExoMinerplusplus\ on the subcategories for both the photometric and spectroscopic dispositions.

\begin{table*}[htb!]
 \centering
\caption{Labels description of ExoFOP's photometric disposition for FP ExoFOP disposition. Only labels with more than 30 TCEs assignment are shown.} 
\label{table:photo_disposition}
\resizebox{.9\linewidth}{!}
{
\begin{threeparttable}
\begin{tabularx}{\linewidth}{@{}Y@{}}
\begin{tabular}{p{1cm}|c|p{3.5cm}|p{11.5cm}|c}
\toprule
& nmnb & Name & Shortened Description & Count\\
\midrule
\rownumber & VPC+ & Verified Achromatic PC & The event has been verified by SG1 to occur within the target star’s follow-up aperture, with no strong filter-dependent depth chromaticity or contamination from nearby Gaia DR2 or TIC stars bright enough to explain the TESS detection. Such planet candidates are usually retired from SG1 but may remain active for light curve collection in other filters. & 88 \\
\rownumber & VPC- & Verified PC, but follow-up aperture is contaminated & The event has been confirmed within the target star's follow-up aperture, but there are other stars from Gaia DR2 or TIC that are bright enough to potentially contaminate the TESS detection. & 42\\
\rownumber & VPC & Verified PC & SG1 verified within target aperture; no bright Gaia DR2 contaminants. & 179 \\
\rownumber & PC	& PC & Planet candidate with no or inconclusive follow-up observations. & 257 \\
\rownumber & NEB & Nearby EB & Nearby EB contamination confirmed by follow-up photometry, usually over two epochs. NEB disposition can be assigned directly if TESS centroid data strongly supports the initial detection. & 808\\
\rownumber & NPC & Nearby PC & The TESS detection is from an event on a nearby star, which might have a potential planet. The star will get a new TOI number with a PC disposition, and the NPC TOI will be retired as a false positive. & 92\\
\rownumber & BEB & Blended EB & The TESS detection likely results from an EB blending. It is classified as BEB due to chromatic depth, lack of RV variation, or correlated bisector span. If followed by '?' (i.e., BEB?), it’s tentative and needs more observations. & 42 \\
\rownumber & EB & Eclipsing Binary	& The TESS detection likely comes from an event at the target star that is too deep for a transiting planet. It may show odd-even depth differences, eccentric EB traits, or indicate a giant star with a stellar companion. & 126\\
\bottomrule
\end{tabular}
\end{tabularx}
\end{threeparttable}
}
\end{table*}

\begin{table*}[!t]
\footnotesize
\centering
\caption{Recall value of \ExoMinerplusplus\ for photometric disposition.}
\begin{threeparttable}
\begin{tabularx}{\linewidth}{@{}Y@{}}
\begin{tabular}{ccc|cccccccc }
\toprule
\multirow{2}{*}{Training} & \multirow{2}{*}{Test} & \multirow{2}{*}{Strategy}  & \multicolumn{8}{c}{FP}  \\
\cline{4-11}
 \multirow{2}{*}{\empty} & \multirow{2}{*}{\empty}& \multirow{2}{*}{\empty} & VPC+ & VPC- & VPC & PC & NEB & NPC & BEB & EB\\
\midrule
\multirow{2}{*}{TESS}  & \multirow{2}{*}{TESS} & Individual  &      0.636  &     \textbf{0.976}  &     0.877  &     \textbf{0.934}  &     0.932  &     0.478  &     \textbf{1.000}  &     \textbf{0.960}   \\
  &  & Aggregate  &   \textbf{0.795}  &     \textbf{0.976}  &     0.872  &     \textbf{0.934}  &     0.934  &     0.337  &     \textbf{1.000}  &     0.952   \ \\
\midrule
\multirow{2}{*}{TESS+\kepler}  & \multirow{2}{*}{TESS} & Individual  &  0.636  &     \textbf{0.976}  &     0.877  &     \textbf{0.934}  &     \textbf{0.939}  &     \textbf{0.533}  &     0.976  &     0.952      \\
  &  & Aggregate & \textbf{0.795}  &    \textbf{0.976}  &     \textbf{0.888}  &     0.930  &     0.937  &     0.424  &     \textbf{1.000}  &     0.952  \\
\bottomrule
\end{tabular}
\end{tabularx}
\end{threeparttable}
\label{table:photo-dis-stats}
\end{table*}

\subsubsection{Photometric Disposition}
\label{sec:photo-dis}

The photometric disposition includes 20 distinct label assignments, ranging from KP and CP to EB and false alarms (FAs), each with varying degrees of label certainty. These subcategories are subsequently mapped to a final ExoFOP disposition using all available information. Some subcategories, such as VPC+ or PC, initially lean toward the exoplanet classification but are later reclassified as part of the FP category in the final ExoFOP disposition following additional follow-up studies, including spectroscopic observations. Since performance metrics for KP and CP have already been reported in Section~\ref{sec:performance}, we now shift our focus to the photometric subcategories mapped to ExoFOP's FP category, the only other ExoFOP category included in our dataset. To ensure a meaningful performance assessment, we limit our analysis to FP subcategories with at least 30 instances.

Table~\ref{table:photo_disposition} lists the photometric disposition labels with more than 30 instances and their counts, and Table~\ref{table:photo-dis-stats} shows the model performance for each category. We would like to emphasize that the counts we see in Table~\ref{table:photo_disposition} only correspond to the FP category in Table~\ref{table:2min_tce_counts_labeled_set}. This is because the only subcategories in photometric disposition that gets into our labels are the confident ones, i.e., KP, CP, and FP. 

With the exception of the NPC sub-category, multi-sector aggregation performs as well as or better than its individual counterpart. For NPC, even the best model (i.e., TESS+\kepler-Individual) demonstrates very low recall. This is primarily due to the challenges of identifying background transits in TESS, which are exacerbated by its large pixel size, wide field of view, short observation windows, and the characteristics of its target stars. Many of these NPC background transits are located within 1 arcsecond of the target star. Furthermore, the uncertainty in difference imaging is greater in TESS compared to \kepler, due to systematic factors such as pointing errors. We will revisit this issue in Section~\ref{sec:TESS_difficulty}, where we summarize why \ExoMinerplusplus\ performs better on \kepler\ than on TESS. The classifier also struggles with the VPC+ category, which we discuss further in Section~\ref{sec:spec-dis}.

\begin{figure}
  \centering
    \subfigure[Photometric Subcategories Count]{\label{fig:photo-count}\includegraphics[width=0.8\columnwidth]{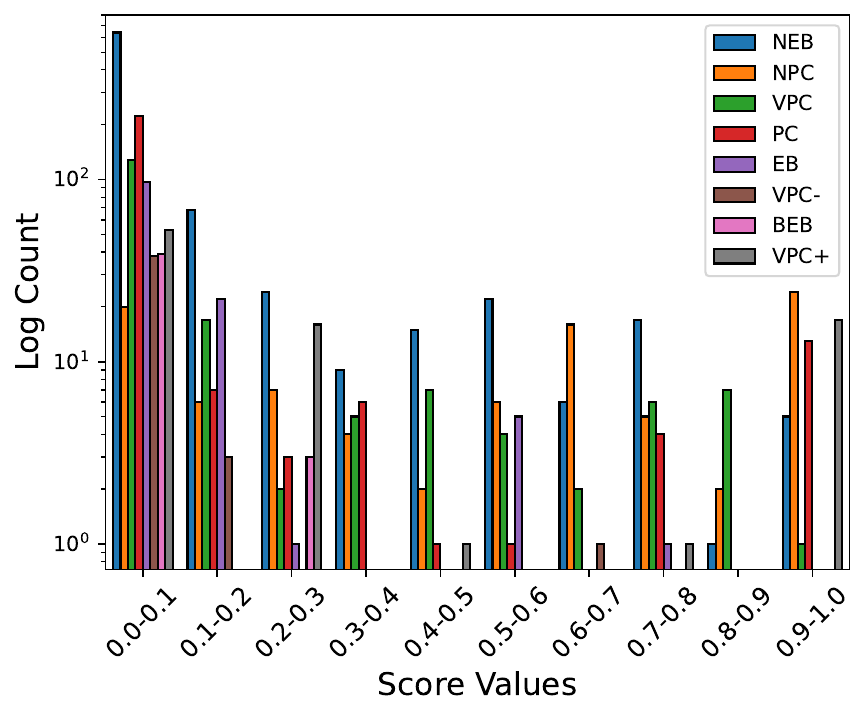}}
    \subfigure[Photometric Subcategories Percentage]{\label{fig:photo-percentage}\includegraphics[width=0.8\columnwidth]{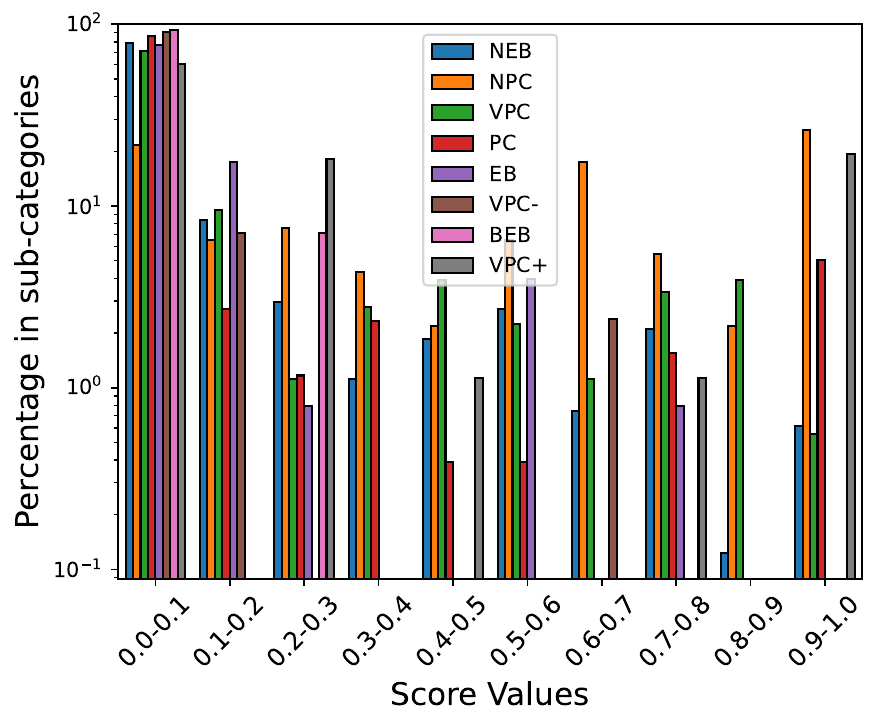}}
\caption{Score histograms of different subcategories in photometric disposition. }
\label{fig:photo-sub-categroies}
\end{figure}

To further examine the model's behavior on these subcategories, we plotted the score histogram in Figure~\ref{fig:photo-sub-categroies}. These histograms reveal that \ExoMinerplusplus\ not only performs inaccurately on the NPC category but also shows significant uncertainty in classifying these cases, even when the model’s disposition is correct (FP), i.e., score $<0.5$. Beyond the NPC class, the model also displays generally low confidence\footnote{Scores are distributed between 0 and 1.} in classifying other subcategories, such as VPC+, PC, and VPC. This is likely because photometric data alone is insufficient for effective labeling of these subcategories, which receive FP labels based on their spectroscopic disposition.

\begin{table*}[htb!]
\centering
\caption{Labels description of ExoFOP's spectroscopic disposition for FP ExoFOP master disposition. Only labels with more than 30 TCEs assignment are shown.} 
\label{table:spec_disposition}
\resizebox{.9\linewidth}{!}
{
\begin{threeparttable}
\begin{tabularx}{\linewidth}{@{}Y@{}}
\begin{tabular}{p{1cm}|c|p{3.5cm}|p{11.5cm}|c}
\toprule
& nmnb &  Name & Description & Count\\
\midrule
\rownumberspec & CRV & Cleared RV & No significant RV variations detected, but expected semi-amplitude is below current sensitivity & 39\\
\rownumberspec & RR & Rapid Rotator & Rapid rotator with too much rotation for DT (generally blindly assigned if Vrot$>$100 km/s) & 37\\
\midrule
\rownumberspec & SB1 & Spectroscopic Binary 1 & Single-lined spectra showing in-phase RV variation too large to be caused by a planet (e.g. only two spectra at opposite quadratures). & 49 \\
\rownumberspec & SEB1  & Spectroscopic EB1 with orbital solution & Single-lined orbital solution with a period and epoch that match the transit ephemeris. & 240 \\
\rownumberspec & SB2& Spectroscopic Binary 2 & Double-lined spectra moving in phase with the photometric orbit (e.g two opposite quad. spectra). & 32\\
\bottomrule
\end{tabular}
\end{tabularx}
\end{threeparttable}
}
\end{table*}

\begin{table*}[!t]
\footnotesize
\centering
\caption{Recall values of \ExoMinerplusplus\ for spectroscopic dispositions.}
\begin{threeparttable}
\begin{tabularx}{\linewidth}{@{}Y@{}}
\begin{tabular}{ccc|ccccc }
\toprule
\multirow{2}{*}{Training} & \multirow{2}{*}{Test} & \multirow{2}{*}{Strategy}  & \multicolumn{5}{c}{FP}  \\
\cline{4-8}
 \multirow{2}{*}{\empty} & \multirow{2}{*}{\empty}& \multirow{2}{*}{\empty}  & CRV & RR & SB1 & SEB1 & SB2\\
\midrule
\multirow{2}{*}{TESS}  & \multirow{2}{*}{TESS} & Individual  &  \textbf{0.872}  &     \textbf{1.000}  &     0.959  &     \textbf{0.796}  &     0.969      \\
  &  & Aggregate &   0.821  &     \textbf{1.000}  &     0.939  &     0.758  &     \textbf{1.000}   \\
\midrule
\multirow{2}{*}{TESS+\kepler}  & \multirow{2}{*}{TESS} & Individual  &  \textbf{0.872}  &     \textbf{1.000}  &     0.980  &     \textbf{0.796}  &     \textbf{1.000}  \\
  &  & Aggregate &   0.821  &     \textbf{1.000}  &     \textbf{1.000}  &     0.771  &     \textbf{1.000}  \\
\bottomrule
\end{tabular}
\end{tabularx}
\end{threeparttable}
\label{table:spec-dis-performance}
\end{table*}

\subsubsection{Spectroscopic Disposition}
\label{sec:spec-dis}
The spectroscopic disposition includes 23 working labels and 18 final labels. As with the photometric disposition, we limit our analysis to labels with more than 30 TCEs that are flagged as FP in ExoFOP. This narrows it down to 5 labels, which are listed in Table~\ref{table:spec_disposition} with their counts. The first two rows represent working dispositions, while the remaining ones are final labels. Table~\ref{table:spec-dis-performance} shows the recall of \ExoMinerplusplus\ for each subcategory.

The total count here is smaller than in the photometric disposition, as more than half of the TCEs with ExoFOP FP labels lack a spectroscopic disposition. The most challenging category for the TESS+\kepler-Aggregate classifier is SEB1, where 55 out of 240 SEB1 cases were misclassified by the TESS+\kepler-aggregate model. Among these, 15 had a photometric disposition of VPC, 17 were VPC+, 18 were PC, and 5 were BEB?. This suggests that, aside from the 5 BEB? cases, which show some FP signatures in the photometric data, the remaining 51 TCEs could not be correctly classified using photometric data alone.

This also explains the poor performance on VPC+, which seems more a result of statistical chance. Although the number of SEB1-labeled VPC+, VPC, and PC TCEs is similar (15-18 instances), the overall count of VPC+ is much lower, leading to poorer performance (higher misclassification rate or lower recall). 

The second most difficult spectroscopic category, CRV, includes 39 TCEs, with 7 misclassified by the TESS+\kepler-aggregate model. Notably, these 7 TCEs were labeled as NEB in the photometric disposition. As discussed in Section~\ref{sec:photo-dis}, the model also struggled with NPC, and given the significantly higher number of NEB-labeled TCEs compared to NPC, this may indicate that the poor NPC performance is coincidental mainly due to low number of total NPCs. Nonetheless, the model finds it difficult to classify NPC and NEB correctly. We discuss the difficulty of classifying background transits for TESS compared to \kepler\ in Section~\ref{sec:TESS_difficulty}.

\begin{figure}
  \centering
    \subfigure[Spectroscopic Subcategories Count]{\label{fig:spec_count}\includegraphics[width=0.8\columnwidth]{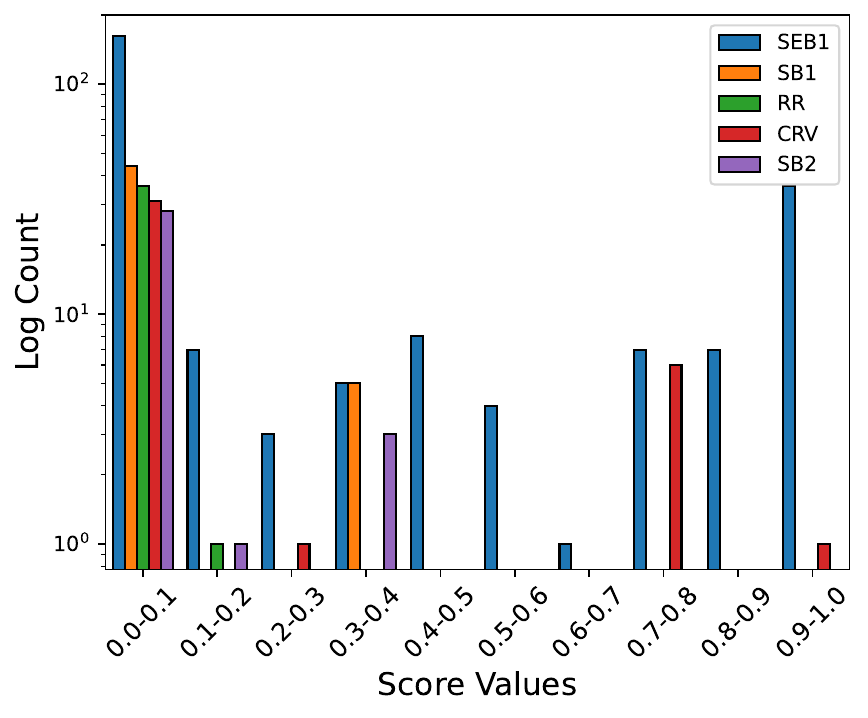}}
    \subfigure[Spectroscopic Subcategories Percentage]{\label{fig:spec_percentage}\includegraphics[width=0.8\columnwidth]{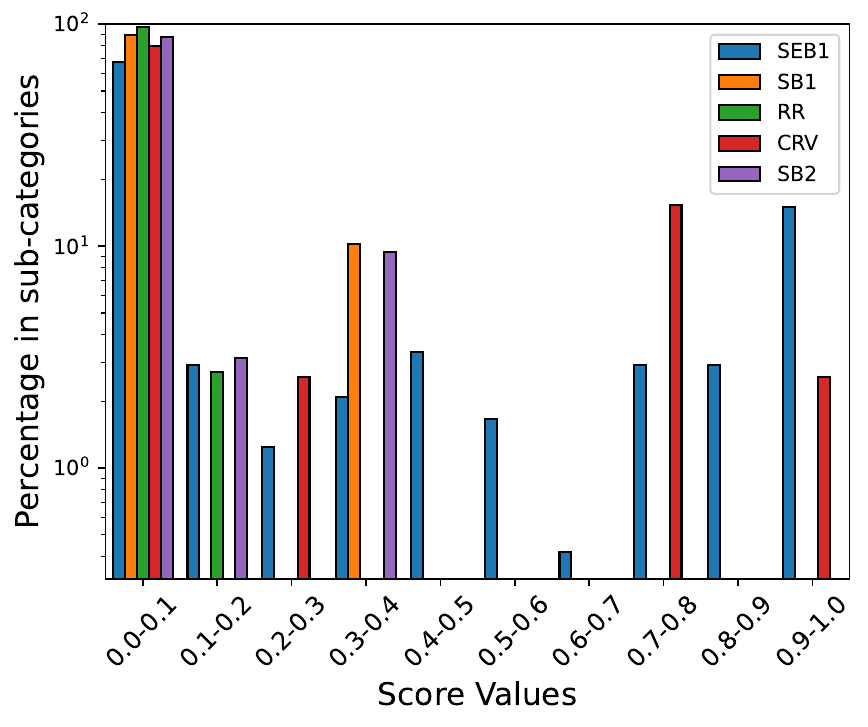}}
\caption{Score histograms of different subcategories in spectroscopic disposition.}
\label{fig:spec-sub-categroies}
\end{figure}

Similar to the photometric disposition, we present the score histograms of \ExoMinerplusplus\ across different spectroscopic subcategories in Figure~\ref{fig:spec-sub-categroies}. Besides the model's poor performance on SEB1, it exhibits greater uncertainty in cases it correctly classifies within spectroscopic binary categories (i.e., SEB1, SB1, and SB2). This may be because spectroscopic binaries are classified based on spectroscopic data, making photometric data alone insufficient for the model to confidently classify them, even though it performs accurately on SB1 and SB2.

\subsection{Difficulty of TESS Data}
\label{sec:TESS_difficulty}

As we mentioned in Section~\ref{sec:exominer++-results}, the lower performance of \ExoMinerplusplus\ on TESS compared to \kepler\ is not solely due to the lower SNR of TESS signals. Instead, this should be studied in terms of the TESS data quality including insufficiency of photometric data and/or imperfect label quality as we discuss below:



\begin{enumerate}
    \item \textbf{Insufficiency of photometric data:} As discussed in Section~\ref{sec:spec-dis}, there are spectroscopic subcategories, such as SEB1, that exhibit clean photometric data (i.e., without any indications of being FPs) and are assigned a non-FP label in the photometric disposition. For instance, among the 241 SEB1 TCEs, 82 are labeled as VPC, 102 as PC, 30 as VPC-, 19 as VPC+, with several other classes having fewer than 5 TCEs. This distribution indicates that photometric data alone is often insufficient to classify these TCEs as FPs. For SEB1, when compared to the photometric labels, \ExoMinerplusplus\ demonstrates better performance than human labelers relying solely on photometric data, suggesting that it can identify patterns not apparent to experts. However, as discussed in Section~\ref{sec:spec-dis}, these spectroscopic subcategories remain challenging for \ExoMinerplusplus.

    The number of known SEB1 cases is higher in TESS compared to \kepler, primarily due to TESS's focus on brighter stars, which facilitates better spectroscopic follow-up. A higher percentage of brighter stars with planets naturally results in more SEB1 detections. For example, TESS SPOC TCEs corresponding to TOI 694.01 are categorized as PC in the photometric disposition and as SEB1 in the spectroscopic disposition. The photometric data shows no indications of being an FP, and specifically no signs of being an EB (i.e., the light curve only exhibits primary eclipses for this likely high mass ratio EB), as illustrated in the DV summary report in Figure~\ref{fig:TOI-694.01}. As noted, \ExoMinerplusplus\ can utilize trend data to detect SEB1 cases exhibiting ellipsoidal variations. However, SEB1 cases with long orbital periods, such as the ones matched to TOI 694.01, lack ellipsoidal variations, making accurate classification based on photometric data particularly challenging. For this specific SEB1 instance, there are 6 TCEs, all of which are misclassified by the model. Notably, \ExoMinerplusplus\ assigns a score of 0.92 to the TCE with the longest transit observation run, i.e., TESS SPOC TCE TIC 55383975-1-S1-65. Similar analysis can be done for other related subcategories such as SEB2, SB1 and SB2.
 

\begin{figure*}
  \centering
    \includegraphics[width=\linewidth]{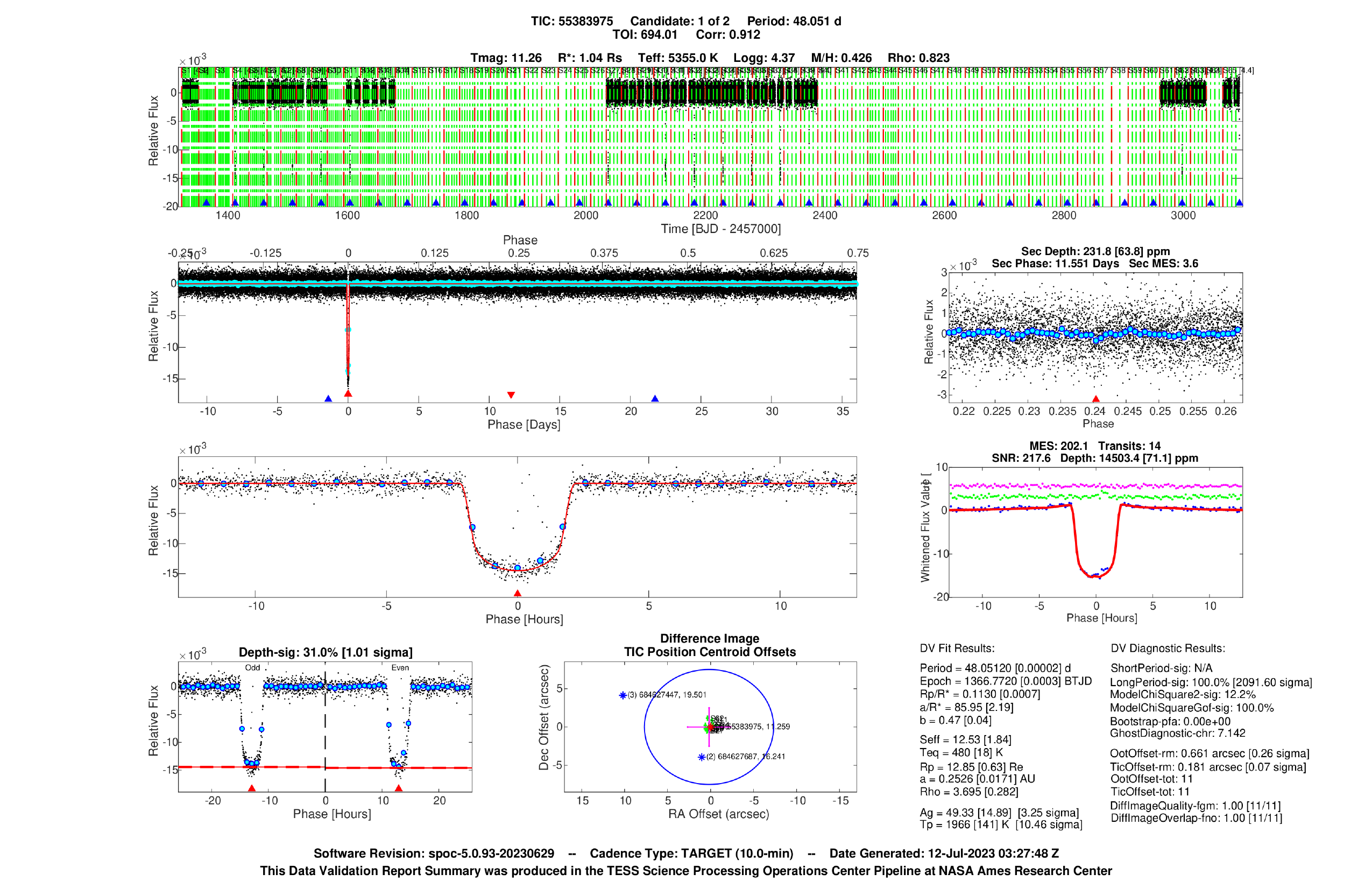}
\caption{SPOC DV summary report for TESS SPOC TCE TIC 55383975-1-S1-65 (TOI 694.01). Photometric disposition is PC however the follow up spectroscopic indicates that this is a SEB1 indeed. }
\label{fig:TOI-694.01}
\end{figure*}

    \item\textbf{Inconclusive diagnostic tests:} Multiple diagnostic tests have been developed to identify FPs~\citep[See--][]{Twicken_2018_DV} originating from various sources. For instance, the weak secondary test aims to distinguish EBs from exoplanets, while the difference image test is designed to detect FPs caused by background transits. However, these tests are inherently imperfect—a challenge also encountered with \kepler. Due to the unique characteristics of the TESS mission, these tests become less reliable in certain regions of the parameter space, as discussed below.

    \begin{enumerate}

    \item \textbf{TCEs for exoplanets with weak secondary:} The weak secondary test is a diagnostic tool used in both \kepler\ and TESS missions to distinguish EBs from exoplanets. It detects faint secondary eclipses in the light curve, which are characteristic of EBs due to the secondary star's contribution to the flux. In \kepler, the high spatial resolution and precise photometry enhance the reliability of the test, allowing clearer identification of weak secondary signals. However, in TESS, the larger pixel scale (21 arcsec vs. 4 arcsec for Kepler), blended light from nearby stars, and shorter observation windows reduce the test's sensitivity and accuracy. This makes it more challenging to detect faint secondary eclipses and differentiate EBs from exoplanets in crowded fields or regions with low signal-to-noise ratios.

\begin{figure*}
  \centering
    \subfigure[Albedo]{\label{fig:spec_count}\includegraphics[width=0.45\linewidth]{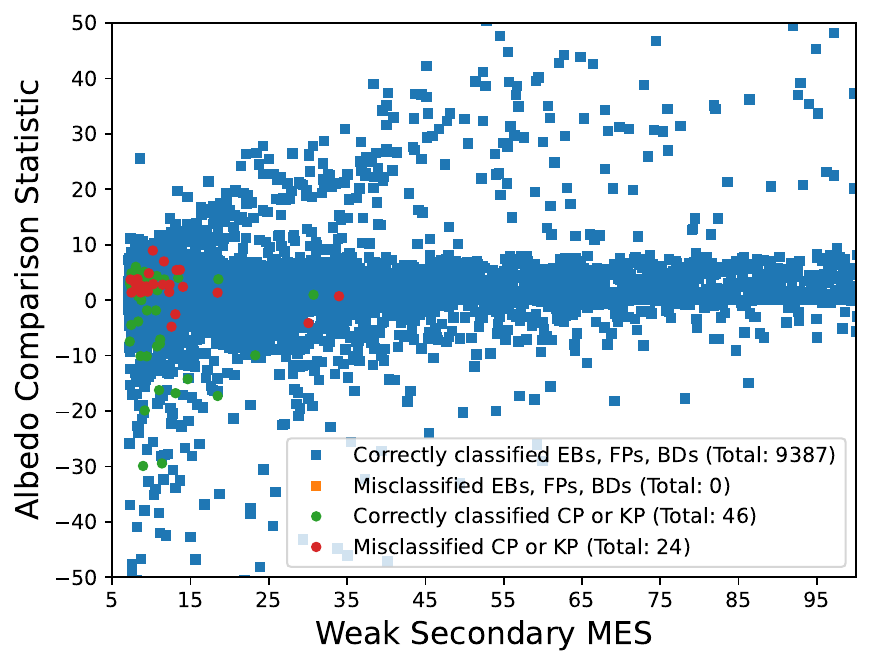}}
    \subfigure[Effective Temperature]{\label{fig:spec_percentage}\includegraphics[width=0.45\linewidth]{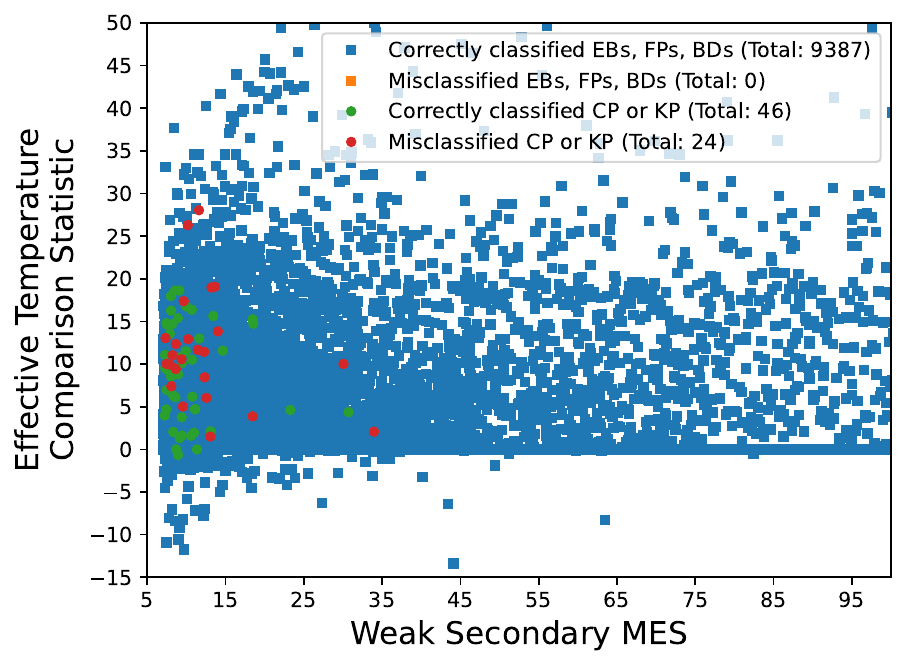}}
\caption{Albedo and planet effective temperature comparison statistics. When weak secondary MES $> 7.1$, the high values of these statistics indicate a FP. However, small values are uninformative.}
\label{fig:weak-secondary}
\end{figure*}
        
    The significant secondary events can also be caused by giant, short-period planets showing reflected light or thermal emission, or inadvertently by transits of other exoplanets in multi-planet systems. Experts differentiate between weak secondary events caused by EBs and those due to planets by examining the secondary geometric albedo, planet effective temperature, and MES of the secondary event~\citep{Twicken_2018_DV, Jenkins2020KeplerHandbook}. A TCE is likely a false positive if the secondary MES exceeds the transit detection threshold 7.1, and the albedo and planet effective temperature comparisons are statistically significant. Otherwise, the test results remain inconclusive.

    Our TESS dataset contains 9,217 EBs, 168 FPs, 2 BDs, and 70 exoplanets (KP+CP) TCEs with weak secondary MES $> 7.1$. \ExoMinerplusplus\ correctly classifies all EBs, FPs, BDs (100\% recall) and 48 exoplanets (65.7\% recall) in this list\footnote{We used TESS+\kepler-Individual for this analysis to assess model performance immediately after training, before post-processing aggregation.}. Figure~\ref{fig:weak-secondary} displays scatter plots of albedo and planet effective temperature comparison statistics against weak secondary MES values for these 9457 (9,217+168+2+70) TCEs. Small values of these statistics provide limited information in distinguishing planets from EBs in the presence of weak secondary, making it difficult for the model to classify TCEs in this region accurately. However, \ExoMinerplusplus\ leverages other branches to correctly classify some exoplanet TCEs with weak secondary in this region.

    In some cases, weak secondary signals arise from other planets in multi-planet systems, like 4600.01 and 1339.02. Earlier \ExoMiner\ versions, lacking the Periodogram and Flux Trend branches, often misclassified these cases. The inclusion of these branches has improved accuracy for such TCEs with weak secondaries. 

    \item \textbf{Misclassified background transits:} As reported in Section~\ref{sec:photo-dis}, the recall for NPCs was sightly higher than 0.5, likely due to the difficulty of distinguishing these TCEs from planets on target stars. TESS’s larger pixel scale, necessitated by its expansive field of view, sacrifices spatial resolution compared to \kepler, which focused on a smaller field with finer spatial detail. The larger pixels in TESS increase the likelihood of blending light from nearby stars (background transits), complicating the distinction between signals originating from the target star and those from nearby sources. Additionally, some TESS fields are highly crowded, such as those along the plane of the Milky Way, further exacerbating this blending issue. 

    This blending significantly complicates the classification of FPs, particularly for background transits, resulting in more errors by \ExoMinerplusplus\ compared to \kepler, where the smaller pixel size and relatively uncrowded target sample allows for clearer photometric signals. For example, in the case of TESS SPOC TCE 77031413-1-S29, the difference image centroid offset points to a background transit source approximately 5.8 arcseconds from the target star—corresponding to just over 0.25 pixel in TESS but nearly 1.5 pixels in \kepler. Consequently, TESS observations are more prone to phantom stars and hierarchical systems, which further complicate the accurate classification of FPs originating from background sources.
    
    The difference image centroid offset statistic, representing the offset divided by its uncertainty, is shown in Figure~\ref{fig:difference-image} for both NPC and NEB subcategories. Notably, some exoplanets have high statistic values (sigma $>5$), complicating the ability to learn from centroid offset alone. A number of the exoplanets with significant offsets orbit saturated stars ($T_{mag} <7$). The change in flux during transit occurs at the top and bottom of saturation bleed trails and the difference images for such targets cannot be reliably centroided. Apparent centroid offsets on saturated stars are not typically considered when vetting TCEs.


    \end{enumerate}

\begin{figure}
  \centering
    \subfigure[NPC versus Exoplanets]{\label{fig:spec_count}\includegraphics[width=\linewidth]{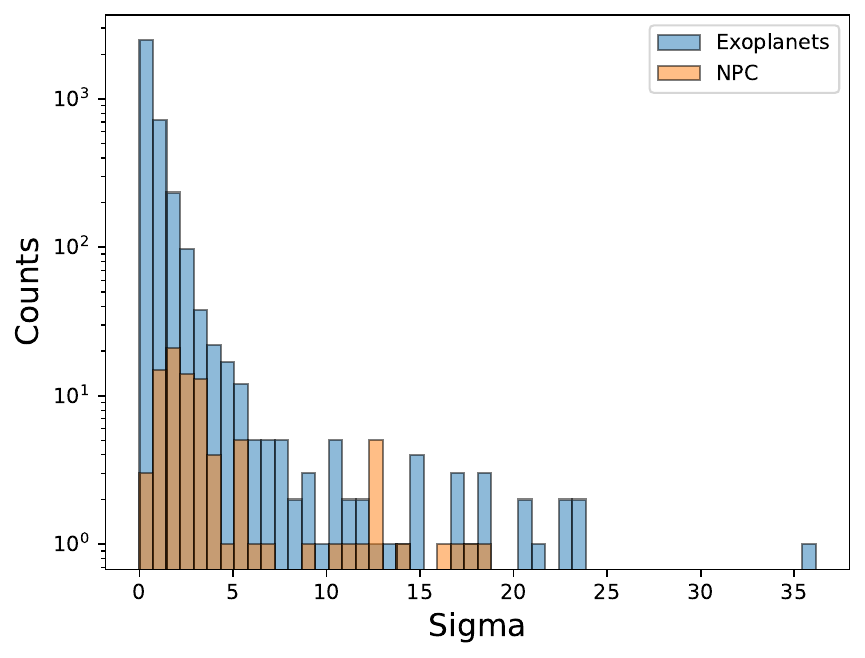}}
    \subfigure[NEB versus Exoplanets]{\label{fig:spec_count}\includegraphics[width=\linewidth]{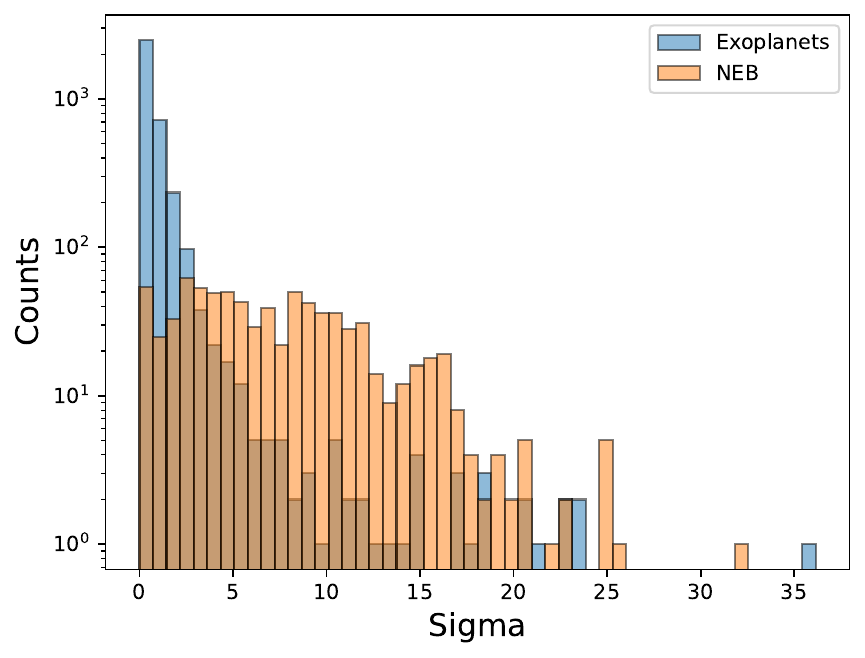}}
\caption{Distribution of centroid offset statistic.}
\label{fig:difference-image}
\end{figure}

    \item\textbf{Incorrect ephemeris and derived parameters:} There are TCEs for which either the calculated ephemeris is incorrect or the derived parameters are inaccurate. The former issue could arise due to the short observational window of TESS, while the latter could result from inaccurate stellar parameters, among other factors. For example, TESS SPOC TCE TIC 337129672-1-S42-43 (TOI 4635.01) has an incorrect period of 49.01 days due to missing values, whereas the correct period is 12 days. The ephemeris information for such cases can be corrected using multi-sector runs. 

    Incorrect ephemeris information may also lead to failed ephemeris matching with TOIs, which in turn results in incorrect labeling, as discussed in the next item.

    \item\textbf{Label quality:} Due to the lack of gold-standard labels for TESS, we relied on surrogate labels from ExoFOP dispositions, the Villanova EB catalog, and TEC's NTPs, which inherently resulted in imperfect labels. While the ExoFOP labels are highly accurate, the Villanova EB catalog, based on manual classification of photometric data, is subject to some inaccuracies. Furthermore, it is important to note that the Villanova EB catalog was constructed using only the first two years of TESS data, and with additional data now available, some dispositions may change. Our model misclassified 15 EBs (37 TCEs) out of 2,557 EBs (12,738 TCEs) from this catalog. In communication with the authors of~\cite{Prsa_2022-EB-catalog}, they acknowledged a higher level of label noise in their catalog. 
    
    We identified five TESS SPOC TCEs---165987272-1-S14-50, 55092869-1-S9, 309792357-1-S1-13, 309792357-1-S1-39, and 309792357-1-S1-36---that were labeled as EBs due to failed ephemeris matching with a known TOI and the sequential nature of our labeling process. Using period matching alone, we determined that the first two are KPs, while the remaining three are CPs. Notably, \ExoMinerplusplus\ correctly classified all five cases. 
    
    Similarly, TEC’s vetting process is not perfect and is subject to labeling errors. For instance, we found TESS SPOC TCEs---280031353-1-S14-26, 288636342-1-S25, and 71347873-1-S20---that were labeled as NTPs due to failed ephemeris matching. These TCEs are actually PCs in ExoFOP and \ExoMinerplusplus\ classifies them as planets.

    Moreover, incorrect periods, such as those being twice or half the true period of a planet, can also lead to misclassification, thereby lowering recall values. The label noise in our datasets not only complicates a thorough evaluation of model performance but can also mislead model training due to systematic errors present in the Villanova EB catalog and TEC.

\end{enumerate}

\subsection{Comparison with Existing TESS Classifiers}
In Section~\ref{sec:existing-classifiers}, we discussed how various existing classifiers can be regarded as different instances of the machine learning classification framework presented in Figure~\ref{fig:architecture}, emphasizing that these classifiers primarily differ in the number and utilization of diagnostic tests they are capable of leveraging. In our previous work~\citep{Valizadegan_2022_ExoMiner}, we quantitatively compared \ExoMiner\ with several existing classifiers on \kepler\ and demonstrated that the incorporation of novel diagnostic tests and the design of \ExoMiner\ led to significantly improved performance.

Aside from the work by~\cite{Tey_2023_astronet}, which integrates some new diagnostic tests, the classifiers employed for TESS~\citep{Yu_2019_AstroNet, Osborn_2020_ExoNet} are largely adaptations of those developed for \kepler. As such, we anticipate similar conclusions for TESS. However, performing a quantitative comparison between \ExoMinerplusplus\ and the classifier proposed by~\cite{Tey_2023_astronet} is challenging for several reasons. First, their source code is not publicly available, limiting direct evaluation. Second, a comparison based on their reported results provides limited insights due to fundamental differences in their classification tasks and datasets. Specifically:

1. The authors in~\cite{Tey_2023_astronet} trained and evaluated their classifier on a dataset constructed through manual visual inspection of photometric data. As we demonstrated in Section~\ref{sec:performance}, there are scenarios where photometric data may appear flawless, yet follow-up studies reveal their FP nature. This implies that their training and evaluation process could have been misled by incorrect labels. 

2. Their classification task involves distinguishing periodic eclipsing signals (encompassing planetary transits and non-contact EBs, both on- and off-target) from contact EBs, single-transit events, and non-transiting phenomena (e.g., stellar variability and instrumental noise). This task is a ``relaxed'' version of the classification task addressed in our work, as it groups planetary transits with other periodic signals.

3. Their dataset consists of TCEs detected by the Quick-Look Pipeline (QLP) from TESS full-frame image (FFI) data. To compare the datasets, one would need to identify the subset of TCEs from the two pipelines that correspond to the same events and obtain the scores generated by the classifier in~\cite{Tey_2023_astronet}, which are not publicly available.

Out of curiosity, we conducted ephemeris matching between the publicly available dataset used in~\cite{Tey_2023_astronet} and our dataset. This matching yielded approximately 3.2k shared TCEs, representing a small fraction of our dataset. This limited overlap can be attributed to the fact that the authors in~\cite{Tey_2023_astronet} focused on a restricted selection of QLP TCEs from the first three years of the \TESSMission. Moreover, only a fraction of their dataset (10\%) was allocated for testing, further reducing the subset of TCEs available for performance comparison.

\subsection{Ablation Experiments for New \ExoMinerplusplus\ Branches}
\label{sec:new_branches}

As described in Section~\ref{sec:exominer++}, we have introduced five new branches to the \ExoMiner\ architecture when designing the new model \ExoMinerplusplus. These branches were added to provide the model with additional information that, according to SMEs and our understanding of the challenges and specifics of the \TESSMission, could be leveraged to improve the performance of the model. To substantiate these hypotheses, we conducted ablation experiments in which we investigated the changes in model's performance due to the inclusion of these new branches. We started by training and evaluating a baseline model (Flux) that uses only the transit- and full-orbit-view flux branches. For each new branch, we trained and evaluated a model that consisted of the branches found in the baseline model plus the branch of interest. Similar to the results obtained for the full \ExoMinerplusplus\ model in Section~\ref{sec:exominer++-results}, we used the same setup described in Section~\ref{sec:hpo}, and trained and evaluated all models on TESS data. The metrics were computed for the `Individual' strategy because we were interested in studying the direct effect of these branches on the model just after the training without any post-processing. 

\begin{table*}[!t]
\centering
\caption{Results of ablation experiments for the new branches added to \ExoMinerplusplus. Models were trained and evaluated only on TESS data. The best performer is highlighted in bold.}
\resizebox{.9\linewidth}{!}{
\begin{threeparttable}
\begin{tabularx}{\linewidth}{@{}Y@{}}
\begin{tabular}{c|cccc|cc|cccc }
\midrule
 \multirow{3}{*}{Model} & \multicolumn{4}{c|}{\multirow{2}{*}{Binary results}} & \multicolumn{6}{c}{Recall for subclasses}  \\
\cline{6-11}
 \multicolumn{1}{c|}{\empty} & \multicolumn{4}{c|}{\empty} & \multicolumn{2}{c|}{Exoplanets}  & \multicolumn{4}{c}{Non-planets}  \\
\cline{2-11}

  & Precision \& Recall & PR AUC & ROC AUC & Accuracy & KP & CP & BD & EB & FP & NTP\\
\midrule
    \multirow{1}{*}{Flux (baseline)}  &   0.811 \&     0.863 &     0.897 &     0.988 &     0.978 &  0.910 &     0.816 &     0.656 &     0.985 &     0.746 &  0.997 \\
\midrule
\midrule
    \multirow{1}{*}{Difference Image}    &  0.850 \&     0.831 &     0.904 &     0.989 &     0.980 &     0.891 &     0.772 &     0.719 &     0.989 &     0.801 &  0.999 \\
\midrule
    \multirow{1}{*}{Flux Trend}    &  0.835 \&     0.876 &     0.923 &    0.991 &     0.981 &     0.917 &     0.835 &  0.656 &     0.986 &     0.772 &  0.998\\
\midrule
    \multirow{1}{*}{Periodogram}   &  0.851 \&     0.842 &     0.911 &     0.990 &     0.980 &     0.893 &     0.793 &     0.750 &     0.989 &     0.790 &  0.999 \\
\midrule
    \multirow{1}{*}{Unfolded Flux}   &  0.839 \&     0.866 &     0.913 &     0.990 &     0.981 &     0.900 &     0.832 &    0.625 &     0.987 &     0.754 &    1.000\\
\midrule
    \multirow{1}{*}{Momentum Dump}   &  0.809 \&     0.867 &     0.897 &     0.989 &   0.978 & 0.916 &     0.818 &     0.656 &     0.984 &     0.738 &     0.998 \\

\bottomrule
\end{tabular}
\end{tabularx}
\end{threeparttable}
}
\label{table:branch_analysis_recall_major_disps}
\end{table*}

\begin{table*}[!t]
\centering
\caption{Recall for photometric and spectroscopic disposition using results from ablation experiments conducted for the new branches added to \ExoMinerplusplus. Models were trained and evaluated only on TESS data. 
}
\begin{threeparttable}
\begin{tabularx}{\linewidth}{@{}Y@{}}
\begin{tabular}{c|cccccccc || ccccc}
 \toprule
\multirow{3}{*}{Model}  & \multicolumn{8}{c||}{FP Photometric}  & \multicolumn{5}{c}{FP Spectroscopic} \\
\cline{2-14}
 & VPC+ & VPC- & VPC & PC & NEB & NPC & BEB & EB  & CRV & RR & SB1 & SEB1 & SB2 \\
\midrule
\multirow{1}{*}{Flux (baseline)} &  0.432 & 0.976 & 0.749 & 0.899 & 0.743 & 0.304 & 0.905 & 0.976  &  0.769 & 0.946 & 0.959 & 0.729 & 0.969\\
\midrule
\midrule
\multirow{1}{*}{Difference Image} &  0.500 & 0.976 & 0.777 & 0.922 & 0.821 & 0.337 & 0.952 & 0.984 &  0.872 & 0.946 & 0.959 & 0.746 & 0.938 \\
\midrule
\multirow{1}{*}{Flux Trend} &  0.523 & 0.976 & 0.821 & 0.911 & 0.756 & 0.283 & 0.952 & 0.984 &  0.795 & 0.973 & 0.939 & 0.792 & 0.969\\
\midrule
\multirow{1}{*}{Periodogram} & 0.648 & 0.976 & 0.888 & 0.914 & 0.759 & 0.315 & 0.952 & 0.976  &  0.846 & 0.919 & 1.000 & 0.808 & 0.938\\
\midrule
\multirow{1}{*}{Unfolded Flux} &  0.443 & 0.976 & 0.777 & 0.918 & 0.752 & 0.283 & 0.929 & 0.960 &  0.744 & 0.946 & 0.918 & 0.729 & 0.938 \\
\midrule
\multirow{1}{*}{Momentum Dump} &  0.443 & 0.976 & 0.760 & 0.895 & 0.731 & 0.239 & 0.905 & 0.968  &  0.590 & 0.946 & 0.959 & 0.704 & 0.906\\
\bottomrule
\end{tabular}
\end{tabularx}
\end{threeparttable}
\label{table:branch_analysis_recall_photo_spec_disps}
\end{table*}

Table~\ref{table:branch_analysis_recall_major_disps} shows the performance metrics for each of these models and their recall values (for a threshold of 0.5) for the multiple subclasses. All models except for the `Momentum Dump' model show an increase in the PR AUC when compared to the baseline model, suggesting that all these branches contribute to an overall better separation between planet and non-planet TCEs.

The top performing branch is `Flux Trend', demonstrating that providing information about the fitted trend is useful to the model. We hypothesized that these data might contain other signals besides the transit event that occur at the time scale of the detected orbital period. As mentioned previously, examples of such cases include ellipsoidal variations of short-period EBs such as the one shown in Figures~\ref{fig:pdcsap_flux_trend_correctly_classified_example} and~\ref{fig:flux_trend_correctly_classified_example}. In this particular case, the baseline model classified incorrectly this EB as a planet with a high score of 0.88. On the contrary, the `Flux Trend' model, which has access to information about the ellipsoidal variations through the phase-folded trend flux time series, assigned a score of 0.09 to this TCE, correctly identifying it as a non-planet.


Focusing on the FP subclasses, the `Periodogram' branch exhibits the highest recall for the brown dwarf subset, leading to three fewer misclassifications relative to the baseline model. This suggests that frequency domain information can be useful to distinguish these objects from their planet counterparts. However, given the small sample size of this subclass, this conclusion should be taken with caution. 

For the EB subclass, both `Difference Image' and `Periodogram' models show the highest increase in recall. While the flux trend data can provide information about a subpopulation of EBs with visible ellipsoidal variations, it seems that there are more information conveyed in the pixel data as well as in the frequency domain that helps in the correct classification of these TCEs. 

For the FP subclass, the `Difference Image' exhibits the highest recall. These results are studied in more detail below when discussing the performance using photometric dispositions for this subclass.

Finally, although the baseline model already shows a high recall for the NTP subclass, the `Unfolded Flux' model shows the highest recall for this subclass among all models evaluated in this experiment. This is expected because the main reason for the inclusion of unfolded flux data is to study the inconsistency of the transits over different periods, which is mainly used to identify NTP transits. 

Table~\ref{table:branch_analysis_recall_photo_spec_disps} shows the recall values for different types of ExoFOP FPs based on their photometric and spectroscopic dispositions. Some of these disposition categories have a small set of TCEs (e.g., BEB, CRV, RR, SB1, and SB2), and so the results should be interpreted with caution. The analysis of the results for the photometric dispositions shows that the `Difference Image' branch is particularly effective at improving the classification of background objects (i.e, NEBs and NPCs), leading to 66 fewer misclassifications out of a total of 900 background objects TCEs. As expected, a significant fraction of these objects shows a clear transit source offset in the difference image data which can be used by the model to infer that these transiting events are not happening on the target. Furthermore, the `Difference Image' branch improves the recall for PC subclass, which suggests that, although these planet candidates do not have any follow-up photometric observations (or if they do, they are inconclusive), the pixel data that are processed through the `Difference Image' branch leads the model to correctly classify more of these instances as non-planet TCEs.

The results for the EB subclass suggest that adding the `Difference Image' and `Flux Trend' branches leads to performance improvement for this subcategory. In a similar fashion, these two branches and the `Periodogram' seem to improve the recall for BEBs. However, given the size of these two sets and the difference in recall compared to the baseline model, these findings need to be taken with caution. 

The `Periodogram' branch also reveals itself significantly effective in improving the classification of VPC+ and VPC subclasses. Interestingly, all models obtain the same recall as the baseline for VPC- subclass. According to the definition in Table~\ref{table:photo_disposition}, VPC- are transiting events that have been confirmed by SG1 to occur within the target star aperture in follow-up observations, but there are known stars contaminating the aperture that are bright enough to be potential sources of the transit. Given this, providing additional difference image data to the model is insufficient to determine whether the transiting event occurs on-target, which explains why we see similar performance for this subset of TCEs across all branch models, including the `Difference Image' model which was designed to help distinguishing detection on neighboring stars. This is a known limitation of \ExoMinerplusplus\ model, and one that we plan to address in future work.

As of now, the model does not have information on nearby stars that potentially contaminate the aperture. We hypothesize that an improved representation of the difference image with additional channels that provide the model information about the location of known background stars in the postage stamp and their brightness values can be leveraged by the model to decrease the scores for crowded field scenarios, thus improving the quality of vetting catalogs produced using \ExoMinerplusplus\ and the validation and follow-up of its planet candidates.

For the VPC subclass, the `Periodogram' data seems to carry useful information that is employed by the model to improve its sensitivity to these observations relative to the baseline model. According to the definition in Table~\ref{table:photo_disposition}, these events have been verified to occur within the target star aperture and no known stars in the vicinity are bright enough to cause the detection. Nonetheless, based on the performance of the `Periodogram' branch model, there is some signature in the frequency domain that makes the model more assertive in classifying these events as false positives. In the future we will conduct explainability studies to investigate which frequencies in the periodogram contribute the most to the model's score for this and other sets of objects.

Regarding the spectroscopic dispositions of ExoFOP's FP TCEs, this analysis shows that the `Periodogram' branch can be effective at distinguishing some types of spectroscopic EBs, including SEB1, which is the largest spectroscopic category (see Table~\ref{table:branch_analysis_recall_photo_spec_disps}), and SB1. Again, the results for SB1 and SB2 should be studied with caution given their smaller sample size. All other branches do not seem to convey any meaningful information that can help the model correctly classify these observations. Rapidly rotating stars (RR subcategory) can induce transit-like events in the photometry data. This type of stellar variability leads to the creation of FP TCEs, especially if these stars have spots. The results in Table~\ref{table:branch_analysis_recall_photo_spec_disps} suggest that the `Flux Trend' branch model helps in the classification of RR TCEs, while all other branches do not outperform the baseline. Despite being a small set of examples, the trend data might capture some of the variability induced by these objects in the light curve time series. As for the CRV disposition, the results for the  `Difference Image' and `Periodogram' models suggest that these branches might help the model in correctly identifying these events. This is interesting  since in these cases no significant radial velocity variations were detected, but the expected value for the semi-amplitude is below the sensitivity threshold. However, we refrain from making final conclusions until we have a larger number of TCEs in this category and a better sense of how these data can help.

All these results point to the usefulness of the new branches added to \ExoMinerplusplus\ to boost the performance of the model. The only exception to this is the `Momentum Dump' branch. The design and inclusion of this branch had been motivated by the desire to provide information to the model about TCEs created by flux level artifacts associated with momentum dump events. However, most of such cases are effectively filtered by the SPOC transit search pipeline\footnote{This occurred early in the TESS mission, but the SPOC pipeline implemented de-emphasis weights to mitigate such occurrences.}, and so it is difficult to find a set of momentum dump TCEs in the data. This means that the model does not see many examples that are relevant to this branch during training, and at the same time it becomes difficult to evaluate the model on such a small sample and provide meaningful statistics.

We note that our ablation experiment does not account for higher-order interactions among branches (e.g., simultaneously adding the `Difference Image' and `Periodogram' branches to the baseline model) due to the increased combinatorial complexity and resource demands. In future work, we plan to perform an in-depth explainability analysis to gain deeper insights into the model's behavior. These insights could not only inform the design of improved models but also provide valuable information to researchers seeking to understand and utilize the model's predictions.

\section{Building a Vetting Catalog for TESS 2-min Data}
\label{sec:catalog}

One of the key outcomes of this study is the construction of a comprehensive vetting catalog for TESS 2-minute cadence data using \ExoMinerplusplus. This catalog serves as a curated resource of transit-like signals, classifying them as likely exoplanets or non-planets. Leveraging the advanced classification capabilities of \ExoMinerplusplus, the catalog aims to provide the TESS and broader exoplanets community with a reliable dataset for follow-up investigations and population studies.

As detailed in Section~\ref{sec:data-preparation}, the labeling process assigned labels to 57,162 TCEs out of a total of 204,729 TCEs, leaving 147,567 unlabeled TCEs, referred to as unknowns (UNKs). To classify these UNKs, we applied the ten TESS+\kepler\ \ExoMinerplusplus\ models trained using 10-fold cross-validation and averaged their scores for final classification. Table~\ref{table:vetting-catalogue} presents the labels and detailed scores for these 147,567 UNK TCEs, of which \ExoMinerplusplus\ identified 7,330 as planet candidates, based on the average score of 10 models and the application of aggregation strategy (Section~\ref{sec:multi-sector-aggregation}) on the top (last column of Table~\ref{table:vetting-catalogue}).

Table~\ref{table:vetting-catalogue-TOI} presents the dispositions and scores for the TOIs matched to the unknown TCEs, using the aggregate strategy to obtain a single disposition/score for the case of multiple TCEs corresponding to the same TOI. Out of 2,806 matched TOIs, \ExoMinerplusplus\ classified 1,868 as planet candidates.

It is important to note that both Table~\ref{table:vetting-catalogue} and Table~\ref{table:vetting-catalogue-TOI} include two distinct columns for TFOPWG dispositions. The first column corresponds to data downloaded in January 2024, which was used to construct the dataset for training and evaluating \ExoMinerplusplus. The second column corresponds to the TFOPWG table downloaded during the final phase of this work in December 2024, which was utilized to build the vetting catalog. 

Over time, some uncertain TFOPWG dispositions (i.e., PC, APC, and FA) have been updated to more certain dispositions such as CPs and FPs, as reported in Table~\ref{table:confusion-matrix-exofops}. 


\begingroup
\setlength{\tabcolsep}{2.5pt} 
\begin{table}[htb]
\centering
\caption{The ExoFOP label changes over an 11-month period, reported in terms of the number of TCEs. The numbers in parentheses represent the corresponding number of TOIs. The labels of some TCEs changed from uncertain labels to certain labels CP and FP.}
\label{table:confusion-matrix-exofops}
\begin{threeparttable}
\begin{tabularx}{\linewidth}{@{}Y@{}}
\begin{tabular}{c|c|ccc}
\toprule
 \multicolumn{2}{c|} {} & \multicolumn{3}{c}{Jan 2024 labels} \\
\midrule
& & PC & APC & FA\\
\multirow{5}{*}{\rotatebox{90}{Dec 2024 labels}} & PC & 7990 (2423) & 0 (0) & 0 (0)\\
& FA  & 1 (1) & 5 (1) & 77 (43) \\
& APC  & 98 (23) & 958 (240) & 0 (0)\\
\cline{2-5}
& FP  & 55 (27) & 87 (13) & 0 (0)\\
& CP & 151 (32) & 0 (0) & 0 (0)\\
\bottomrule
\end{tabular}
\end{tabularx}
\end{threeparttable}
\end{table}
\endgroup

The difference between these two TFOPWG tables provides an opportunity to compare the performance of the \ExoMinerplusplus\ TCE catalog with the TOI TFOPWG uncertain dispositions, such as PC, APC, and FA, and to understand the expected value of \ExoMinerplusplus\ over time. However, it is important to note that this analysis should be interpreted with caution due to the small sample size of changes between January 2024 and December 2024. 

The December 2024 TFOPWG table has 151 new CP TCEs and 142 new FP TCEs, that were labeled with uncertain labels (PC, APC, and FA) in January 2024 table. These TCEs were labeled as unknown in our data set and were not used for training or evaluation of the model. We can now use them to provide some insight into what to expect from \ExoMinerplusplus\ over time. \ExoMinerplusplus\ correctly classified 147 out of 151 CP TCEs and 95 out of 142 FP TCEs (precision=0.76, recall=0.97). In comparison, the TFOPWG January 2024 dispositions labeled all 151 CPs and 55 FPs as PCs, with the remaining FPs labeled as APC, resulting in precision=0.50, recall=1.0. This demonstrates that \ExoMinerplusplus\ significantly enhances the efficiency of the planet search process by providing more accurate classifications, achieving higher precision at the modest cost of recall.

Finally, we were interested in identifying whether there are new TOIs discovered by \ExoMinerplusplus. Out of a total of 7,330 UNK TCEs classified as PC, 6,322 TCEs were already matched to 1,868 existing TOIs using the procedure described in Section~\ref{sec:data-preparation}, leaving 1,008 TCEs without a TOI match. However, as discussed earlier, our ephemeris matching process may occasionally fail to associate a TCE with known TOIs. Furthermore, there are Community TOIs~(CTOIs)\footnote{\url{https://exofop.ipac.caltech.edu/tess/view_ctoi.php}} that we should exclude from consideration. 

To address this, we performed an aggressive period matching by removing any TCEs for a target star that had a TOI or CTOI with a similar period, using a 1\% threshold for matching ($|P_{\mathrm{TOI}} - P_{\mathrm{TCE}}| < 0.01 \times P_{\mathrm{TOI}}$). This process excluded 512 TCEs matched to existing TOIs and 69 matched to existing CTOIs, leaving 427 new TCEs classified as PC by \ExoMinerplusplus. 

Subsequently, we conducted another level of period-based ephemeris matching among these 427 TCEs, resulting in a total of 288 unique events. To generate a more conservative list of CTOIs, we applied two additional vetoes: (1) ensuring that all 10 models trained using 10-fold cross-validation classified the TCEs with the longest sector run of these events as PC (score $> 0.5$), and (2) requiring that each TCE exhibited at least three observed transits. This refinement resulted in 91 unique events that are potentially new CTOIs. 

The SMEs on the team then manually examined these 91 potential CTOIs, ultimately rejecting 41 due to various reasons, the majority being related to target stars that were determined to be duplicates, or artifacts in TIC-8. This left us with 50 new CTOIs, which are reported in Table~\ref{table:new-TOI-catalog}.

By making these catalogs publicly available, we aim to support ongoing and future studies, ranging from follow-up investigations of the most promising candidates to detailed atmospheric characterizations and population-level analyses of exoplanets. This catalog represents a significant advancement in the automated vetting of TESS 2-minute cadence data, bridging the gap between detection and validation in the search for worlds beyond our solar system.

\begingroup
\begin{table}[htb]
\centering
\caption{Scores and dispositions of TESS+\kepler-Aggregation model for UNK TCEs. This table describes the available columns. The full table is available online.}
\label{table:vetting-catalogue}
\begin{threeparttable}
\begin{tabularx}{\linewidth}{@{}Y@{}}
\begin{tabular}{p{3cm}p{5cm}}
\toprule
Column & Description \\
\midrule
uid & Unique ID that includes TIC ID, planet number, and sector run\\
target\_id & TIC ID\\
tce\_plnt\_num & TCE planet number\\
toi & TOI number\\
tce\_period & TCE period\\
tce\_duration & TCE duration\\
tce\_prad & TCE planet radius (Earth Radii)\\
mes & TCE MES\\
TFOPWG Disposition (January 2024) & One of multiple TFOPWG Dispositions downloaded in January 2024 if available for a TCE\\
TFOPWG Disposition (December 2024) & One of multiple TFOPWG Dispositions  downloaded in December 2024 if available for a TCE\\
score\_fold\_i & for $i\in[0,9]$, this is the score of \ExoMinerplusplus\ model trained for fold i\\ 
\ExoMinerplusplus\ score & Average score of 10 \ExoMinerplusplus\ models\\ 
\ExoMinerplusplus\ score std & Standard deviation of scores of 10 \ExoMinerplusplus\ models\\ 
\ExoMinerplusplus\ aggregate score & Aggregate scores computed on the top of tess-individual mean score\\ 
\ExoMinerplusplus\ label & `PC' if `\ExoMinerplusplus\ score'  $> 0.5$, `FP' otherwise\\ 
\ExoMinerplusplus\ aggregate label & `PC' if `\ExoMinerplusplus\ aggregate score' $> 0.5$, `FP' otherwise\\ 
DV full report & The URL to the DV full report in the MAST\\
DV summary report & The URL to the DV summary report in the MAST\\
DV mini report & The URL to the DV mini report in the MAST \\
\bottomrule
\end{tabular}
\end{tabularx}
\end{threeparttable}
\end{table}
\endgroup

\begingroup
\begin{table}[htb]
\centering
\caption{Scores and dispositions of TESS+\kepler-Aggregation model for TOIs that could be matched to TCEs using ephemeris matching procedure in~\ref{sec:data-preparation}. This table describes the available columns. The full table is available online.}
\label{table:vetting-catalogue-TOI}
\begin{threeparttable}
\begin{tabularx}{\linewidth}{@{}Y@{}}
\begin{tabular}{p{3cm}p{5cm}}
\toprule
Column & Description \\
\midrule
uid & Unique ID that includes TIC ID, planet number, and sector run\\
target\_id & TYIC ID\\
tce\_plnt\_num & TCE planet number\\
toi & TOI ID\\
tce\_period & TCE period\\
tce\_duration & TCE duration\\
tce\_prad & TCE planet radius (Earth Radii)\\
mes & TCE MES\\
TFOPWG Disposition (January 2024) & One of multiple TFOPWG Dispositions  downloaded in January 2024 if available for a TOI\\
TFOPWG Disposition (December 2024) & One of multiple TFOPWG Dispositions  downloaded in December 2024 if available for a TOI\\
score\_fold\_i & for $i\in[0,9]$, this is the score of \ExoMinerplusplus\ model trained for fold i\\ 
\ExoMinerplusplus\ score & Average score of 10 \ExoMinerplusplus\ models\\ 
\ExoMinerplusplus\ score std & Standard deviation of scores of 10 \ExoMinerplusplus\ models\\ 
\ExoMinerplusplus\ label & `PC' if `\ExoMinerplusplus\ score'  $> 0.5$, `FP' otherwise\\ 
DV full report & The URL to the DV full report in the MAST\\
DV summary report & The URL to the DV summary report in the MAST\\
DV mini report & The URL to the DV mini report in the MAST \\
\bottomrule
\end{tabular}
\end{tabularx}
\end{threeparttable}
\end{table}
\endgroup

\begin{table*}[htb!]
\centering
\footnotesize
\caption{List of 50 new CTOIs.}
\label{table:new-TOI-catalog}
\resizebox{.85\linewidth}{!}{
\begin{threeparttable}
\begin{tabularx}{\linewidth}{@{}Y@{}}
\begin{tabular}{clrcccccc}
\toprule
 Number & UID & Target ID & Planet Number & Period &  Duration & Planet Radius (Earth Radii) & MES & \ExoMinerplusplus\ Score \\
\midrule
 \rownumberctoi & 55655482-1-S14-60 &  55655482 &  1 &       3.96 &       0.11 &      13.58 &     229.30 &      0.993 \\ 
 \rownumberctoi & 207109417-1-S1-36 &  207109417 &  1 &       0.29 &       0.02 &       1.71 &      12.96 &      0.984 \\ 
 \rownumberctoi & 292547242-1-S39 &  292547242 &  1 &       7.14 &       0.06 &       7.09 &      10.15 &      0.983 \\ 
 \rownumberctoi & 46298321-1-S42-46 &  46298321 &  1 &       1.11 &       0.04 &       6.08 &      18.05 &      0.981 \\ 
 \rownumberctoi & 1715469667-1-S14-55 &  1715469667 &  1 &       6.50 &       0.23 &       7.85 &     103.10 &      0.976 \\ 
 \rownumberctoi & 233535738-2-S14-60 &  233535738 &  2 &       7.93 &       0.15 &       2.36 &      10.24 &      0.976 \\ 
 \rownumberctoi & 158561812-1-S14-55 &  158561812 &  1 &       3.69 &       0.03 &       6.08 &       9.59 &      0.973 \\ 
 \rownumberctoi & 241514551-2-S1-65 &  241514551 &  2 &       8.12 &       0.07 &       6.14 &       8.48 &      0.972 \\ 
 \rownumberctoi & 96202086-3-S14-60 &  96202086 &  3 &       0.63 &       0.02 &       0.89 &      10.09 &      0.970 \\ 
 \rownumberctoi & 396572386-1-S58 &  396572386 &  1 &       4.68 &       0.11 &      12.23 &      62.59 &      0.965 \\ 
 \rownumberctoi & 198178859-1-S14-55 &  198178859 &  1 &       7.31 &       0.14 &       1.34 &       8.04 &      0.956 \\ 
 \rownumberctoi & 66013259-1-S59 &  66013259 &  1 &       2.92 &       0.07 &      12.43 &      10.41 &      0.942 \\ 
 \rownumberctoi & 417829948-1-S14-50 &  417829948 &  1 &       8.95 &       0.12 &       2.17 &       7.41 &      0.942 \\ 
 \rownumberctoi & 224603921-1-S14-55 &  224603921 &  1 &      26.35 &       0.20 &       1.39 &       8.68 &      0.941 \\ 
 \rownumberctoi & 459928783-1-S14-50 &  459928783 &  1 &      11.07 &       0.11 &       2.96 &       8.53 &      0.936 \\ 
 \rownumberctoi & 309155144-1-S38 &  309155144 &  1 &       4.51 &       0.03 &       2.06 &       7.26 &      0.934 \\ 
 \rownumberctoi & 51079186-1-S13 &  51079186 &  1 &       5.29 &       0.11 &       2.75 &      10.18 &      0.928 \\ 
 \rownumberctoi & 270174158-1-S1-13 &  270174158 &  1 &       8.73 &       0.15 &       2.36 &       7.80 &      0.922 \\ 
 \rownumberctoi & 423454257-1-S1-65 &  423454257 &  1 &      12.60 &       0.11 &       1.28 &       8.18 &      0.917 \\ 
 \rownumberctoi & 193607307-1-S14-55 &  193607307 &  1 &       3.62 &       0.15 &       0.77 &       7.28 &      0.906 \\ 
 \rownumberctoi & 18018496-1-S14-50 &  18018496 &  1 &       6.25 &       0.09 &       1.95 &       8.25 &      0.906 \\ 
 \rownumberctoi & 96966437-1-S55 &  96966437 &  1 &       3.81 &       0.11 &      15.40 &      76.58 &      0.906 \\ 
 \rownumberctoi & 18068144-1-S14-50 &  18068144 &  1 &      14.25 &       0.15 &       3.05 &       8.28 &      0.890 \\ 
 \rownumberctoi & 5892614-1-S42-46 &  5892614 &  1 &      20.44 &       0.33 &      23.02 &      15.65 &      0.874 \\ 
 \rownumberctoi & 61710094-1-S1-36 &  61710094 &  1 &       1.84 &       0.06 &       1.91 &       8.58 &      0.873 \\ 
 \rownumberctoi & 20291519-1-S14-55 &  20291519 &  1 &       0.83 &       0.02 &       2.61 &      12.09 &      0.842 \\ 
 \rownumberctoi & 390201695-1-S14-26 &  390201695 &  1 &      10.88 &       0.19 &       3.32 &      10.38 &      0.839 \\ 
 \rownumberctoi & 239633605-1-S14-60 &  239633605 &  1 &       3.96 &       0.15 &       1.59 &       7.28 &      0.836 \\ 
 \rownumberctoi & 301969042-1-S41 &  301969042 &  1 &       0.88 &       0.09 &       3.10 &       8.27 &      0.829 \\ 
 \rownumberctoi & 258776466-1-S14-55 &  258776466 &  1 &       7.41 &       0.08 &       1.59 &      11.06 &      0.828 \\ 
 \rownumberctoi & 176218374-1-S1-46 &  176218374 &  1 &      13.28 &       0.12 &       2.16 &       7.28 &      0.827 \\ 
 \rownumberctoi & 117357458-1-S42-46 &  117357458 &  1 &       6.92 &       0.15 &       9.28 &       8.20 &      0.809 \\ 
 \rownumberctoi & 354103297-1-S17 &  354103297 &  1 &       2.78 &       0.11 &       3.59 &       9.55 &      0.784 \\ 
 \rownumberctoi & 147476037-1-S60 &  147476037 &  1 &       4.24 &       0.16 &       4.27 &       7.32 &      0.777 \\ 
 \rownumberctoi & 178645961-1-S1-65 &  178645961 &  1 &      23.53 &       0.57 &       0.90 &       8.22 &      0.756 \\ 
 \rownumberctoi & 97931135-1-S1-65 &  97931135 &  1 &       2.49 &       0.11 &       1.05 &       9.73 &      0.748 \\ 
 \rownumberctoi & 402313695-1-S1-46 &  402313695 &  1 &      14.86 &       0.12 &       1.89 &       8.74 &      0.741 \\ 
 \rownumberctoi & 278862747-1-S1-65 &  278862747 &  1 &      13.58 &       0.21 &       4.02 &      11.60 &      0.734 \\ 
 \rownumberctoi & 297262361-1-S14-60 &  297262361 &  1 &       7.78 &       0.14 &       0.79 &       8.56 &      0.732 \\ 
 \rownumberctoi & 295679570-1-S50 &  295679570 &  1 &       6.37 &       0.12 &       2.81 &       7.32 &      0.706 \\ 
 \rownumberctoi & 35228717-1-S42-46 &  35228717 &  1 &      11.68 &       0.19 &       4.29 &       8.52 &      0.687 \\ 
 \rownumberctoi & 257816591-1-S42-43 &  257816591 &  1 &       5.27 &       0.05 &       8.85 &       8.83 &      0.682 \\ 
 \rownumberctoi & 363459149-1-S42-46 &  363459149 &  1 &      18.24 &       0.12 &      18.20 &     124.83 &      0.682 \\ 
 \rownumberctoi & 63068354-1-S14-55 &  63068354 &  1 &      20.49 &       0.18 &       2.94 &      14.63 &      0.679 \\ 
 \rownumberctoi & 370977908-1-S1-65 &  370977908 &  1 &      17.09 &       0.22 &       2.39 &       7.32 &      0.678 \\ 
 \rownumberctoi & 196778101-1-S17 &  196778101 &  1 &       2.78 &       0.11 &       3.94 &       8.14 &      0.666 \\ 
 \rownumberctoi & 191702922-1-S58 &  191702922 &  1 &       3.32 &       0.14 &       3.52 &       7.37 &      0.642 \\ 
 \rownumberctoi & 410393041-1-S1-65 &  410393041 &  1 &      26.84 &       0.10 &       2.58 &       8.91 &      0.640 \\ 
 \rownumberctoi & 302609939-1-S1-46 &  302609939 &  1 &      49.62 &       0.13 &       2.04 &       8.96 &      0.635 \\ 
 \rownumberctoi & 274212635-1-S31 &  274212635 &  1 &       0.24 &       0.02 &       3.79 &      17.53 &      0.586 \\ 
\bottomrule
\end{tabular}
\end{tabularx}
\end{threeparttable}
}
\end{table*}

\section{Conclusion}
\label{sec:conclusion}

In this work, we introduced \ExoMinerplusplus, a machine learning-based model tailored for the classification and ranking of transit signals in TESS 2-minute cadence data. By leveraging multi-source learning, \ExoMinerplusplus\ exhibits robust performance in distinguishing planetary signals from false positives across a variety of challenging scenarios.

The model outputs both discrete classifications of TCEs as planets or non-planets and a continuous confidence score, providing a dual utility. This enables not only precise classification but also the development of a ranking-based vetting catalog for TESS 2-minute data. Such a catalog serves as a valuable resource for the astronomical community, streamlining the identification and prioritization of promising planetary candidates while minimizing the need for manual vetting.

Our analysis underscores several inherent challenges in classifying TESS data, including label noise, the effects of larger pixel sizes, crowded stellar fields, and systematic uncertainties. Despite these challenges, \ExoMinerplusplus\ demonstrates exceptional performance, particularly in addressing difficult subcategories. Its success stems from a combination of multi-source learning and innovative diagnostic tests, enhanced by advanced deep learning-based feature extraction capabilities.

\ExoMinerplusplus\ marks a significant advancement in automated exoplanet discovery, complementing human expertise with scalable, data-driven methodologies. Future work could extend this approach to other data sources, such as TESS Full-Frame Images (FFIs), and improve the handling of blended signals and misclassified subcategories. These advancements will pave the way toward fully automated validation of TESS exoplanets. As TESS continues to deliver an abundance of observational data, tools like \ExoMinerplusplus\ will be pivotal in maximizing the mission’s contributions to exoplanetary science.

\section*{Acknowledgments}
HV and MM are supported through TESS XRP 2022 contract 22-XRP22\_2-0173, NASA Academic Services Mission (NAMS) contract number NNA16BD14C as well as the Intelligent Systems Research and Development-3 (ISRDS-3) Contract 80ARC020D00100. DC and JT are supported through NASA Cooperative Agreement 80NSSC21M0079. VK acknowledges support from the youth scientific laboratory project, topic FEUZ-2020-0038. G.N. thanks for the research funding from the Ministry of Science and Higher Education program the ``Excellence Initiative - Research University" conducted at the Center of Excellence in Astrophysics and Astrochemistry of the Nicolaus Copernicus University in Toru\'n, Poland. The postdoctoral fellowship of KB is funded by F.R.S.-FNRS grant T.0109.20 and by the Francqui Foundation. 

We would like to thank multiple people who directly or indirectly contributed to this work. This paper includes data collected by the \kepler\ mission and obtained from the MAST data archive at the Space Telescope Science Institute (STScI). Funding for the \kepler\ mission was provided by the NASA Science Mission Directorate.  Resources supporting this work were provided by the NASA High-End Computing (HEC) Program through the NASA Advanced Supercomputing (NAS) Division at Ames Research Center for the production of the Kepler SOC and the TESS SPOC data products and for training our deep learning model, \ExoMinerplusplus. 

Funding for the TESS mission is provided by NASA's Science Mission Directorate. KAC and CNW acknowledge support from the TESS mission via subaward s3449 from MIT. This paper made use of data collected by the TESS mission and are publicly available from the Mikulski Archive for Space Telescopes (MAST) operated by the Space Telescope Science Institute (STScI). We acknowledge the use of public TESS data from pipelines at the TESS Science Office and at the TESS Science Processing Operations Center. This work makes use of observations from the LCOGT network. Part of the LCOGT telescope time was granted by NOIRLab through the Mid-Scale Innovations Program (MSIP). MSIP is funded by NSF. This paper is based on observations made with the Las Cumbres Observatory’s education network telescopes that were upgraded through generous support from the Gordon and Betty Moore Foundation. This paper is based on observations made with the MuSCAT3/4 instruments, developed by the Astrobiology Center (ABC) in Japan, the University of Tokyo, and Las Cumbres Observatory (LCOGT). MuSCAT3 was developed with financial support by JSPS KAKENHI (JP18H05439) and JST PRESTO (JPMJPR1775), and is located at the Faulkes Telescope North on Maui, HI (USA), operated by LCOGT. MuSCAT4 was developed with financial support provided by the Heising-Simons Foundation (grant 2022-3611), JST grant number JPMJCR1761, and the ABC in Japan, and is located at the Faulkes Telescope South at Siding Spring Observatory (Australia), operated by LCOGT. This research has made use of the Exoplanet Follow-up Observation Program (ExoFOP; DOI: 10.26134/ExoFOP5) website, which is operated by the California Institute of Technology, under contract with the National Aeronautics and Space Administration under the Exoplanet Exploration Program.

This material is based upon work supported by the NASA under Agreement No.\ 80NSSC21K0593 for the program ``Alien Earths''. The results reported herein benefitted from collaborations and/or information exchange within NASA’s Nexus for Exoplanet System Science (NExSS) research coordination network sponsored by NASA’s Science Mission Directorate.

This work makes use of observations from the ASTEP telescope. ASTEP benefited from the support of the French and Italian polar agencies IPEV and PNRA in the framework of the Concordia station program and from OCA, INSU, Idex UCAJEDI (ANR- 15-IDEX-01) and ESA through the Science Faculty of the European Space Research and Technology Centre (ESTEC). This research also received funding from the European Research Council (ERC) under the European Union's Horizon 2020 research and innovation program (grant agreement No. 803193/BEBOP) and from the Science and Technology Facilities Council (STFC; grant No. ST/S00193X/1).

We would also like to acknowledge the use of OpenAI’s ChatGPT~\citep{openai2024} in editing portions of this manuscript to improve clarity and writing quality.

\bibliography{AI_bib,ExoPlanet,ExoPlanetV2}{}
\bibliographystyle{aasjournal}


\appendix
\section{Details of optimized Architecture for ExoMiner++}
\label{sec:optimized_exominer}

We provide in detail the architecture and optimization parameters of \ExoMinerplusplus:
\begin{itemize}
    \item The `Difference Image' branch has three convolutional blocks each with three convolutional layers.  The three convolutional blocks have convolutional layers with 4, 8, and 16 filters, respectively. All convolutional layers have filters of size $3\times3\times1$ followed by a maxpooling layer with $2\times2\times1$.
    \item All other convolutional branches have two convolutional blocks each with three convolutional layers. The two convolutional blocks have convolutional layers with 8 and 16 filters, respectively. Transit-view convolutional branches have filters of size $6$ and $4$ for the convolutional and maxpooling layers, respectively. Full-orbit branches (and the `Periodogram' branch) have filters of size $5$ and $8$ for the convolutional and maxpooling layers, respectively.
    \item All convolutional layers use `same' padding (i.e., the feature map output is preserved), a stride of 1, and their weights are initialized randomly following He initialization~\citep{he2015delving}. All maxpooling layers use `valid' padding and a stride of 1.
    \item The Fully Connected (FC) layer at the end of each branch has 4 neurons and a dropout rate of 12.11\% (applied only during training).
    \item The final FC block has four FC layers, each with 512 neurons and with a dropout rate of 2.15\% (again, applied only during training). 
    \item All convolutional and FC layers are followed by a rectified linear unit activation~\citep[ReLU]{fukushima1969visual}.
    \item Learning rate was set to $4.18\mathrm{e}{-5}$.
\end{itemize}

\begingroup
\setlength{\tabcolsep}{2.5pt} 
\begin{table}[htb]
\centering
\caption{Hyperparameter Optimizer Parameters. Refer to~\cite{hpo-Falkner-2018} for the details regarding the meaning of these parameters for the BOHB optimizer.}
\label{table:hpo_parameters}
\begin{threeparttable}
\begin{tabularx}{\linewidth}{@{}Y@{}}
\begin{tabular}{lr}
\toprule
Parameter & Value(s) \\
\midrule
Budget (number of training epochs) & 6, 12, 25, 50\\
$\eta$ & 2\\
Top-n (\%) & 15\\
Random fraction & 0.3 \\
Number of samples & 64 \\
Bandwidth factor & 3\\
Minimum bandwidth & 0.001 \\
\bottomrule
\end{tabular}
\end{tabularx}
\end{threeparttable}
\end{table}
\endgroup

\begingroup
\setlength{\tabcolsep}{2.5pt} 
\begin{table}[htb]
\centering
\caption{Values/ranges of tested hyperparameters for \ExoMinerplusplus. All hyperparameters are integers except for the learning rate and the dropout rate hyperparameters which are real values.}
\label{table:hyperparameters_range}
\begin{threeparttable}
\begin{tabularx}{\linewidth}{@{}Y@{}}
\begin{tabular}{lrr}
\toprule
Hyperparameter & Value Ranges & Optimized Value\\
\midrule
Learning rate & $[1\mathrm{e}{-6}, 1\mathrm{e}{-1}]$ & $4.2\mathrm{e}{-6}$\\
Number of conv blocks (transit view) & [1, 5]& $2$\\
Number of conv blocks (full-orbit view) & [1, 5]& $2$\\
Number of conv layers per block & [1, 3] & $3$\\
Initial number of kernels & $2^x$ for integer $ x \in [2, 6]$ & $3$\\
Kernel size (transit view) & [1, 8] & $6$\\
Kernel size (full-orbit view) & [1, 8] & $5$\\
Pool size (transit view) & $2^x$ for integer $ x \in [1, 4]$ & $4$\\ 
Pool size (full-orbit view) & $2^x$ for integer $ x \in [1, 4]$ & $8$\\ 
Number of FC neurons in conv branch FC layer & [1, 4] & $3$\\
Dropout rate for conv branch FC layer & $1\mathrm{e}{-3}, 0.2$ & $0.12$\\
Number of FC layers in FC block & [1, 4] & $4$\\
Number of neurons in FC block layers & $2^x$ for integer $ x \in [5, 9]$ & $512$\\ 
Dropout rate for FC block layers & $1\mathrm{e}{-3}, 0.2$ & $0.02$\\
\bottomrule
\end{tabular}
\end{tabularx}
\end{threeparttable}
\end{table}
\endgroup

\end{document}